\documentclass[12pt,preprint]{aastex}

\slugcomment{c2d Lupus Synthesis \today}

\shorttitle{Lupus synthesis paper}
\shortauthors{Mer\'{\i}n et al.}

\begin{document}


\title{The Spitzer c2d Survey of Large, Nearby, Interstellar Clouds.
XI. \\Lupus Observed With IRAC and MIPS}


\author{Bruno Mer\'{\i}n\altaffilmark{1,2}, 
Jes J{\o}rgensen\altaffilmark{3},
Loredana Spezzi\altaffilmark{4}, 
Juan M.~Alcal\'a\altaffilmark{4}, 
Neal J. Evans II\altaffilmark{5}, 
Paul M. Harvey\altaffilmark{5}, 
Nicholas Chapman\altaffilmark{6},
Tracy Huard\altaffilmark{7},
Ewine F. van Dishoeck\altaffilmark{2},
Fernando Comer\'on\altaffilmark{8}
}

\altaffiltext{1}{Research and Scientific Support Department, European Space Agency (ESTEC), PO Box 299, 2200 AG Noordwijk, The Netherlands }
\altaffiltext{2}{Leiden Observatory, Leiden University, PO Box 9513, 2300 RA Leiden, The Netherlands }
\altaffiltext{3}{Argelander-Institut f\"ur Astronomie, University of Bonn,
Auf dem H\"ugel 71, 53121 Bon, Germany}
\altaffiltext{4}{INAF - Osservatorio Astronomico di Capodimonte, via Moiariello 16, I-80131, Naples, Italy}
\altaffiltext{5}{Department of Astronomy, University of Texas at Austin, 
	    1 University Station C1400 Austin, TX 78712-0259, USA}
\altaffiltext{6}{Astronomy Department, University of Maryland, College Park, 
	    MD 20742, USA}
\altaffiltext{7}{Smithsonian Astrophysical Observatory, 60 Garden Street, MS42, Cambridge, MA 02138, USA}
\altaffiltext{7}{European Southern Observatory, Karl-Schwarzschild-Strasse 2, 
85748 Garching bei M\"unchen, Germany}

\begin{abstract}
 We present c2d Spitzer/IRAC observations of the Lupus I, III and IV dark clouds 
  and discuss them in combination with optical and near-infrared and c2d MIPS data.  
  With the Spitzer data, the new sample contains 159 stars, 4 times larger than the previous
  one. It is dominated by low- and very-low mass stars and it is complete down to
  M $\approx$ 0.1M$_\odot$. We find 30-40\% binaries with separations between
  100 to 2000 AU with no apparent effect in the disk
  properties of the members. A large majority of the objects are Class II or Class III
  objects, with only 20 (12\%) of Class I or Flat spectrum
  sources. The disk sample is complete down to ``debris''-like systems in stars as
    small as M $\approx$ 0.2 M$_\odot$ and includes sub-stellar objects with
    larger IR excesses. The disk fraction in Lupus is 70 -- 80\%, consistent
    with an age of 1 -- 2 Myr. However, the young population contains 20\% optically
    thick accretion disks and 40\% relatively less flared disks. A growing variety of inner 
    disk structures is found for larger inner disk clearings for equal disk masses. 
    Lupus III is the most centrally populated and rich, followed by Lupus I with
    a filamentary structure and by Lupus IV, where a very high density core with little 
    star-formation activity has been found. We estimate star formation rates in Lupus 
    of 2 -- 10 M$_\odot$ Myr$^{-1}$ and star formation efficiencies of a few percent, 
    apparently correlated with the associated cloud masses.
    \end{abstract}


\section{Introduction}

The Lupus dark cloud complex is one of the main low-mass star forming
regions within 200 pc of the sun. Located in the Scorpius-Centaurus OB
association, it consists of a loosely connected group of dark clouds
and low-mass pre-main sequence stars located between galactic
coordinates 334$^{\rm o} < l < 352^{\rm o}$ and latitudes $+5^{\rm o}
< b < +25^{\rm o}$ (RA $\sim$ 16$^h$20$^m$ -- 15$^h$30$'$ and DEC
$\sim$ -43$^{\rm o}$00$'$ -- -33$^{\rm o}$00$'$ )
\citep{Krautter1991}.  The low ecliptic latitudes ($\sim$ -33 -- -41)
of the clouds guarantee a low abundance of asteroids. Given its large
size ($\sim$ 20$^{\rm o}$) and proximity, it has been subject of many
studies. Large-scale $^{\rm 12}{\rm CO}$ ($J = 1 \to 0$) millimeter
maps of the whole complex give total molecular gas masses of several
times 10$^4 M_\odot$ \citep{Murphy1986,Tachihara2001} and a relatively
small spread of cloud velocities among the different sub-clouds
($\leq$ 3 km s$^{-1}$ \citealt{Vilas-Boas2000}). It hosts four active
star-forming regions, including the rich T Tauri association in Lupus
3, plus five other looser dark clouds with signs of moderate
star-formation activity \citep{Hara1999}. Objects in all
evolutionary phases, from embedded Class I objects to evolved Class
III stars, are found in the Lupus clouds.  A comprehensive review about the
Lupus clouds by \cite{Comeron2007a} provides more details on previous
observations and is used throughout this work to assess the new results
from the Spitzer data. The Lupus dark cloud is one of the five large
nearby star-forming regions observed as part of the Spitzer Legacy
Project ``From Molecular Cores to Planet-forming Disks'' (c2d)
\cite{Evans2003}. From the several Lupus subclouds
\citep{The1962,Murphy1986}, only Lupus I, III and IV were observed by
c2d and are discussed in this paper.

Optical and near-infrared (near-IR) ground-based observations of Lupus
identified three regions, denominated as Lupus I, II and III, with
large numbers of classical T Tauri stars
\citep{Henize1954,The1962}. \cite{Schwartz1977} made a catalog of
H$\alpha$ emitting young stars which contained the vast majority of
the objects in the clouds known until today. His observations also showed
that Lupus 3 (roughly consistent with Cambr\'esy's Lupus III) is one
of the most active star-forming regions in the southern sky, followed
by Lupus I, II and IV and that a large number of the stars in these
clouds are low-mass stars. In addition to these classical T Tauri
stars, \cite{Krautter1997} reported a much larger number of weak-line
T Tauri stars with X-ray emission, which are believed to be older
stars not physically bound to the dark clouds but rather belonging to
a more nearby structure (\citealt{Wichmann1999}, see also
\citealt{Cieza2007} for a confirmation of this conclusion from c2d
observations). Finally, there is evidence for different distances to
the different subclouds: {\sl Hipparcos} parallaxes and extinction
source counts yield reasonable distance estimates of 150 $\pm$ 20 pc
for Lupus I and IV and 200 $\pm$ 20 pc for Lupus III (Comer\'on 2008
in prep. and references therein), which we assume for this work.


Following previous c2d standards, observational results from the
Spitzer Space Telescope (\citealt{Werner2004}) Infrared Array
Camera (IRAC, \citealt{Fazio2004}) and from the Multiband Imaging
Photometer for Spitzer (MIPS, \citealt{Rieke2004}) are reported
separately for each cloud, followed by a ``synthesis'' paper which
combines all the c2d Spitzer and complementary observations made of
the region (see e.g. \citealt{Harvey2006,Harvey2007a,Harvey2007b} for
Serpens or \citealt{Young2005,Porras2007}, and \citealt{Alcala2008}
for Chamaeleon II). For Lupus, the observational results obtained with
MIPS have been reported by \cite{Chapman2007} (Paper I) and this paper
describes the IRAC observations and analyzes the complete data set for
the region. In that sense, this paper merges the contents of the IRAC
and ``synthesis'' papers of the c2d observations in Lupus: it
  presents the 3.6 to 8.0 $\mu$m IRAC observations of the clouds for
  the first time and combines it with all previously available
  information from the optical to the millimeter.

  All the observations analysed in this study are presented in
  \S~\ref{observations}, with a discussion of the complementary 
  observations and stellar multiplicity in the
  optical in \S~\ref{binaries} and a detailed description of the
  Spitzer IRAC observations in \S~\ref{IRAC_BDP} and \ref{diffcounts}. 
  These data are used
  to construct a high-reliability catalog of young stars in the clouds
  in \S~\ref{YSOc_list}. This catalog is composed of Young Stellar
  Objects (YSO) identified with the Spitzer data (selected in
  \S~\ref{yso_sel}) and previously known Pre-Main Sequence stars (PMS,
  discussed in \S~\ref{pms_sel}). The rest of the paper is organized
  in two main sections, which deal with the individual sources
  (\S~\ref{disks}) and with the global cloud properties
  (\S~\ref{clouds}), respectively. In the first one, the disk
  populations are studied with color-color diagrams
  (\S~\ref{colorcolor}), multi-wavelength Spectral Energy
  Distributions (SEDs, \S~\ref{SEDs}), stellar and disk luminosity
  functions (\S~\ref{star_luminosities} and
  \ref{fractional_luminosities}), and finally with a new `2D'
  classification system of the SEDs (\S~\ref{2D_param}). The Spitzer
  observations of outflows and Herbig-Haro objects in the region are
  also described in \S~\ref{outflow_sources}. The second section
  describes the structures of the clouds with the help of
  Spitzer-derived extinction maps (\S~\ref{extinction}), the spatial
  distribution of the YSOs in the clouds and a nearest-neighbor
  analysis of their clustering properties
  (\S~\ref{spatial_distribution}). Finally, the Star Formation Rates
  and Star Formation Efficiencies are computed and discussed in
  \S~\ref{SFR} and \ref{SFE}, respectively. A complete summary
  of this work is given in \S~\ref{summary}.

\section{Observations and data analysis}
\label{observations}

The observations discussed in this paper come from the Spitzer Space
Telescope's Infrared Array Camera (IRAC hereafter) and Multiband
Imaging Photometer for Spitzer (MIPS) observations of Lupus I, III and
IV made by the c2d Spitzer Legacy Program \citep{Evans2003}, together
with an optical coordinated survey of the three clouds (F. Comer\'on
et al., in preparation) and data from the literature. Detailed
information about the MIPS observations can be found in
\cite{Chapman2007}. 

\subsection{Complementary data and multiple visual systems}
\label{binaries}

An optical survey of the three clouds was performed with the
Wide-Field Imager (WFI) attached to the ESO 2.2m telescope, at La
Silla Observatory in Chile. The areas observed in the $R_C$ (0.652
$\mu$m), $I_C$ (0.784 $\mu$m), and $z_{WFI}$ (0.957 $\mu$m) optical
bands in Lupus I, III and IV were defined to overlap completely with
the areas observed with Spitzer shown in Figures \ref{lupI} to
\ref{lupIV}.  The observational strategy, data reduction and source
extraction are described in detail elsewhere (F. Comer\'on et al., in
preparation). These observations were complemented
with photometry from the {\sl NOMAD} optical and near-IR catalog
\citep{Zacharias2005}, which contains $B$, $V$ and $R_C$ magnitudes
and from the $J$, $H$ and $K_S$ 2MASS near-IR all-sky catalog
\citep{Cutri2003}. This paper also includes the c2d MIPS observations of the Lupus
clouds. Figures \ref{spa_distr_I} to \ref{spa_distr_IV} also show the
areas mapped with MIPS at 24, 70 and 160 $\mu$m as part of the c2d
project. They were defined to provide the maximum overlap and minimum
observing time and cover completely the areas mapped with IRAC. 
The MIPS data acquisition, reduction and
source extraction is presented and discussed in detail in
\citet{Chapman2007}.

The optical images were inspected to search for visual binaries in
the sample. Table \ref{binaries_tab} reports all
apparent companions detected in the images. The WFI images in $Rc$,
$Ic$ and $z_{WFI}$ bands were inspected for all sources. Those
observations were performed in service mode over several nights in two
different observing seasons and have a range of seeing values, but an
average seeing of 1.5$^{''}$ limits the smallest separation detectable
to $\sim$ 0.7$^{''}$ in our case. The range of separations for which
we report the presence of possible companions is 0.7 to 10$^{''}$ (140
to 2000 AU in Lupus III and 105 to 1500 AU at the distance of Lupus I
and IV).

There are several pairs of stars with disks which are binaries. Inspection 
of the optical
images indicates that the probability of projection effects might be
small given the relatively low number of optical sources at distances
smaller than 10$^{''}$ in most of the examined stars. We recover all
binaries listed by \cite{Ghez1997} which fell in our field except HR
5999, which was too bright to allow a shape analysis. We also recover
the two objects in common with a yet unpublished AO/ADONIS survey of
multiplicity in Lupus which probe at angular distances as close as
0.2$^{''}$ (A. Knockx et al., in prep.). The total binary fractions of
41$\pm$12\% (7/17), 29$\pm$9\% (39/124) and 44$\pm$13\% (8/18) in the
three clouds, respectively, compare well with the numbers given for
Taurus and Ophiuchus in a similar range of angular distances (e.g.
\citealt{Padgett1997}).

In any case, given the difficulty in apportioning the excess among the
companions with separations smaller than 2.0$^{''}$, in which the IRAC
fluxes could have been merged, and those with separations smaller than
4.0$^{''}$, which could have merged fluxes in MIPS, we set those
fluxes as upper limits in the SEDs for the disk evolution studies.
More discussion on the effects of binarity in the disks is presented
in \S~\ref{fractional_luminosities}.

\subsection{Spitzer IRAC data}
\label{IRAC_BDP}

Lupus I, Lupus III and Lupus IV were observed with all IRAC bands
(3.6, 4.5, 5.8 and 8.0 $\mu$m) on the 3rd and 4th of September of 2004
as part of the c2d Spitzer Legacy Program (P.I.: N. Evans, program ID:
174) and as part of GTO observations (P.I.: G.  Fazio, program ID:
6). The total observed areas were defined to cover all the
regions with extinctions $A_V \ge 2$ for Lupus III and IV and $A_V \ge
3$ for Lupus I \citep[see Fig.~2 of][]{Evans2003} as measured in the
extinction maps by \cite{Cambresy1999}. The combined mosaics of the three regions overlap in
all IRAC bands with areas of 1.39, 1.34 and 0.37 deg$^2$ in Lupus I,
Lupus III and Lupus IV, respectively. Furthermore 4 off-cloud regions,
each of $\sim$ 0.08 deg$^2$ were observed in low-extinction regions
with a range of galactic latitudes for statistical comparison with the
on-cloud fields. Each of them contains a 0.05 deg$^2$ overlap between
all IRAC bands. All c2d maps were observed in two epochs at least 6
hours apart to guarantee the production of highly reliable catalogs of
sources clean of asteroids and transient artifacts. The GTO
observation of Lupus III was only observed once. For more information
about the c2d mapping strategies, methods and results, consult
the delivery documentation \citep{Evans2007}.

Table \ref{obslog} shows the details of the individual pointings for
all three clouds and the off cloud regions. Figures \ref{spa_distr_I} to
\ref{spa_distr_IV} show the different coverages of the c2d IRAC, MIPS 
and optical mosaic areas overlaid on the optical extinction map of Lupus
from \cite{Cambresy1999}.

\begin{table}
\caption{c2d Spitzer IRAC Observations summary in Lupus}             
\begin{tabular}{l c c c c}        
\hline\hline                 
Field & Position ($\alpha$,$\delta$)$_{J2000}$ & Date (2004) & AOR epoch 1 & AOR epoch 2 \\\hline
\multicolumn{5}{c}{\sl Lupus I c2d Observations} \\
LupI\_1 & (15:45:09.0, -34:23:11.0) & Sept 3 & 0005717248 & 0005719808 \\
LupI\_2 & (15:43:26.0, -34:06:27.0) & Sept 3 & 0005717504 & 0005720064 \\
LupI\_2a& (15:42:23.0, -34:00:55.0) & Sept 3 & 0005717760 & 0005720320 \\
LupI\_3 & (15:41:16.0, -33:43:33.0) & Sept 3 & 0005718272 & 0005720832 \\
LupI\_4 & (15:40:08.0, -33:35:16.0) & Sept 3 & 0005718528 & 0005721088 \\
LupI\_5 & (15:38:34.0, -33:20:18.0) & Sept 3 & 0005718784 & 0005721344 \\
LupI\_6 & (15:39:09.0, -34:23:11.0) & Sept 3 & 0005717248 & 0005721600 \\
LupI\_7 & (15:40:34.0, -34:36:34.0) & Sept 3,4 & 0005719296 & 0005721856 \\
LupI\_8 & (15:38:08.0, -34:39:27.0) & Sept 3 & 0005719552 & 0005722112 \\
LupI\_9 & (15:42:18.0, -34:33:50.0) & Sept 3 & 0005737472 & 0005737728 \\
LupI\_10 & (15:40:55.0, -34:06:31.0) & Sept 3,4 & 0005739008 & 0005739264 \\
\multicolumn{5}{c}{\sl Lupus III c2d Observations} \\
LupIII\_1 & (16:12:28.0, -38:05:43.0) & Sept 4 & 0005724416 & 0005725696 \\
LupIII\_2 & (16:10:35.0, -37:50:04.0) & Sept 4 & 0005724672 & 0005725952 \\
LupIII\_3 & (16:11:59.0, -38:58:30.0) & Sept 4 & 0005724928 & 0005726208 \\
LupIII\_4 & (16:10:32.0, -39:07:18.0) & Sept 4 & 0005725184 & 0005726464 \\
LupIII\_6 & (16:07:24.0, -39:07:03.0) & Sept 4 & 0005725440 & 0005726976 \\
LupIII\_7 & (16:12:45.0, -38:31:31.0) & Sept 4 & 0005737984 & 0005738240 \\
LupIII\_8 & (16:10:38.0, -38:35:24.0) & Sept 4 & 0005738496 & 0005738752 \\
\multicolumn{5}{c}{\sl Lupus III GTO \& c2d Observations} \\
LupIII\_5 & (16:08:55.8, -39:08:33.2) & Sept 4 & 0003652608 & 0005726720 \\
\multicolumn{5}{c}{\sl Lupus IV c2d Observations} \\
LupIV\_1 & (16:02:46.0, -41:59:49.0) & Sept 4 & 0005728768 & 0005729536 \\
LupIV\_2 & (16:01:27.0, -41:41:00.0) & Sept 4 & 0005729024 & 0005729792 \\
LupIV\_3 & (16:00:43.0, -42:02:37.0) & Sept 4 & 0005729280 & 0005730048 \\
\multicolumn{5}{c}{\sl Off-Cloud c2d Observations} \\
LupOC\_2 & (16:13:00.0, -34:00:00.0) & Sept 4 & 0005731072 & 0005735680 \\
LupOC\_6 & (16:40:00.0, -40:30:00.0) & Sept 4 & 0005732096 & 0005736704 \\
LupOC\_7 & (15:50:00.0, -41:00:00.0) & Sept 3 & 0005732352 & 0005736960 \\
LupOC\_8 & (16:07:00.0, -43:00:00.0) & Sept 4 & 0005732608 & 0005737216 \\
\hline                                   
\end{tabular}
\label{obslog}
\end{table}


The IRAC images were processed by the Spitzer Science Center (SSC)
using the standard pipeline version S13 to produce the Basic
Calibrated Data (BCDs). These images were then processed and combined
into mosaics and source catalogs by the c2d pipeline version
2007/January. \cite{Harvey2006} and \cite{Jorgensen2006} describe in
detail the processing of the IRAC images for Serpens and Perseus,
respectively. The observations presented here were processed in
exactly the same way so we refer the reader to those articles and to the
Delivery Documentation for details about the data
reduction. Here we will just describe briefly the basic steps carried
out to produce the presented images and catalogs.

In total 63384, 138270 and 42737 sources were detected with at least
IRAC in Lupus I, III and IV, respectively. Table \ref{detection_stat}
summarizes the number of sources detected in the three surveyed areas
with S/N of at least 5. This corresponds to selecting all sources with
detections of quality `A' or `B' in any of the IRAC bands from the
delivered c2d catalogs. Most of the sources in this
catalog can be well fitted with reddened stellar atmospheres from
field stars. IRAC bands 1 and 2 are the most sensitive in all three
catalogs and produce an average of 7 times more high quality
detections than those in IRAC bands 3 and 4. About 300, 400, and 350
sources in Lupus I, III and IV respectively of the four band
detections are found to be extended in one or more bands (classified
with the ``image type'' 2 in the c2d catalogs). Our final catalogs
contain 2776, 7539 and 2403 four-band IRAC sources in the three Lupus
clouds, respectively (see Table \ref{detection_stat}).

\begin{table}[h]
  \caption{Total numbers of IRAC sources detected with S/N $\ge$ 5$\sigma$ in Lupus }\label{detection_stat}
\begin{tabular}{llll}\hline\hline
                                    & Lupus I   & Lupus III & Lupus IV \\\hline
Detection in at least one IRAC band & 63384   & 138270 & 42737  \\
Detection in all four IRAC bands       & 2834      & 7638      & 2418  \\
Detection in three IRAC bands           & 3012      & 6559     & 2001  \\
Detection in two IRAC bands           & 29301     & 67039    & 21125 \\
Detection in one IRAC band            & 28237     &  57034   & 17193 \\
\\
Detection in 2MASS only$^a$         & 0         & 7        & 1  \\
Detection in IRAC only              & 56219     &  122166  & 37904 \\
Detection in 4 IRAC bands and not 2MASS$^a$ &136  & 268 & 80 \\\hline
{\sl Excluding extended sources} &&&\\
Four band detections   &   2776 & 7536 & 2403 \\
Four band detections with 2MASS association$^a$ & 2637 & 7244  & 2262 \\
Detected in IRAC1+2 and 2MASS$^a$ & 6913 & 15702 & 4581 \\\hline
\end{tabular}

$^a$ A source is counted as detected in 2MASS if it has a S/N of at least
10 in both $H$ and $K_s$.
\end{table}

\begin{figure}
   \centering
   \includegraphics[angle=0,width=17cm]{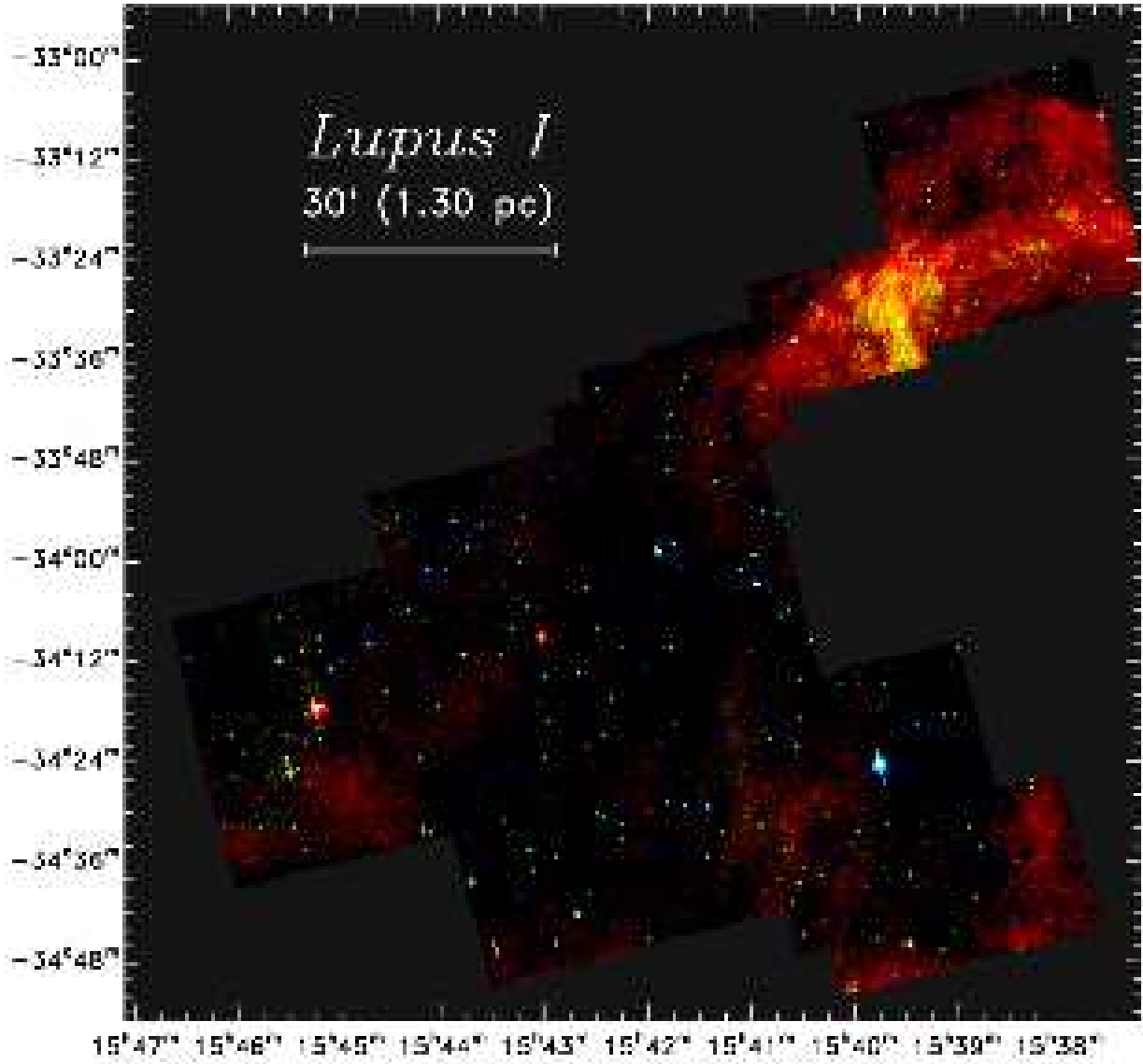}
   \caption{Color-composite image of the mapped area in Lupus I by c2d.  The
     color mapping is blue for IRAC2 at 4.5 $\mu$m, green for IRAC4 at
     8.0 $\mu$m and red for MIPS1 at 24 $\mu$m. The bright red
       emission in the North-West of the cloud is produced by cold
       dust.}  \label{lupI}
   \end{figure}

\begin{figure} \centering
   \includegraphics[angle=-90,width=17cm]{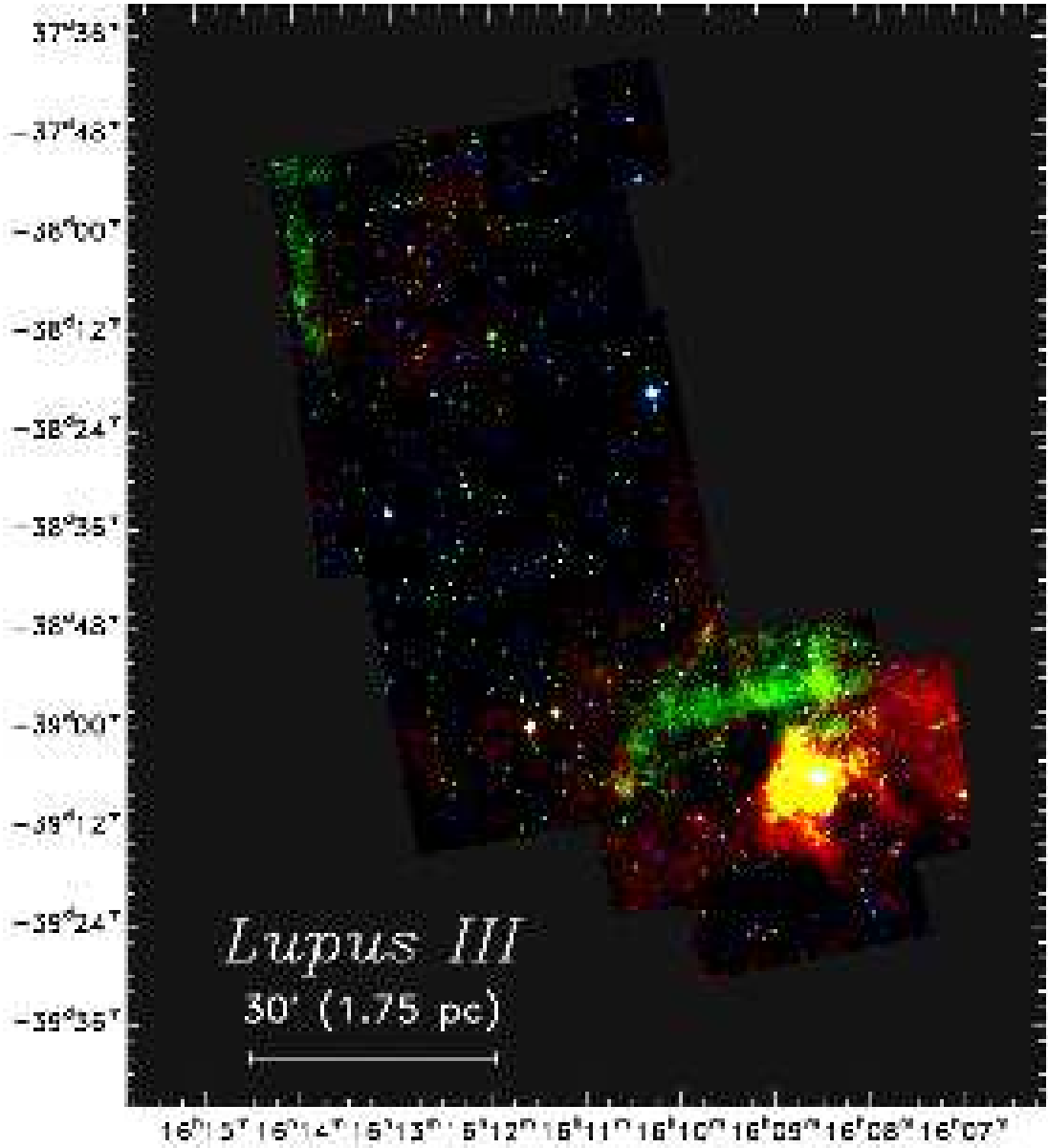}
   \caption{Color-composite image of the mapped area in Lupus III by c2d.  The
     color mapping is the same as in Figure \ref{lupI}. The figure
     shows the very active star-forming core, which contains the two
     Herbig Ae/Be stars HR 5999 and HR 6000 at RA $\sim$ 16$^h$09$^m$
     and DEC $\sim$ -39$^{\rm o}$10$'$ and a stream of PAH emission to
     the North of the core (RA $\sim$ 16$^h$10$^m$ -- 16$^h$08$^m$ and
     DEC $\sim$ -39$^{\rm o}$).}  \label{lupIII} \end{figure}

\begin{figure} \centering
   \includegraphics[angle=-90,width=17cm]{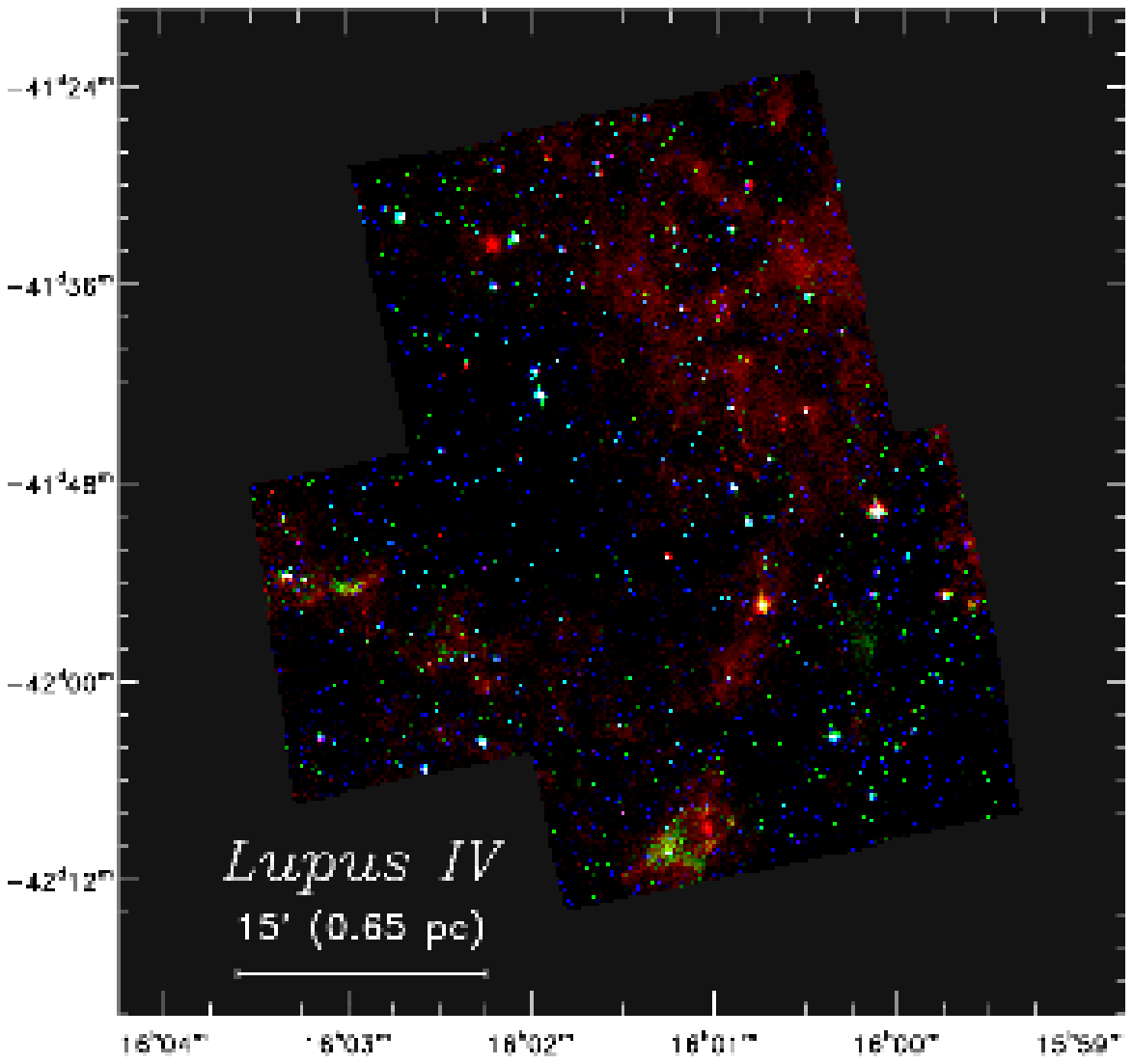}
   \caption{Color-composite image of the mapped area in Lupus IV by c2d.  The
     color mapping is the same as in Figure \ref{lupI}. The figure
     shows the remnant cloud structure emitting at long wavelengths
     and many young stellar objects covering a large range of
     evolutionary types.  The very red source in the South of the
     mosaic is the unrelated Planetary Nebula PN
     G336.3+08.0.}  \label{lupIV} \end{figure}

Figures \ref{lupI} to \ref{lupIV} show RGB color-composite images of
Lupus I, III and IV respectively, with IRAC2 band at 4.5 $\mu$m in
blue, IRAC4 band at 8 $\mu$m in green and MIPS1 band at 24 $\mu$m in
red. The images show the overlapping areas between the mosaics done
with both instruments and include the GTO-imaged area in the core of
Lupus III at RA $\sim$ 16$^h$09$^m$ and DEC $\sim$ -39$^{\rm o}$10$'$
\citep{Allen2007}. The images show a large and dense concentration of
cold dust in the North-West of Lupus I and South of Lupus III. The
cloud emissions in Lupus are consistent with the Cambr\'esy's
extinction maps, but now shown in much more detail. A general 8 $\mu$m
emission gradient coming from the nearby Galactic plane was
compensated in the color scales to bring up the cloud structure. The
bright green emission to the North-East of the core in Lupus III is
likely produced by interstellar PAH molecules, which emit strongly at
8 $\mu$m, possibly illuminated by the two bright Herbig Ae/Be stars HR
5999 and HR 6000 in the core.

\subsection{Differential source counts}
\label{diffcounts}

\begin{figure}
\resizebox{\hsize}{!}{\includegraphics{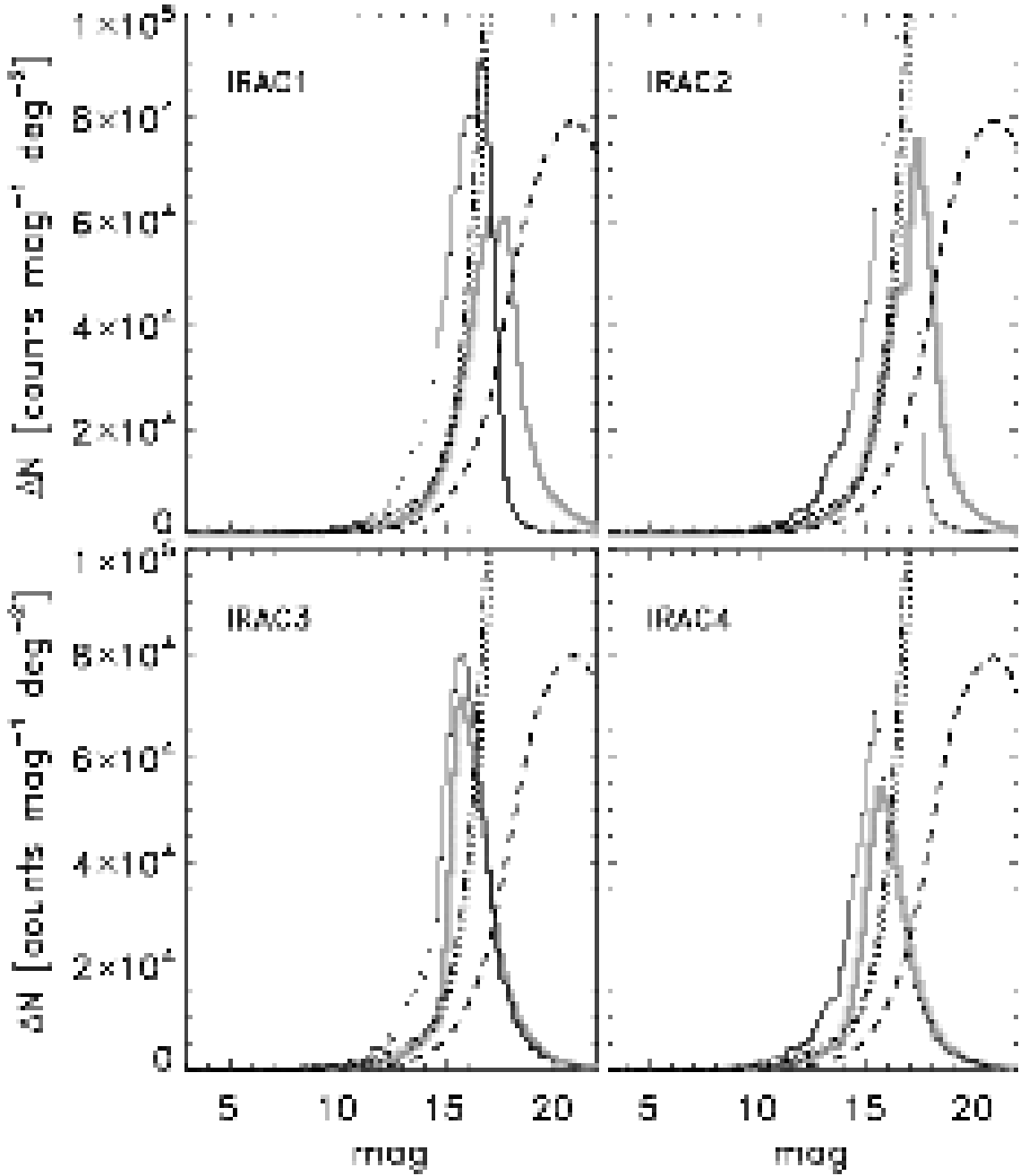}}
\caption{Differential source counts for the on- and off-cloud regions
  (grey and black, respectively). The predictions from the Wainscoat
  models are shown with the dashed lines for each of the three parts
  of the cloud (Lupus IV in dot-dashed line, Lupus III in dotted line
  and Lupus I with dashed line).} \label{diffcount}
\end{figure}

\begin{figure}
\resizebox{\hsize}{!}{\includegraphics{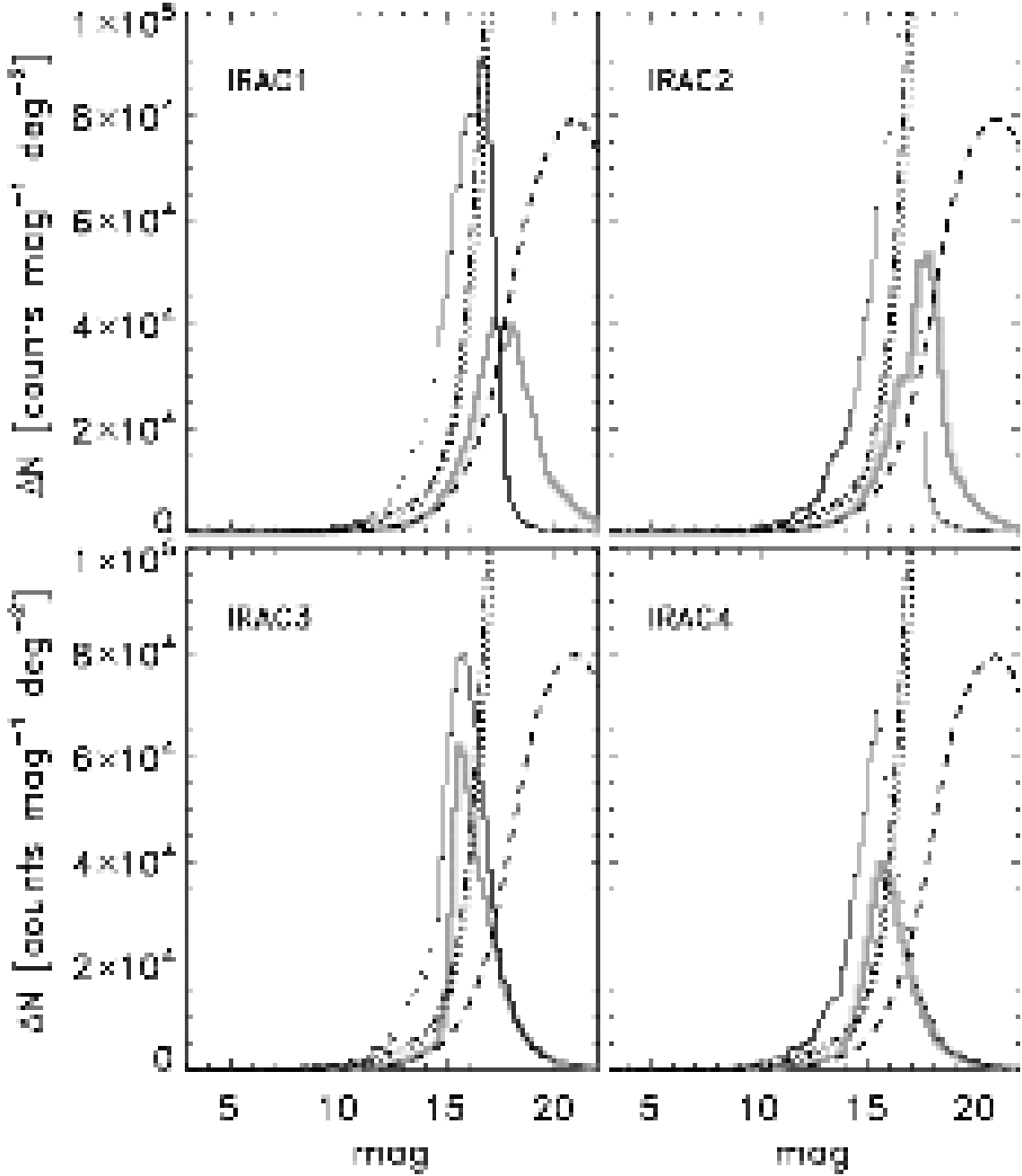}}
\caption{As in Fig.~\ref{diffcount} but here only for the Lupus I
  cloud.} \label{diffcount1}
\end{figure}

\begin{figure}
\resizebox{\hsize}{!}{\includegraphics{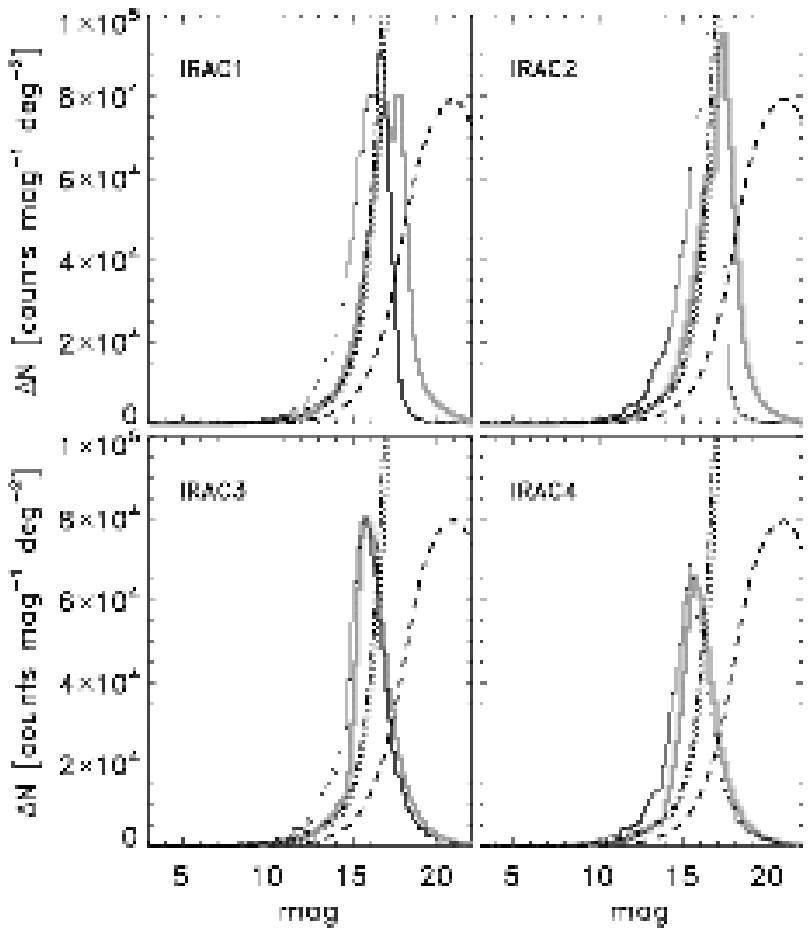}}
\caption{As in Fig.~\ref{diffcount} but here only for the Lupus III
  and IV clouds.} \label{diffcount2}
\end{figure}

Given the proximity to the Galactic plane and center (see
Table~\ref{cloudinfo}), the Lupus clouds are expected to show a
relatively high number of background stars compared to the other
clouds surveyed by c2d. Figures \ref{diffcount} to \ref{diffcount2}
show the differential source counts for the Lupus regions. Since Lupus
I, at $+17^\circ$, is further from the Galactic plane than Lupus III
and IV (at 8--9$^\circ$) and likewise the two off-cloud fields with
both IRAC and MIPS observations are at $+4^\circ$ and $+12^\circ$
closer to the Galactic plane than the on-cloud fields on average, we
expect larger differences between the counts of individual regions
compared to the other c2d clouds. Figures \ref{diffcount1} and
\ref{diffcount2} therefore separate the Lupus I cloud from the III and
IV clouds. The observed differential source counts are compared to the
predictions of the Galactic source counts by \cite{Wainscoat1992}. In
all plots a steady increase is seen in number of sources per magnitude
bin up to a break at about 16--16.5~magnitudes in IRAC bands 1 and 2
and 15--15.5 in bands 3 and 4 corresponding roughly to the sensitivity
levels of the surveys.

\begin{table}
  \caption{Ecliptic and galactic coordinates for the centers of the mapped IRAC regions and their areas.}\label{cloudinfo}
\begin{tabular}{lccc}\hline\hline
  & $(\alpha,\delta)$ & $(l,b)$ & area  \\
  &  deg              & deg     & deg$^2$ \\\hline
  Lupus I    & 235.4 ; $-$34.2   & 338.6 ; 16.7 & 1.391 \\
  Lupus III  & 242.7 ; $-$38.6   & 340.2 ;  9.4 & 1.336 \\
  Lupus IV   & 240.4 ; $-$41.8   & 336.7 ;  8.2 & 0.374 \\
  OC2        & 243.2 ; $-$34.1   & 343.7 ; 12.4 & 0.051 \\
  OC6        & 250.0 ; $-$40.6   & 342.7 ;  4.0 & 0.051 \\ \hline
\end{tabular}
\end{table}

Some interesting differences are seen between the shorter wavelength
IRAC bands 1 and 2 compared to the longer wavelength bands 3 and 4. In
the shorter bands the match between the on- and off-cloud counts is
poor with the off-cloud fields showing significantly more sources in
given magnitude bins than do the on-cloud fields; consistent with
their proximity to the Galactic plane. The on-cloud source counts are
clearly higher in the Lupus III and IV regions than the Lupus I
region. In each plot the observed on-cloud source counts at IRAC bands
1 and 2 are traced reasonably well by the Wainscoat model
predictions. These effects all suggest that the source counts in the
shorter IRAC bands are dominated by background stars as expected. In
the longer IRAC bands the situation is slightly different: both the
off-cloud and on-cloud source counts are here seen to exceed the
predictions of the Wainscoat models at faint magnitudes. Also the
difference between the different on-cloud regions internally or
compared to the off-cloud regions are found to be smaller than in IRAC
bands 1 and 2. This is likely due to the longer wavelength source
counts starting to be dominated by the extra-galactic background,
which naturally is independent of the Galactic location.

Compared to other c2d clouds, the differential source counts towards
the Lupus clouds show a similar behaviour to that seen in Serpens,
where the source counts exceed the galactic models in IRAC bands 3 and
4 and match better in IRAC bands 1 and 2 \citep{Harvey2006}. Both
clouds are the closest to the direction of the galactic centre, with
the Lupus off-cloud fields being the closest of the whole c2d data set
(Table \ref{cloudinfo}). The source counts in the Chamaleon II (Cha II
hereafter) and Perseus clouds, well away from the Galactic plane, show
good matches with the Wainscoat models at all bands
(\citealt{Porras2007} and \citealt{Jorgensen2006}, respectively). This
illustrates the strong dependence of the source counts on galactic
coordinates and suggests the presence of a larger number of mid-IR
sources close to the Galacic center than those predicted by the
Wainscoat models. Of course, the excess of mid-IR source counts
towards the Lupus clouds could also be partially attributed to the
presence of YSO's in these clouds in cases where the excesses are
larger in the on-cloud than in the off-cloud regions. This would also
imply a larger number of YSO's in Lupus III and IV than in Lupus I.


\section{Young stellar objects and pre-main-sequence stars in Lupus}
\label{YSOc_list}

This section presents and discusses the complete list of Young Stellar
Objects (YSOs hereafter) and Pre-Main Sequence (PMS) stars in the
three Lupus clouds, obtained from the new Spitzer observations. The
final list of objects is obtained by merging the IR-excess sources
from Spitzer (YSOs) with the list of all other known young stars in
the clouds from previous optical and near-IR surveys inside the
Spitzer covered area (PMS stars). These two c2d standards were
introduced in  \citet{Harvey2007a} and \citet{Alcala2008}. The full 
list of YSOs and PMS stars is given in Table \ref{YSO_list1}.

\subsection{Selection of YSOs with IRAC and MIPS data}
\label{yso_sel}

We call an object a Young Stellar Object candidate
(YSOc) if it appears in the catalog of such objects delivered in the
c2d delivery \citep{Evans2007}. This
sample of objects was also visually inspected to subtract suspect
galaxies and to add obvious YSOs missed by the c2d criteria either
because they were too faint at short wavelengths, but associated with
millimeter emission, or because they were saturated in the Spitzer
images.  The objects in the resulting list are called simply YSOs;
however some caution should still be exercised.  

\begin{figure}
\centering \includegraphics[width=12cm]{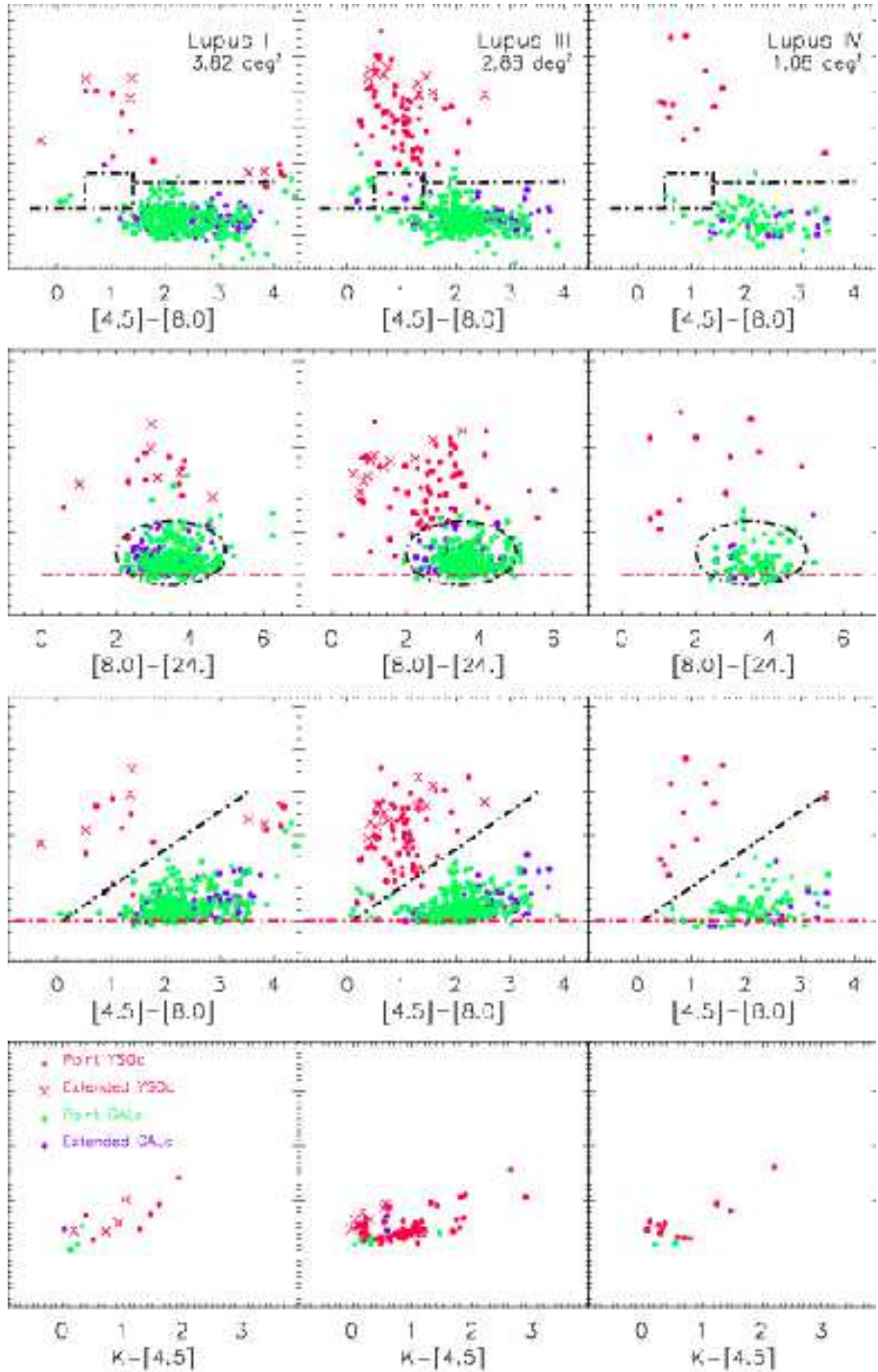}
\caption{Color-magnitude and color-color diagrams for the Lupus I
  (left), III (center) and IV (right) clouds. The black dot-dashed
  lines show the ``fuzzy'' color-magnitude cuts that define the YSO
  candidate criterion \citep{Harvey2007a} in the various
  color-magnitude spaces. The red dot-dashed lines show hard limits,
  fainter than which objects are excluded from the YSO
  category.} \label{YSO_sel}
\end{figure}

Complete details on the YSO selection method can be found in the c2d
Delivery Documentation \citep{Evans2007} and in \citet{Harvey2007a},
but in short, it consists in the definition of an empirical
probability function which depends on the relative position of any
given source in several color-color and color-magnitude diagrams where
diffuse boundaries have been determined to obtain an optimal
separation between young stars and galaxies. For that comparison, the
SWIRE catalog \citep{Surace2004} was ``extincted'' and resampled to
match as accurately as possible the c2d sensitivity limits for each
star forming cloud and to provide the statistical color distributions
expected for the background galaxies in our fields. The filter also
includes a flux cut-off to exclude bright galactic post-AGB stars in
the background which resemble Class III objects in the cloud. Figure 
\ref{YSO_sel} shows the Spitzer color-color
and color-magnitude diagrams used to select the YSOs in Table
\ref{YSO_list1}.


The application of this method in the Lupus
catalogs yields 18, 69 and 12 YSO candidates in Lupus I, III and IV
respectively (shown in Figure \ref{YSO_sel} as red dots and crosses
for point- and extended sources, respectively). Out of these, only 4
(22 \%), 26 (38 \%) and 2 (16 \%) were already known from previous
ground-based optical and near-IR observations. This shows that the
space observatory, with its highly sensitive detectors, has multiplied
by 5, 4, and 6 the number of known YSOs in the three clouds
respectively, capturing very low-mass young objects which escaped
detection due to low sensitivities and also getting the stars with
moderate excess only in the mid-IR which were not identified as excess
sources in near-IR surveys. 

The optical, IRAC and MIPS images of all the objects in the sample
were examined to confirm the point-source nature and to search for
blended features or long-period visual binaries. In Lupus I we
identified two galaxies, clearly resolved in the optical data, which
corresponded to two sources each in the Spitzer bands. Interestingly,
all of them were classified as 'YSOc\_PAH\_em' in the Spitzer catalog
of the region, due to their strong 8 $\mu$m emission due to PAHs (see
the delivery document for a description of these classes). No other
resolved galaxies were found in the optical images of all the other
YSO candidates in Lupus III and IV. However, a fake source was found
in the extended IRAC emission around the embedded object IRAS
15356-3430 (see also \S~\ref{outflow_sources}). These sources were taken
off our YSO list and reduced the list of YSOs in Lupus I from 18 to
13. 

\subsection{Sample of PMS and PMS candidate stars in Lupus}
\label{pms_sel}

We will use the term Pre-Main Sequence (PMS) star for other objects 
added to the list whose youth had already been confirmed using other 
observational techniques, mostly optical spectroscopy. 
If an object has not been spectroscopically confirmed as young but it was 
selected by its optical and near-IR photometry as such it will be labeled as 
a PMS candidate. 

We have added to our list the PMS candidates found by
\citet{Chapman2007} in the MIPS-only covered area. Because the MIPS coverage was much more extensive than the IRAC
coverage, there are sources that have clear excesses over photospheres
based on 2MASS and MIPS data, but for which we have no IRAC data
\citep{Chapman2007}. They would have been classified as YSOs, but
 since they cannot be tested against the galaxy
filter, we do not include them as YSOs or even YSOc. They are listed
as PMS candidates. The embedded class 0 
source Lupus3MMS, detected by \cite{Tachihara2007} at 1.2 mm, was 
added to the YSO sample in Lupus III. This object was detected at all 
IRAC and MIPS bands, but with too
poor S/N in the shorter wavelengths to be classified as a YSO
automatically. 

Also, we added all
previously known members of Lupus cited in Comer\'on (2008, in prep.),
which come mostly from the list of H$\alpha$-selected PMS stars in
Lupus of \citet{Schwartz1977} after the revisions of
\citet{Krautter1991} , \citet{Hughes1994} and \citet{Comeron2003}. This only added to our YSO
list the PMS stars without any detectable IR excess with good enough S/N.
It must be noted that the multi-color
criteria described in the previous section already recovered the 28
Classical T Tauri stars in the cloud listed in Comer\'on 2008 (in
prep.)  with clear IR excess in the Spitzer wavelength range and
good detections in all IRAC and MIPS1 bands and did not select the
other 13 from that list that did not show detectable IR
excesses. Interestingly, the only two bona-fide CTTs with IR excess
but not selected as YSOs by our criteria are in high background
emission regions and therefore have bad quality detection in one or
more Spitzer bands.

The candidate PMS stars from the previous optical and near- and mid-IR
surveys of \citet{Nakajima2000}, \citet{LopezMarti2005},
\citet{Allers2006}, and \citet{Allen2007} were also added to our list
of objects. These sources are labeled as PMS candidates in the 
``PMS'' column of Table \ref{YSO_list1} and their respective 
original references are given in column ``Sel''. The
percentages of candidates from these studies that show IR excess in
the Spitzer bands are 6/18 (33 \%), 5/15 (33 \%), 1/3 (33 \%), and
16/16 (100 \%), respectively. According to this, the H$\alpha$
narrow-band imaging criteria by \citet{LopezMarti2005} have a success
rate equal to that of the optical and near-IR deep photometric surveys
of \citet{Nakajima2000} and \citet{Allers2006}. Overall, the small abundance
of mid-IR excess sources in these samples is surprising
since the all three estimators used to select them are traditional proxies for 
disk mass accretion towards the central star and should be correlated
with the presence of a disk (see \S~\ref{disks} for a discussion on the 
disks properties). Finally, all the candidates selected
by \citet{Allen2007} in the Lupus III core show mid-IR excess since
IRAC data from the GTO observations were used to select them but those
not classified as YSOs by our criteria fall in the galaxy areas of the
color magnitude diagrams and have SEDs difficult to explain with star
plus disk models. This comparison with other
samples of YSO candidates gives us confidence on the robustness of the
c2d color criteria for selecting a reliable set of YSOs in the
clouds. In the following sections, we will analyze separately the YSO
sample and the total list objects.

\begin{table}[h]
\begin{center}
\center{\caption{Total Number of Stars and YSOs in the
       Lupus Clouds Organized by SED Class.
       \label{YSO_classes}}}

\begin{tabular}{lcccccc}
\tableline  \tableline
Lada Class  & \multicolumn{3}{c}{YSOs}          & \multicolumn{3}{c}{Total} \\
           & Lupus I  &  Lupus III  &  Lupus IV   & Lupus I  & Lupus III   & Lupus IV \\
\tableline
I   &  2 (15 \%)  &  2 (3 \%)  &  1 (8 \%)  &  2 (12 \%)  &  5 (4 \%)  &  1 (6 \%) \\
Flat   &  3 (23 \%)  &  6 (9 \%)  &  1 (8 \%)  &  3 (18 \%)  &  8 (6 \%)  &  1 (6 \%) \\
II   &  6 (47 \%)  & 41 (59 \%)  &  5 (42 \%)  &  8 (47 \%)  & 56 (45 \%)  & 11 (61 \%) \\
III   &  2 (15 \%)  & 20 (29 \%)  &  5 (42 \%)  &  4 (23 \%)  & 55 (44 \%)  &  5 (28 \%) \\
Total & 13 &  69 &  12 &  17 & 124 &  18 \\
\tableline
\end{tabular}
\end{center}
\tablecomments{The percentages for each class in the table are calculated with respect to
the total numbers at the bottom of each column. The `Total' population consists of the YSOs 
plus the PMS stars and candidates (\S~\ref{YSOc_list}).}
\end{table}

\subsection{Class distribution of the sample}

The complete sample contains 17, 124
and 18 objects in Lupus I, III and IV, respectively. Table
\ref{YSO_classes} shows the number of objects per cloud and
their respective SED classes, as defined by \cite{Lada1984} and
extended by \cite{Greene1994} (i.e. Class I for objects with
$\alpha_{(K-24 \mu m)} > 0.3$, Class Flat for $0.3 > \alpha_{(K-24 \mu
  m)} \geq -0.3$, Class II for $-0.3 > \alpha_{(K-24 \mu m)} \geq
-1.6$, and finally Class III for $\alpha_{(K-24 \mu m)} < -1.6$). The
table also shows the total number of YSOs, as defined above, for each
cloud to facilitate comparison to other c2d clouds where the PMS
sample will be obviously different. The number of sources shows that Lupus 
in general
presents a moderate star formation activity compared with the other
clouds surveyed by the Spitzer c2d program, only larger than that
measured in Cha II \citep{Alcala2008}. It also shows that amongst
them, the most active regions are Lupus III, followed by Lupus
I and IV, which present a lower amount of YSOs.

   \begin{figure}
   \centering
   \includegraphics[angle=90,width=15cm]{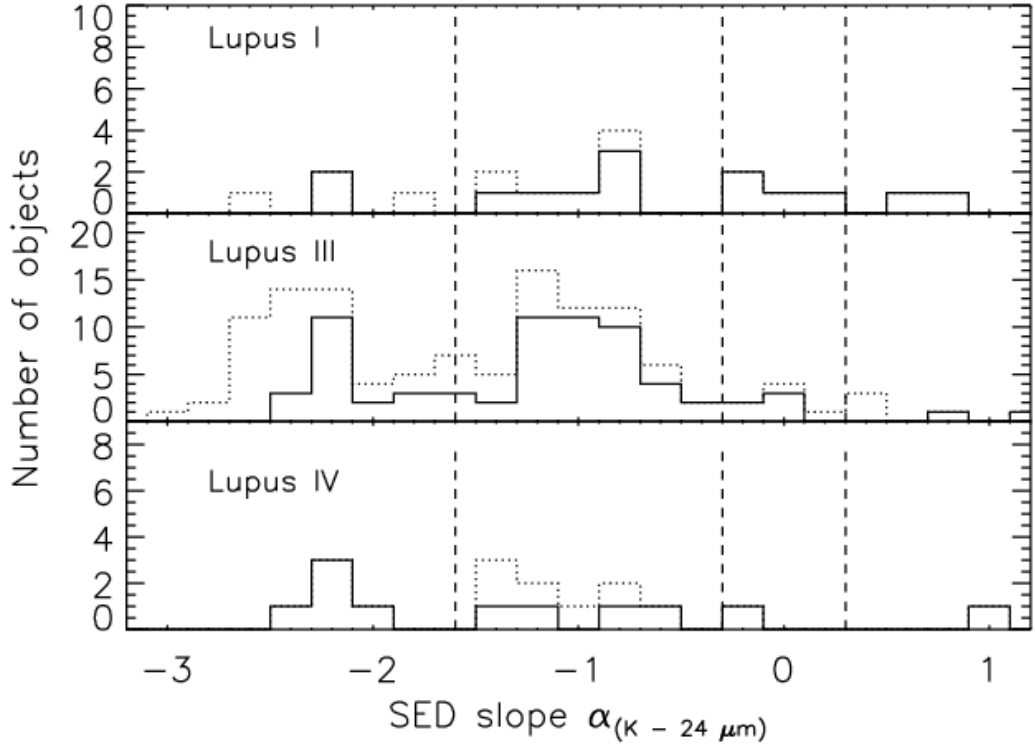}
   \caption{Distribution of objects in $\alpha_{(K-24 \mu m)}$ for
   Lupus I, III and IV. Thin lines show the YSO population and dotted
   lines the total sample of stars. The intervals defining the Lada
   classes are marked with vertical dashed lines. The comparison illustrates
   that Lupus I is a region with a small number of YSOs, many of them
   being early classes, Lupus III contains all kinds of objects, with a
   great number of Class III sources while Lupus IV contains almost no
   early class sources and relatively more late classes.}
   \label{alpha_hist}
   \end{figure}

   Figure \ref{alpha_hist} shows the distribution of $\alpha_{(K-24
     \mu m)}$ values in the three clouds. It illustrates that Lupus
   III shows a larger variety of objects, from Class I to Class III,
   while Lupus I has a much smaller number of YSOs relatively rich in
   Class I and Flat spectrum objects, and Lupus IV shows a clear lack of early class
   objects. The figure also shows that the addition of spectroscopic
   or photometric member candidates to the YSO samples mostly adds
   Class III objects in Lupus III, whose association with the cloud
   cannot be confirmed until optical spectroscopy reveals youth
   signatures. This is a selection effect, since Lupus III is the most
   studied region of the three and illustrates that the difference in
   percentages between the YSO and the total sample is only driven 
   by different available data and not based on physical characteristics
   of the stars. For this reason, we will concentrate
   in the YSO sample on the following discussion, which has been
   selected as genuine IR-excess objects and therefore likely belong
   to the star-forming clouds. The analysis below is only valid as a rough indicator
   of the relative abundances of objects within Lupus. A more solid
   statistical analysis which combines all the YSO samples in all c2d
   clouds will be presented elsewhere (N. Evans et al., in
   preparation).

One interesting difference is that the percentage of Class I and Flat
sources compared to the number of Class II and III sources is
particularly high in Lupus I, compared to Lupus III and IV. Assuming
that the classes correspond to a succession of evolutionary phases
from the cores to the disks, this suggests that Lupus I, the least
active cloud of the three in terms of YSO abundance, is in a less 
evolved phase of evolution, while Lupus III and Lupus IV are respectively 
more evolved than Lupus I.

It is also possible to study the later phases of disk evolution: the
percentage of Class III vs Class II plus III (i.e. N(III)/(N(II) +
N(III))) YSOs in the clouds (50$\pm$25\% in Lupus IV, 32$\pm$16\% in
Lupus III and 25$\pm$12\% in Lupus I) suggests that star formation
started longer ago in Lupus IV than in any of the other clouds. The
errors are dominated by the actual number of Class III objects, or
alternatively, the completeness of the sample for this kind of
objects. Given our selection criteria based on presence of IR excess,
the sample will be biased against the Class III objects with little
excesses. However, we can estimate the Class III completeness level to
be of the order of 50\% from the number of spectroscopically confirmed
young stars from the `Sz' catalog which were observed to have
detectable IR excess in the Spitzer bands (namely, 13 out of 28). This
uncertainty dominates the total error budget of these ratios and let
us only conclude significatively that Lupus IV has a larger percentage of
Class III objects than any of the other two clouds. Actually, this is
the largest number of Class III vs Class II+III sources in any of the
c2d studied star-forming clouds
(e.g. \citealt{Harvey2007a,Alcala2008}). Comparing Lupus III and IV,
which have a similar density of YSOs, these numbers also suggest that
the major star-formation activity already took place (and is about to
start again, see \S~\ref{extinction}) in Lupus IV, while it is currently
taking place in Lupus III, mostly in its very dense star-forming
core. 

 We can confirm our Class III completeness ratio of 50\% given
above with the list of XMM detections in 
the Lupus III core by \cite{Gondoin2006}, which is deeper than
the ROSAT All Sky Survey by \cite{Krautter1997}. That survey 
covers a circular area of 30$'$ in diameter around the Lupus III core and found 
102 X-ray source detections, out of which 25 are associated with 
optical and IR counterparts. A cross-match of
our total list of objects and the list of X-ray detections with a match
range of 4 arcsec yielded 24 matches, 13 of which have SEDs indicating
the presence of a disk and the remaining 11 showing no IR excess. 
All stars with associated X-ray emission were labeled in Table \ref{YSO_list1}
with a reference to \cite{Gondoin2006}.
Assuming that the presence of X-ray emission is a signature of youth, this 
yields an X-ray disk fraction in the core in Lupus III of 54\% which is of the order of the 
Class III completeness we derived from the independent measurement
with the H$\alpha$-selected `Sz' catalog above.

The full list of PMS stars is given in Table \ref{YSO_list1}: columns
2 to 5 give their c2d and previous names of the objects and their
coordinates in the Spitzer catalog, column 6 gives the selection
source as described above, a flag determining the status of membership
spectroscopic confirmation (PMS status) of the object is given in
column 7, the Spitzer-derived class based on the SED slope
$\alpha_{(K-24 \mu m)}$ is given in column 8 and the references to the
objects are given in column 9.


\section{Individual sources properties}

\label{disks}

The properties of the circumstellar material surrounding the young
stars in Lupus can be studied with the optical to mid-infrared
emission in several different ways. We are specifically interested in
estimating the overall disk fraction, the amount of circumstellar dust, 
and the morphology of the circumstellar disks (flared versus flat) which
is likely to represent some stage in the
formation of planetary systems. The special interest that the new Spitzer data bring to the
  analysis comes from the fact that the IRAC and MIPS data, due to
  their wavelength coverage, probe the region in the disks around 
  low-mass stars between $\sim$ 0.1 and 5 AU, where planet formation
  may take place. Therefore, large samples of Spitzer SEDs of
  star-forming regions with ages in the critical time-frame between 1
  and 10 Myr can be used to study statistically the initial conditions
  for planet formation.

\subsection{Color-Color and Color-Magnitude diagrams}
\label{colorcolor}

\begin{figure}[!ht]
\epsscale{1.0}
\plotone{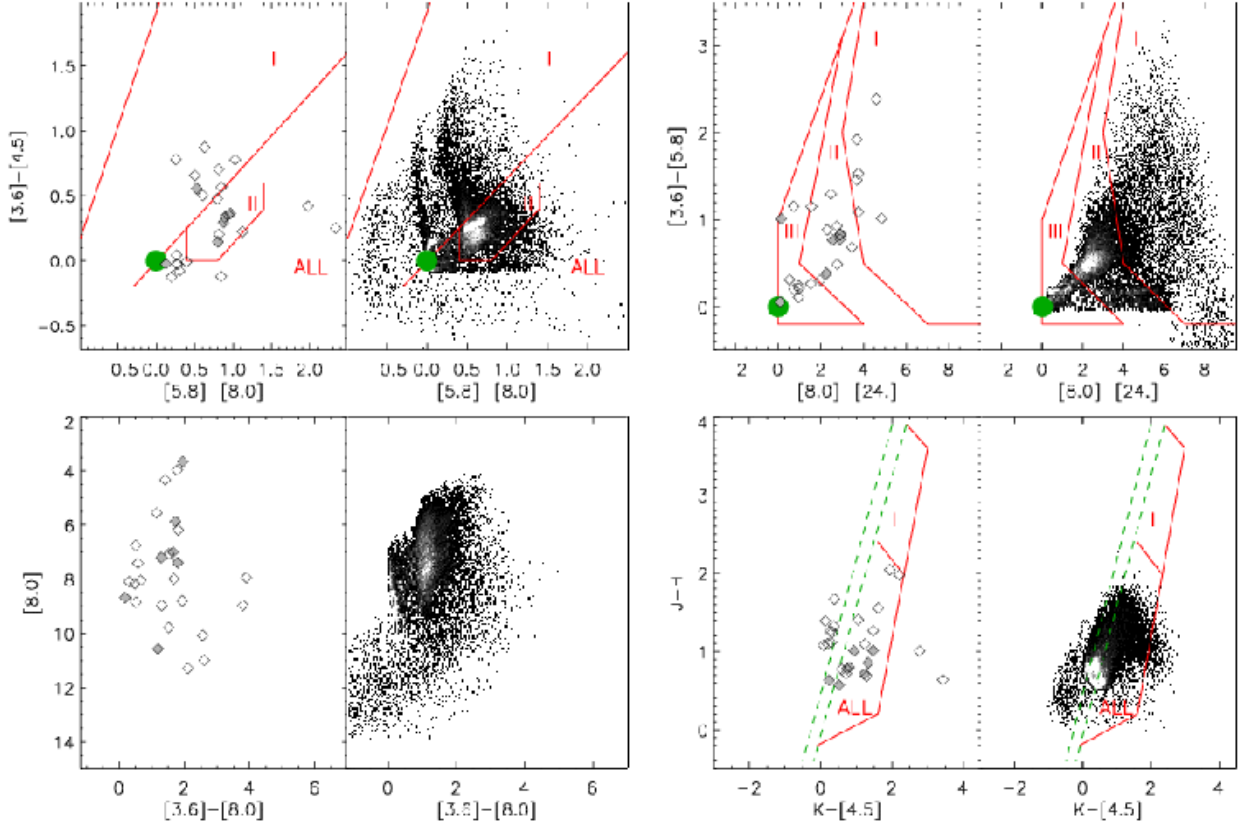}
\caption{The Color-color and color-magnitude diagrams for the
 spectroscopically confirmed PMS objects (gray filled diamonds) and
 candidates (open diamonds) with Spitzer and 2MASS data reported in
 Table~\ref{YSO_list1} in the Lupus I and IV clouds are plotted in the
 left panels of each diagram. The colors derived from the SED models
 by \citet{Robitaille2006} are plotted (in gray-scale intensity
 representing the density of points) in the right panels of each
 diagram.  The areas corresponding to the Stages I, II, and III as
 defined by \citet{Robitaille2006} are also indicated in each
 diagram. The label 'ALL' mark the regions where models of all
 evolutionary stages can be present. The green circle represents
 normal reddening free photospheres. The PMS stars in
 Lupus I and IV fall in regions corresponding to Stage I to III sources.
 \label{cccmd1}}
\end{figure}

\begin{figure}[!ht]
\epsscale{1.0}
\plotone{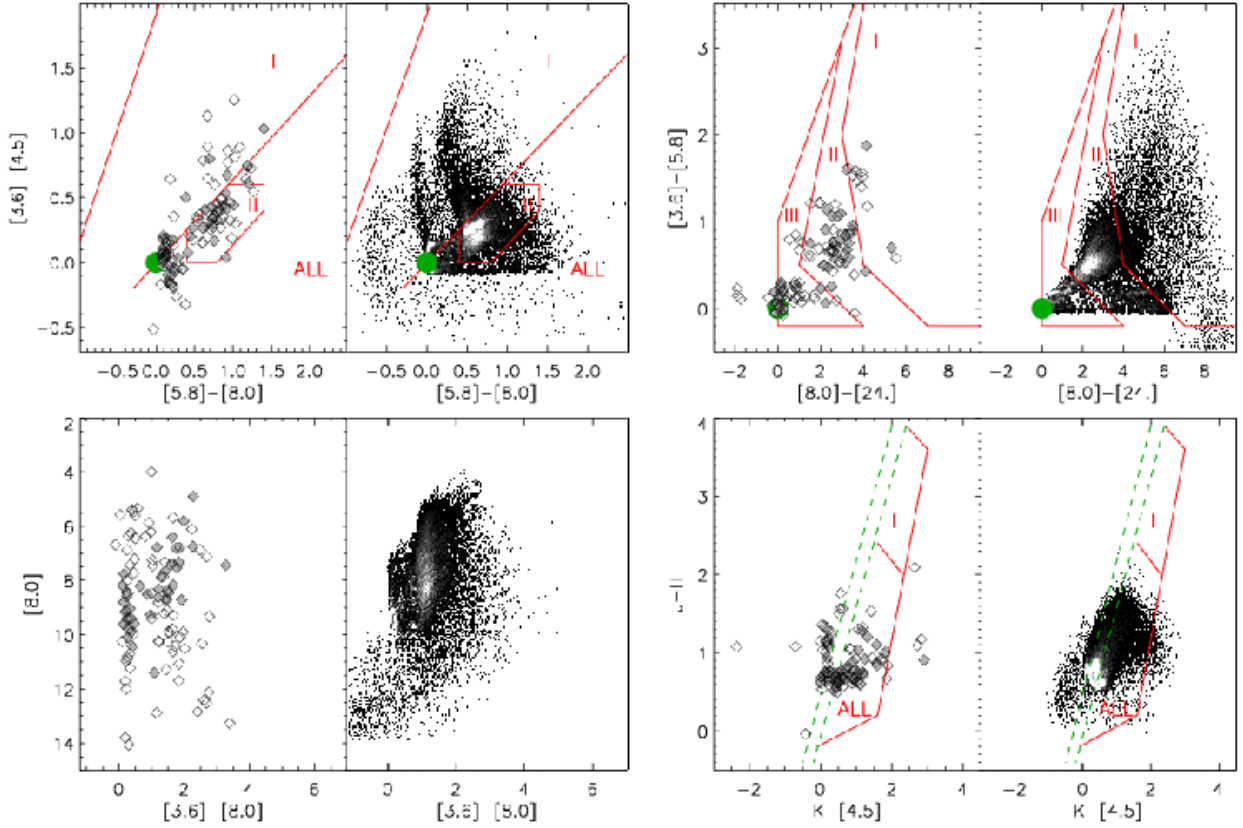}
\caption{Same as figure \ref{cccmd1} but with the YSOs in 
Lupus III. \label{cccmd2}}
\end{figure}

Infrared color-color (CC) and color-magnitude (CM) diagrams are good
diagnostic tools for the investigation of circumstellar matter around
YSOs in a statistical way \citep[][and references
therein]{Hartmann2005,Lada2006}. In recent years, several authors have
produced grids of YSO models and computed their colors in the Spitzer
bands to allow direct comparison with the observations
\citep{Whitney2003,Allen2004,Robitaille2006}.  Both previous data sets
and the models roughly agree in the spatial distributions of the
objects of different Lada classes in the different CC and CM diagrams
although the interplay between the presence or absence of envelopes
and disks together with the whole possible range of system
inclinations and total interstellar extinctions makes the problem
highly degenerate. Therefore it is impossible to associate a position
in those diagrams with a physical configuration for individual
objects. \cite{Robitaille2006} computed a very large grid of YSOs
with a broad range of physical parameters, and inclination angles that
can be used to make statistical analyses. In discussing the
evolutionary stages of our models, they adopt a "Stage" classification
analogous to the Class scheme, but referring to the actual
evolutionary stage of the object, based on its physical properties
(e.g., disk mass or envelope accretion rate) rather than properties of
its SED (e.g., slope). Stage 0 and I objects have significant
infalling envelopes and possibly disks, Stage II objects have
optically thick disks (and possible remains of a tenuous infalling
envelope), and Stage III objects have optically thin disks. Figures
\ref{cccmd1} and \ref{cccmd2} compare three CC and CM diagrams of the
Lupus YSO sample with the synthetic colors of a model cluster from
\cite{Robitaille2006}.  To make these comparisons, we have scaled the
models to 150 pc in the first case to compare them with the YSOs in
Lupus I and IV and to 200 pc to compare them with those in Lupus III.

The comparison between the position of the Lupus YSOs and those of the
model grids in the [3.6]-[4.5] vs [5.8]-[8.0] CC diagram shows several
early Class~I and Flat source SEDs in Lupus I and IV (Figure
\ref{cccmd1}), redder in [3.6]-[4.5] than the Stage II objects from
Robitaille et al. and a large majority of Class~II sources in Lupus
III (Figure \ref{cccmd2}). The position of the maximum density of such
kind of objects is also consistent with that predicted by
\citet{Allen2004} with a grid of physical disk models of
\citet{Dalessio2005}. The [8.0] vs [3.6]-[8.0] CM diagram of Lupus III
(Figure \ref{cccmd2}) shows a bimodal distribution with a large number
of objects with photospheric [3.6]-[8.0] colors and a range of 8.0
\micron\ magnitudes and another one with a similar range of
luminosities and a range of color excesses in the IRAC bands
compatible with the presence of moderate to large mid-IR excess.  No
special difference is found between the distributions of
spectroscopically confirmed PMS objects (grey diamonds) and candidate
YSOs in any of the diagrams, which suggests that both populations
contain indeed the same kind of objects.

\subsection{Spectral Energy Distributions}
\label{SEDs}

\begin{figure}[!ht]
\epsscale{0.9}
\plotone{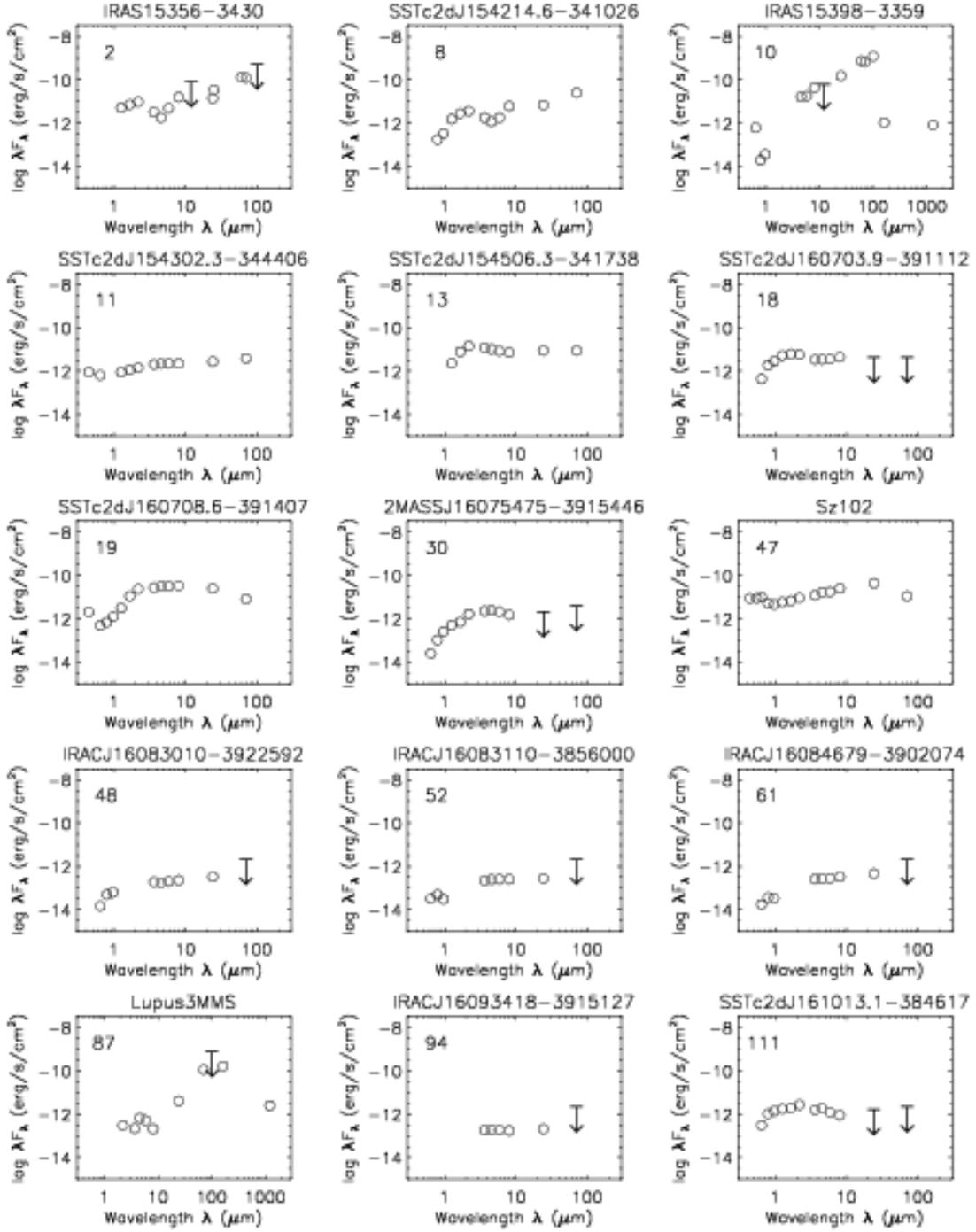}
\caption{Spectral energy distributions of the Class~I and Flat SED PMS
       objects and YSOs ordered by RA and labeled with their
       name and ID as in Table \ref{YSO_list1}. Only the observed fluxes are
       plotted with open dots and upper limits with arrows. The new Spitzer data are shown in grey
       to distinguish them from previous data.
       \label{seds_I}}
\end{figure}

We have constructed SEDs for each source, similar to what has been
done for other star-forming regions \citep[e.g.,
][]{Hartmann2005,Lada2006,Sicilia-Aguilar2006}. These allow the full
set of multiwavelegth data to be displayed for each source. Figure 
\ref{seds_I} shows the SEDs of all Class I and `Flat' sources
in the sample. For all objects, the c2d catalog provides the 2MASS
near-IR magnitudes and the IRAC1-4 and MIPS1-3 fluxes. We have added
all the complementary data described in \S~\ref{binaries} plus IRAS
fluxes from the PSC and 1.3 mm fluxes from \cite{Nuernberger1997}. The
open dots are the observed fluxes and the Spitzer data have been highlighted
in grey to compare with previous observations. 

Figure \ref{seds_II} shows all the SEDs for the Class II and III
objects in the Lupus sample for which there was sufficient data to
make a good fit to a stellar SED and characterize the stars and disks
separately. Two different
procedures were applied for the SED fits: for the objects with known
spectral types, marked as `PMS' in Table \ref{YSO_list1}, we obtained
the best-fit visual extinction $A_V$ by fitting all photometry between
$V$ and $J$ to the stellar NEXTGEN models \citep{Hauschildt1999}. This
method provides very good agreement with the traditional use of the
$R_C-I_C$ color, as compared with the results by \citet{Alcala2008}.
The $\chi^2$ minimization is extended to try all spectral types
between A0 and M9 and all visual extinctions $A_V$ between 0 and 30
magnitudes for the new YSOs for which there is no spectral type
available in the literature. This technique is similar to that used by
Spezzi et al. (2007b, in preparation) with the c2d Cha II sample, and
provides agreement within $\pm$ 200 K in the effective temperature
when applied to objects with known spectral types and moderate
extinctions. Once the spectral type and extinction are computed for
the sample, the observed fluxes are dereddened with the extinction law
by \citet{Weingartner2001} with $R_V = 5.5$ and fitted to the
appropriate NEXTGEN
stellar atmosphere models.  The sample of SEDs of stars from the
Schwartz catalog (Sz) was also compared with the SED fits made by
\citet{Hughes1994} and there is in general good agreement in the
resulting dereddened stellar and disk energy distributions. The
resulting spectral types and extinctions, together with their
corresponding references are given in Table \ref{sed_results}.

  The SEDs show considerable variety. A useful classification
  system has to rely on disk models which are again degenerate in several
  parameters as e.g. the inclination angle of the disk or the degree
  of dust processing of the disk. Therefore, we will not attempt a
  full characterization of the SED types in this section but rather
  define four types of objects and give their frequencies in the
  sample. To help with the classification, we have plotted with a
  dashed line the median SED of the Classical T Tauri stars (CTTs)
  from Taurus \citep{Dalessio1999} normalized to the optical
  dereddened fluxes of all the low-mass stars. Based on this
  comparison benchmark, we can identify the systems whose colors from
  optical to the millimeter follow closely the decaying slope of a
  classical accreting optically thick disk around a low-mass star (e.g.  Sz100 or
  SSTc2d J160901.4-392512). We will call these systems `T'-type, for T
  Tauri. We then call `L'-type (from `low' IR excess) all those
  objects where the IR excess is clearly smaller than the median SED of
  a CTTs (e.g. ACK2006-19). These are what \citet{Lada2006} call the `anemic' disks. We
  call `H'-type objects, those systems with higher IR fluxes than
  those of the median CTTs SED (e.g. 2MASS J16081497-3857145). 
  Finally, we will call `E'-type from
  {\sl empty} the spectrally confirmed young stars in the clouds where
  no IR excess at all is detected in the Spitzer bands (e.g. Sz 67). These are
  systems with a very little amount of cold dust or ``debris''-like
  disks (see also some in \citealt{Sicilia-Aguilar2006} and
  \citealt{Lada2006}).

  These classifications are given in Table \ref{sed_results} for
  each system. Considering the whole sample of YSOs and PMS stars and
  candidates in Lupus, 22$\pm$10\% of them are
  `T'-Type, 39$\pm$18\% are `L'-type, 6$\pm$2\% are `H'-type, and
  19$\pm$8\% are `E'-type. The remaining 20\% are the Class I
  and Flat SED sources. The large errors in this percentages come
  from the difficulty in classifying the borderline cases if we
  take into account the implicit uncertainties in the fitting process
  of the dereddened photometry to the stellar
  photospheres and from the completeness estimation for disk-less members
  presented in \S~\ref{pms_sel}. We can consider a negligible amount of
  `H'-type objects, likely due to source variability at different
  wavelegths and problems with the stellar normalization. From the
  remaining sample, approximately 40\% show `L'-type SEDs, 20\%
  show `T'-type SEDs and another 20\% show `E'-type SEDs. The large
  pertentage of `L'-type objects in Lupus is consistent with that
  found in Serpens \citep{Harvey2007a} and suggests more evolved inner
  disks in these objects compared to the initial inner disk
  confguration in the T Tauri stars. This percentages contrasts interestingly with 
  the 30\% of `T'-type and 20\% of `L'-type SEDs found in the older 3 Myr star-forming 
  cluster IC 348 by \cite{Lada2006}. These different disk populations in different
  clouds could be related to different environmental conditions (e.g. clustered
  vs extended star-formation), to a different stellar population (abundance of low-mass vs 
  high-mass stars) or alternatively to a faster evolutionary time-scales for
  the less flared `L'-type disks. More samples of disks will be needed to
  pursue this research, which is outside the scope of this paper.
  
  One fifth of the sample are
  `E'-type stars and this gives an overall disk fraction for the Lupus clouds 
  of 68\% to 81\%, depending on whether the assume a 50\% completeness
  ratio for diskless stars of or not. This range of values matches
  well with other disk fractions calculated for 1 to 2 Myr-old  star forming regions
  with near and mid-IR data \citep{Haisch2001,Sicilia-Aguilar2006}, 
  including Cha II \citep{Alcala2008}.  The `E'-type objects in Lupus show 
  photospheric IRAC and MIPS fluxes which 
  indicate that the stars do not have any circumstellar dust out
  to a distance of at least 50 AU, depending on the stellar luminosity
  (see e.g. HR 6000, Sz 119 or Sz 124). Once spectroscopically confirmed 
  as young members of the clouds and therefore with ages $<$ 3 Myrs, they 
  will provide very valuable constraints to the disk dissipation time-scales and
  mechanisms and will be the subject of a separate study.

  Also interestingly, the Spitzer data of a subsample of the `L'-type
  objects show a very peculiar SED, with almost photospheric IRAC
  and MIPS1 fluxes typical of Class III sources, and a MIPS2 excess
  flux at 70 $\mu$m comparable to those of the CTTs. This subgroup has
  been labeled `LU' in Table \ref{sed_results} and contains the
  objects SSTc2d J154240.3-341343, Sz 91, and Sz 111. It represents
  a 2$\pm$1\% of the young population in Lupus. Similar objects
  have been found in Serpens \citep{Harvey2007a} and Cha II
  \citep{Alcala2008} and seem to be extremely rare. These objects,
  called cold disks \citep{Calvet2005,Brown2007}, are interpreted as
  optically thick disks with large inner holes of several to tens of
  AUs, where potentially planets could be currently forming. However,
  both the objects in Lupus and the star ISO-ChaII 29 in Cha II
  \citep{Alcala2008} are different from previously known systems of
  this kind in that the long wavelength at which they show the sizable
  excess (70 $\mu$m) implies larger inner holes of up to 70 AU, while
  previously known objects show excess already at 24 $\mu$m, implying
  inner holes smaller than 50 AU.

\begin{figure}[!ht]
\epsscale{0.9}
\plotone{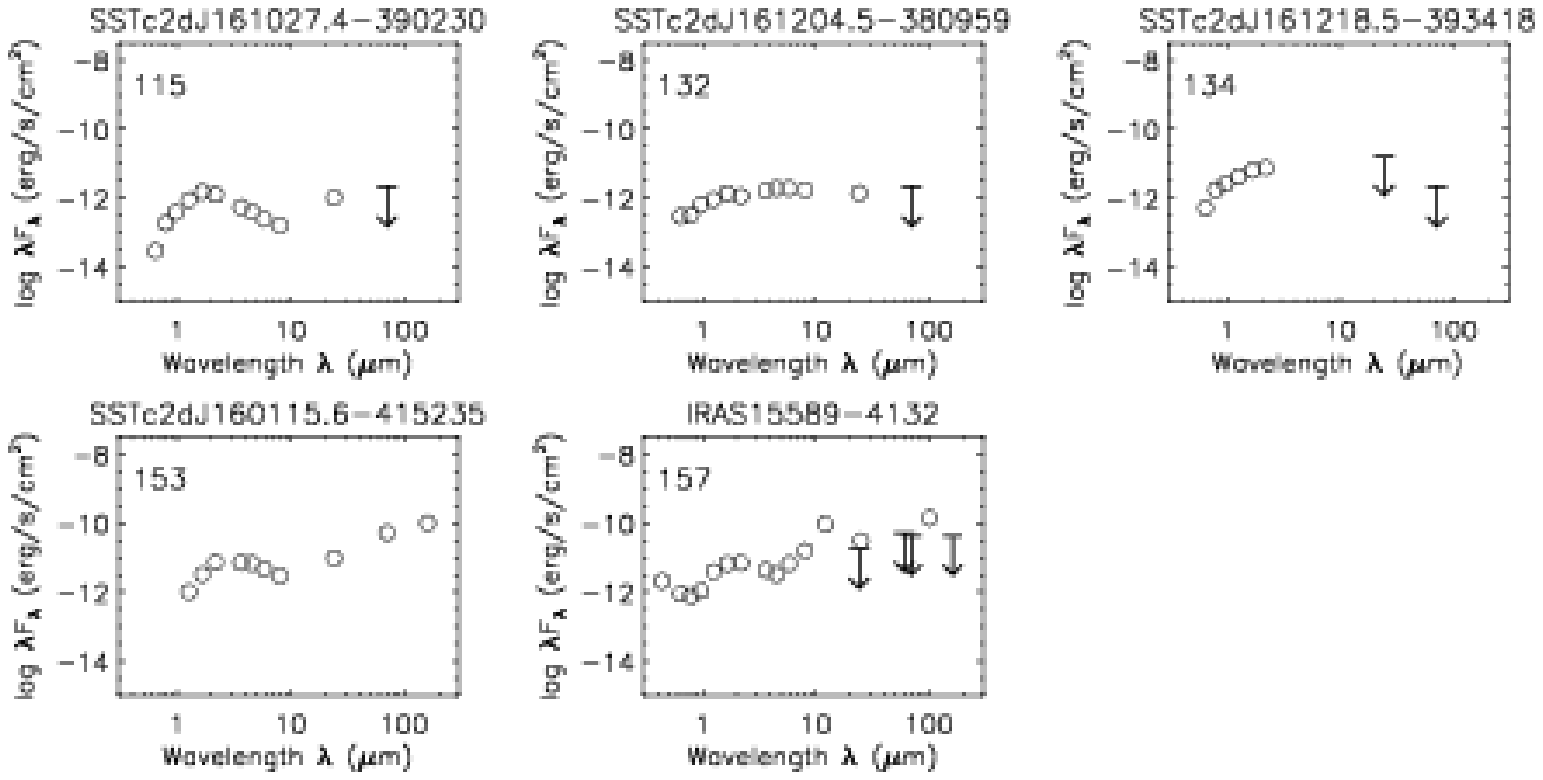}\\
Fig.~\ref{seds_I} - Continued
\end{figure}

\begin{figure}[!ht]
\epsscale{0.9}
\plotone{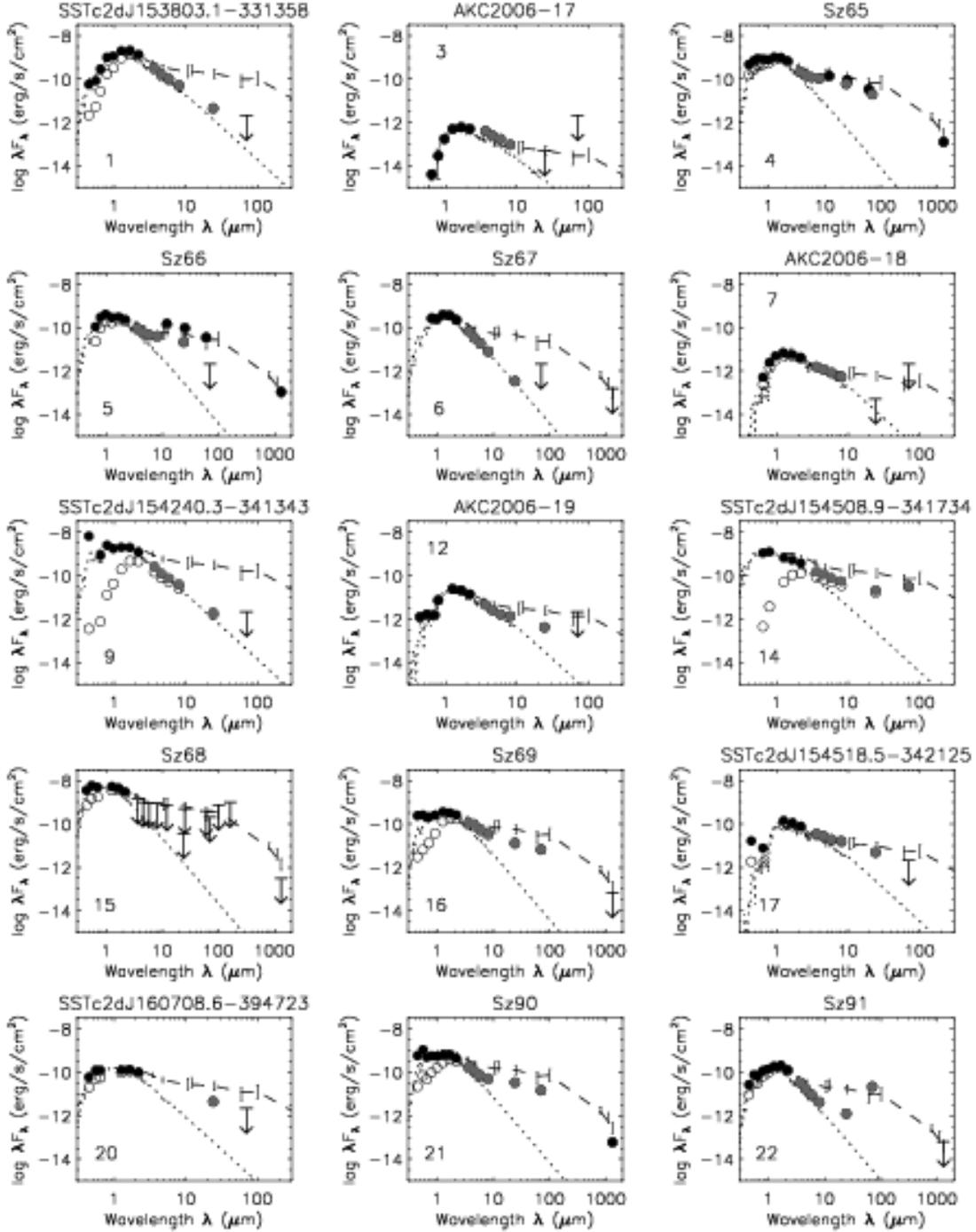}\\
\caption{Spectral energy distributions of the Class~II and Class~III
       PMS objects and candidates ordered by RA and labeled with their
       names and ID as in Table \ref{YSO_list1}. The observed, dereddened
       fluxes and upper limits are represented with open and solid dots and arrows,
       respectively. The plotted error bars are usually smaller than
       the symbols. NEXTGEN stellar models for the spectral type of
       each star are shown in dotted lines. For all low-mass objects,
       the median SED of the T Tauri stars in Taurus from
       \cite{Dalessio1999} is shown normalized to the stellar
       photosphere in dashed line for comparison.  \label{seds_II}}
\end{figure}


\begin{figure}[!ht]
\epsscale{0.9}
\plotone{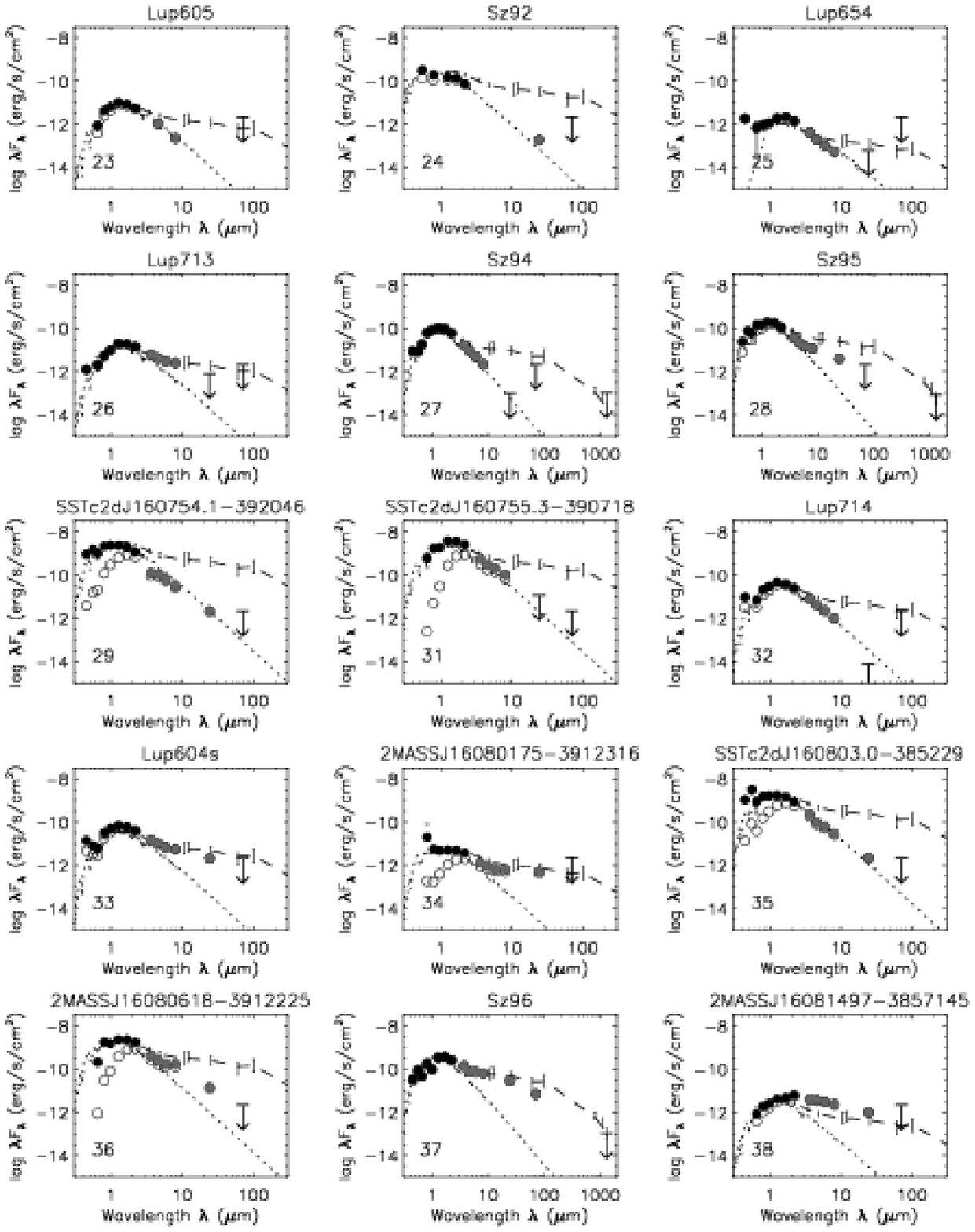}\\
Fig.~\ref{seds_II} - Continued. 
\end{figure}


\begin{figure}[!ht]
\epsscale{0.9}
\plotone{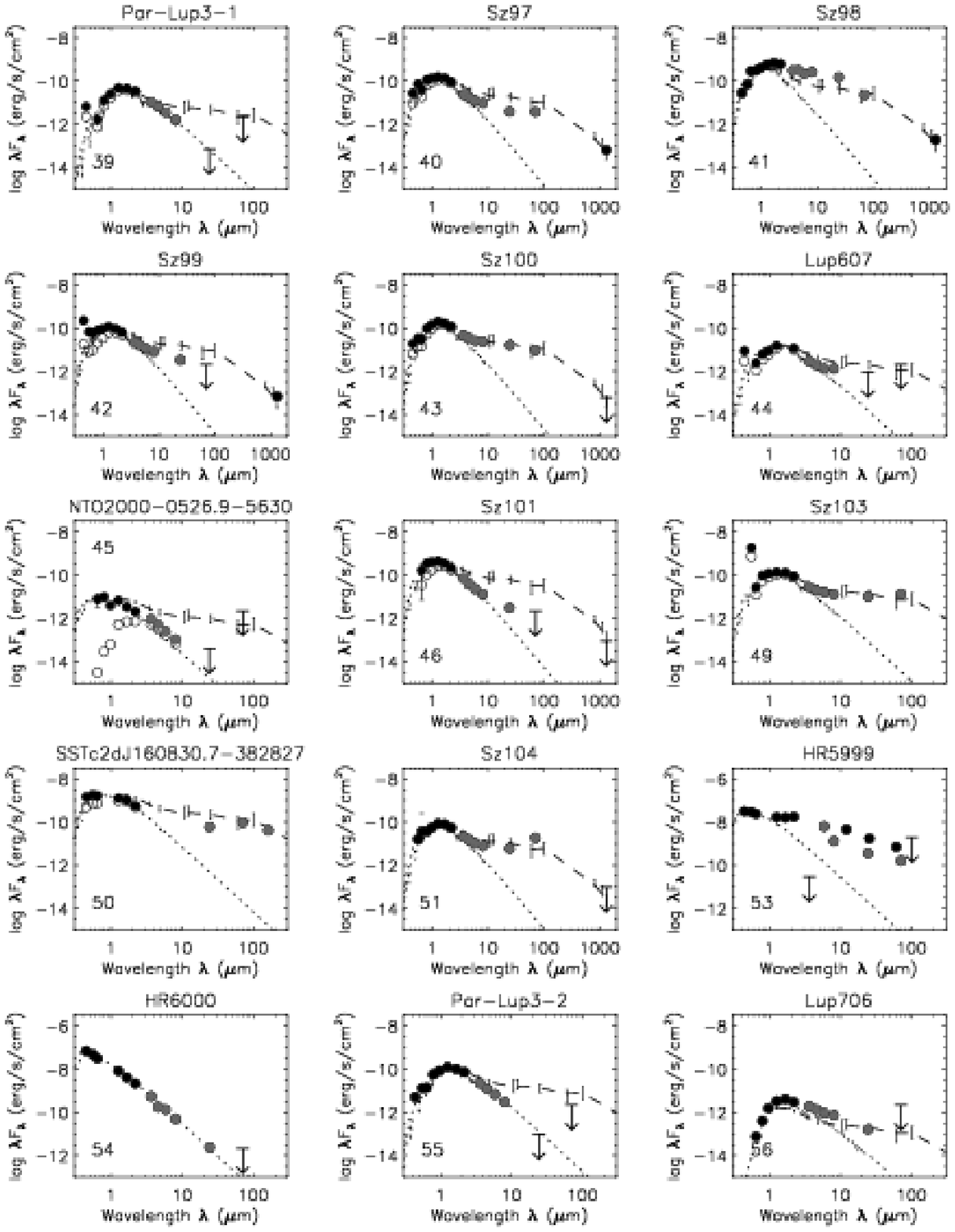}\\
Fig.~\ref{seds_II} - Continued.
\end{figure}


\begin{figure}[!ht]
\epsscale{0.9}
\plotone{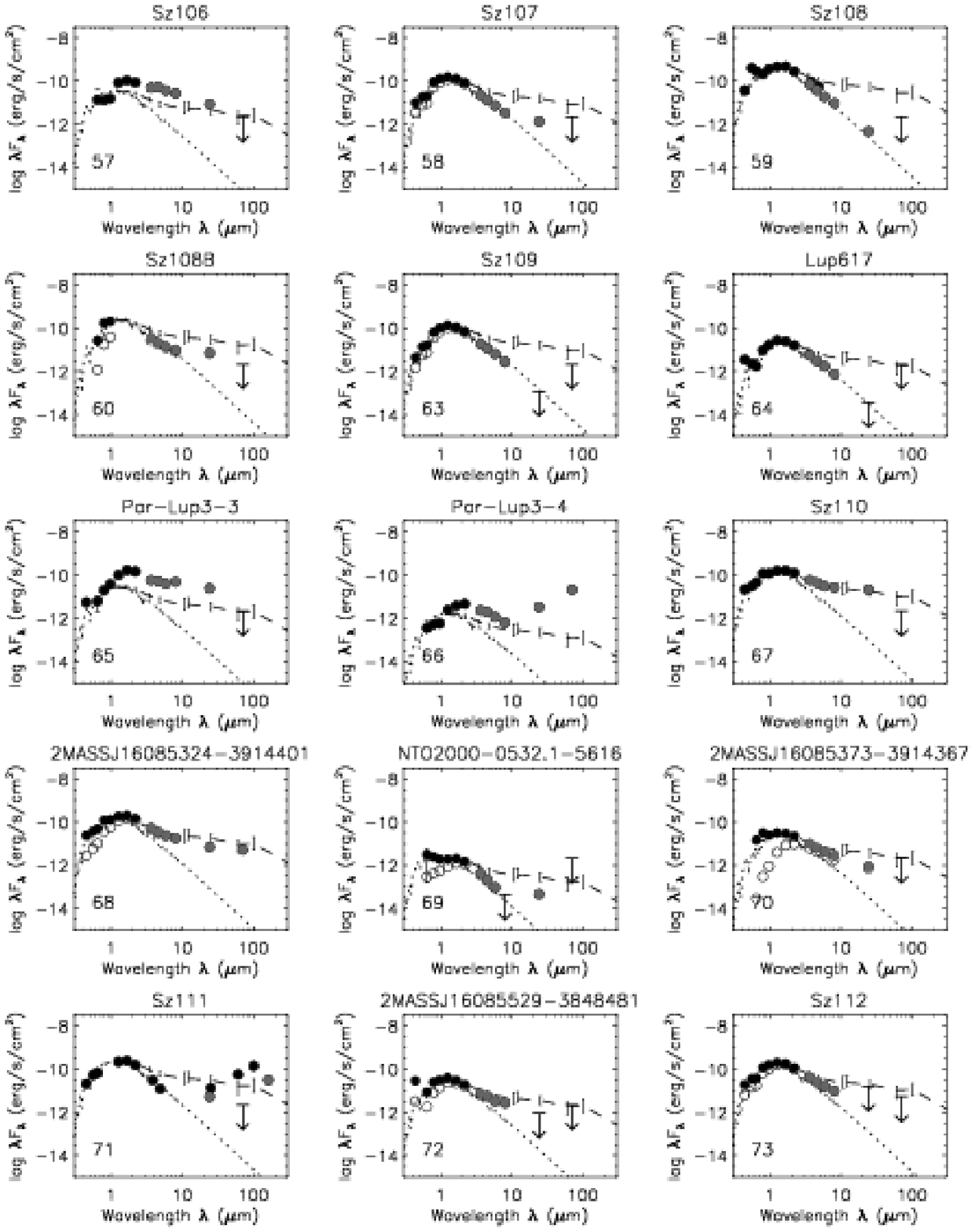}\\
Fig.~\ref{seds_II} - Continued.
\end{figure}


\begin{figure}[!ht]
\epsscale{0.9}
\plotone{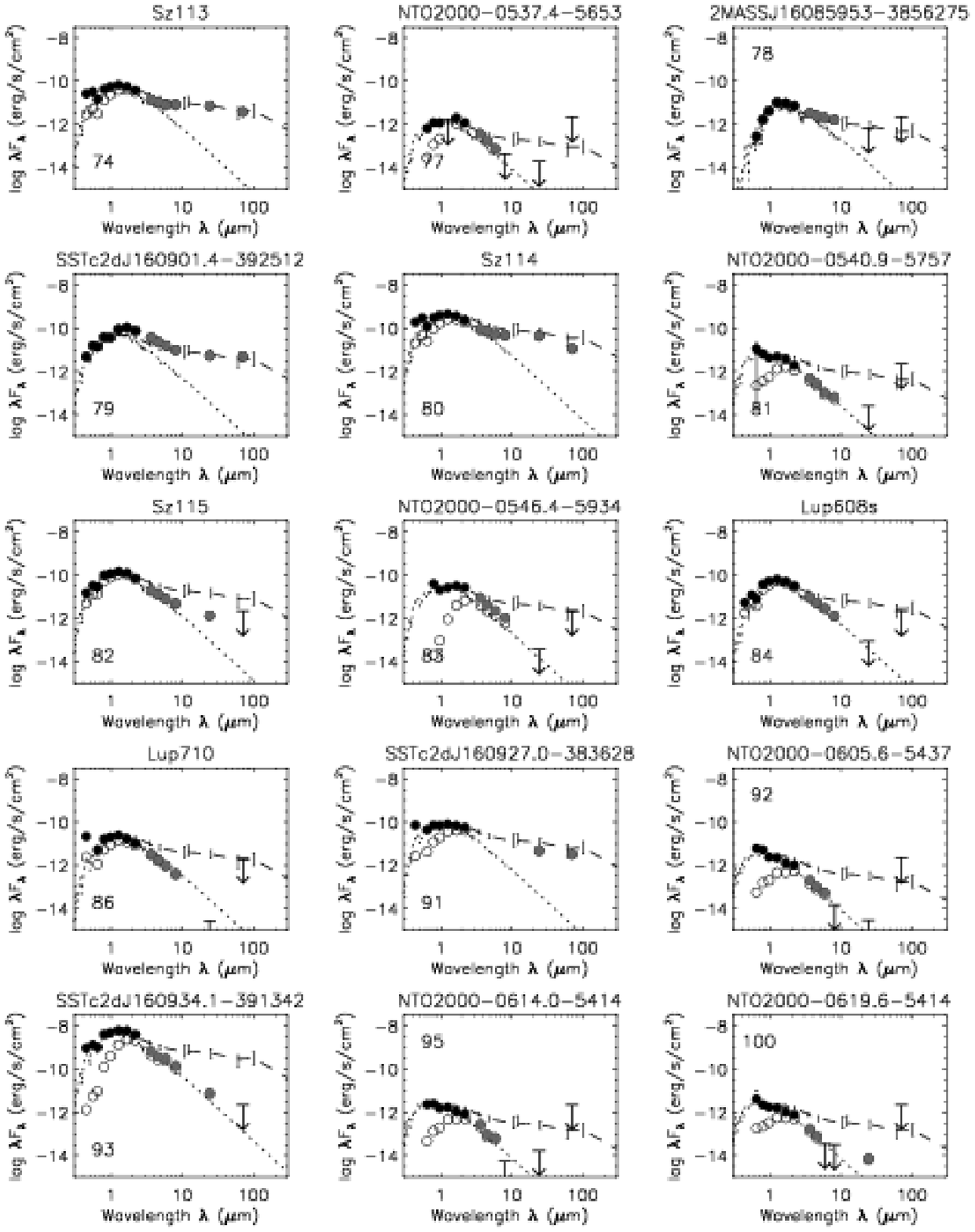}\\
Fig.~\ref{seds_II} - Continued.
\end{figure}


\begin{figure}[!ht]
\epsscale{0.9}
\plotone{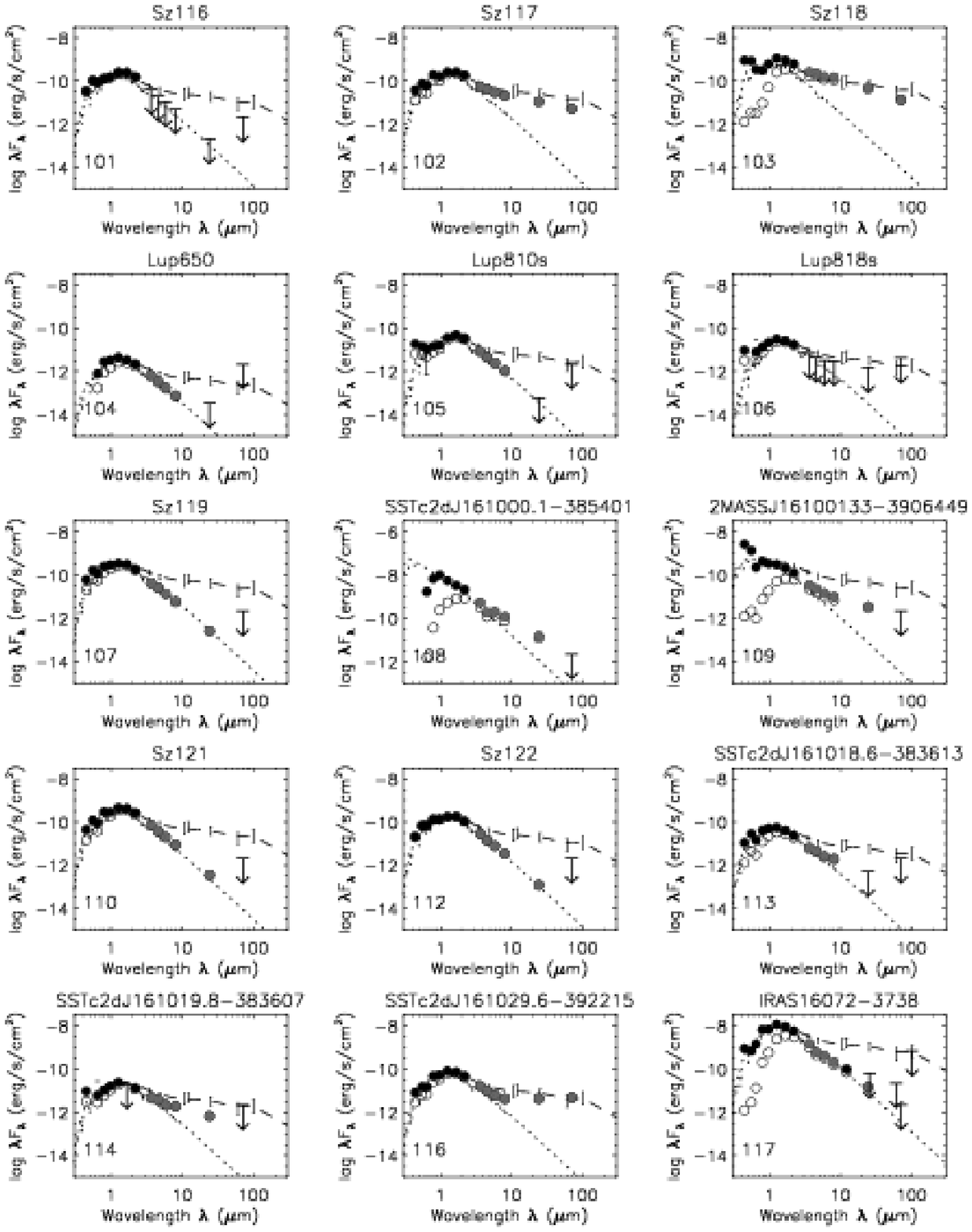}\\
Fig.~\ref{seds_II} - Continued.
\end{figure}


\begin{figure}[!ht]
\epsscale{0.9}
\plotone{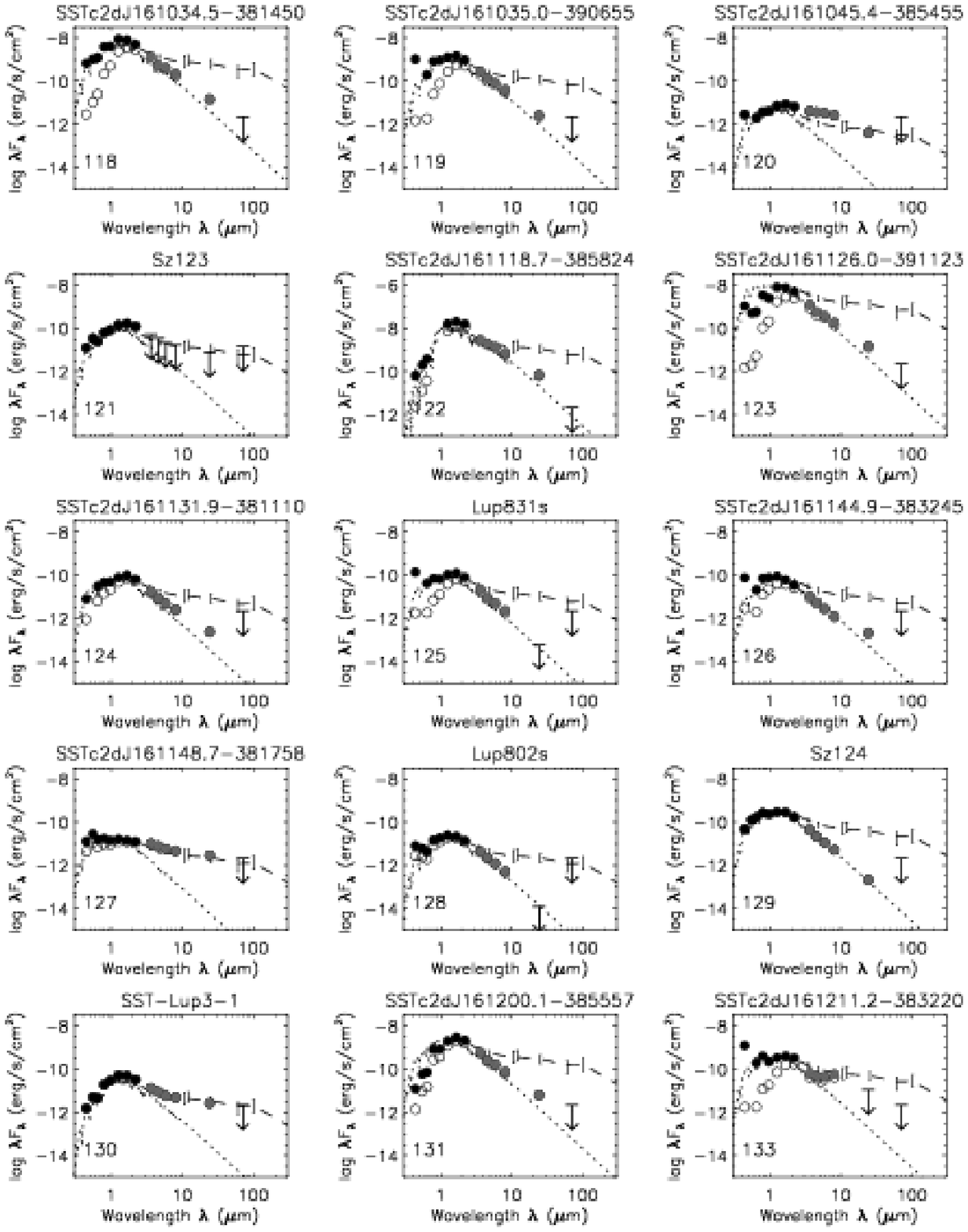}\\
Fig.~\ref{seds_II} - Continued.
\end{figure}


\begin{figure}[!ht]
\epsscale{0.9}
\plotone{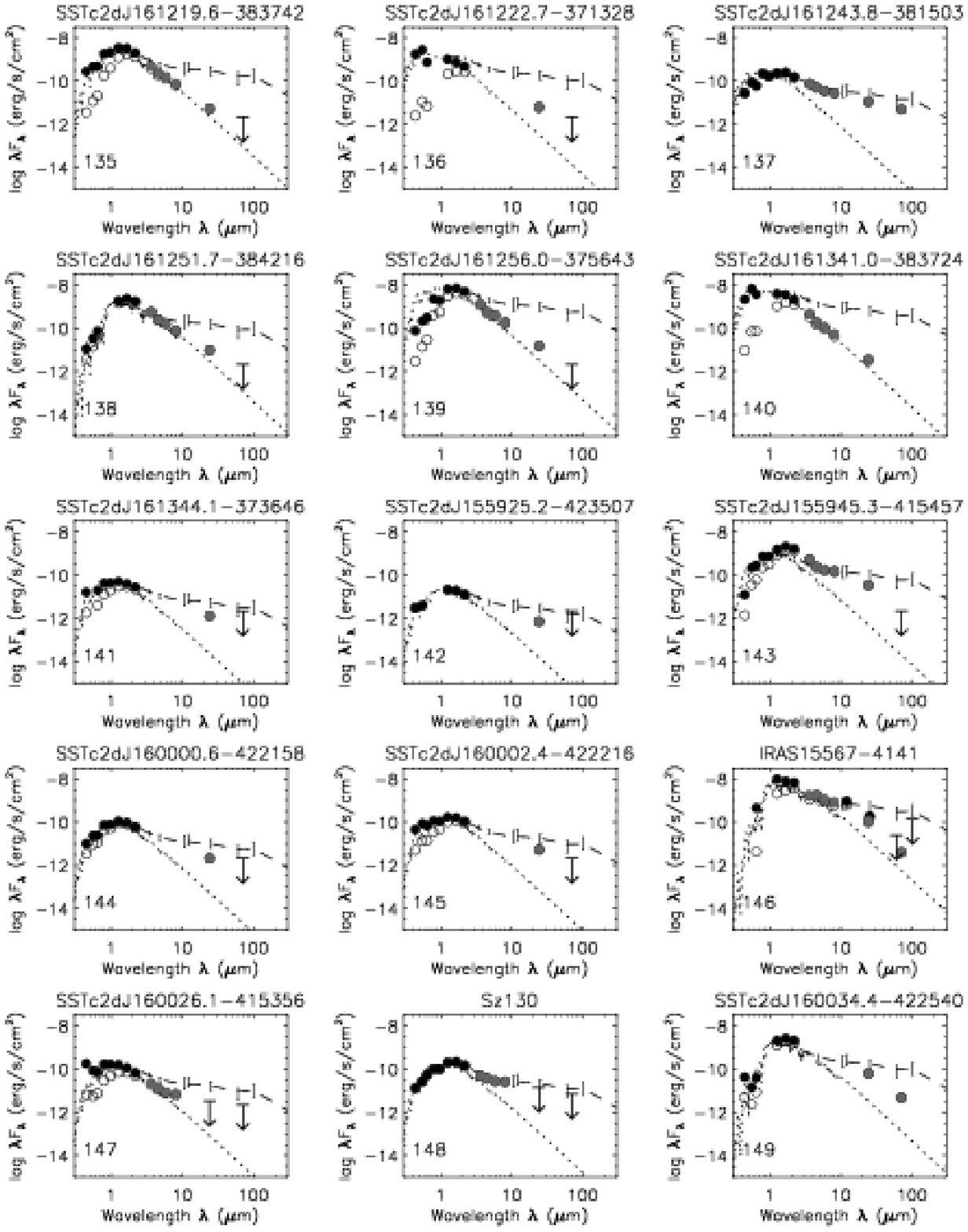}\\
Fig.~\ref{seds_II} - Continued.
\end{figure}


\begin{figure}[!ht]
\epsscale{0.9}
\plotone{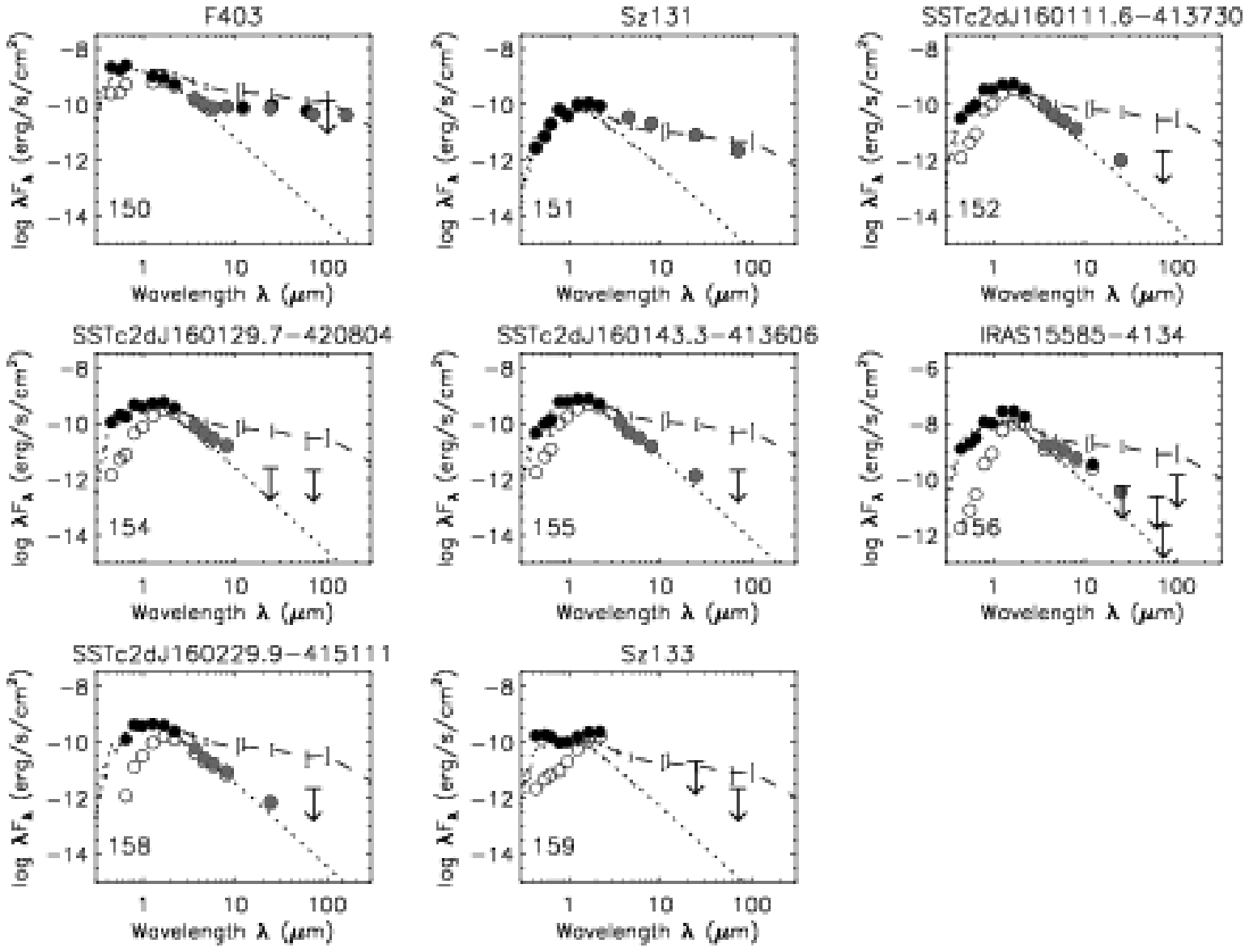}\\
Fig.~\ref{seds_II} - Continued.
\end{figure}

\clearpage

\subsection{The luminosity function in Lupus}
\label{star_luminosities}


A basic step in the characterization of the PMS population is the
determination of its degree of completeness in the effective
temperature and luminosity scales, which might be correlated in those
objects that formed more or less simultaneously.  Both parameters can
be accurately determined from the spectral type and the SEDs shown in
the previous section for Class II and III objects. In order to calculate the best possible stellar
luminosities for the sample, we integrate the NEXTGEN stellar model
normalized to the dereddened optical fluxes (dotted curves in Figure
\ref{seds_II}), and assume distances of 150 pc for Lupus I and IV and 
200 pc for Lupus III. The computed stellar luminosities can be found in Table
\ref{sed_results}. The complete emission of the objects is obtained in
the bolometric luminosity, where we integrate under all the observed
fluxes of the SED, the corresponding bolometric flux is then converted
to luminosities with the same distances quoted above for the different
clouds.

\begin{figure}[!ht]
\epsscale{0.7}
\includegraphics[angle=90,scale=.60]{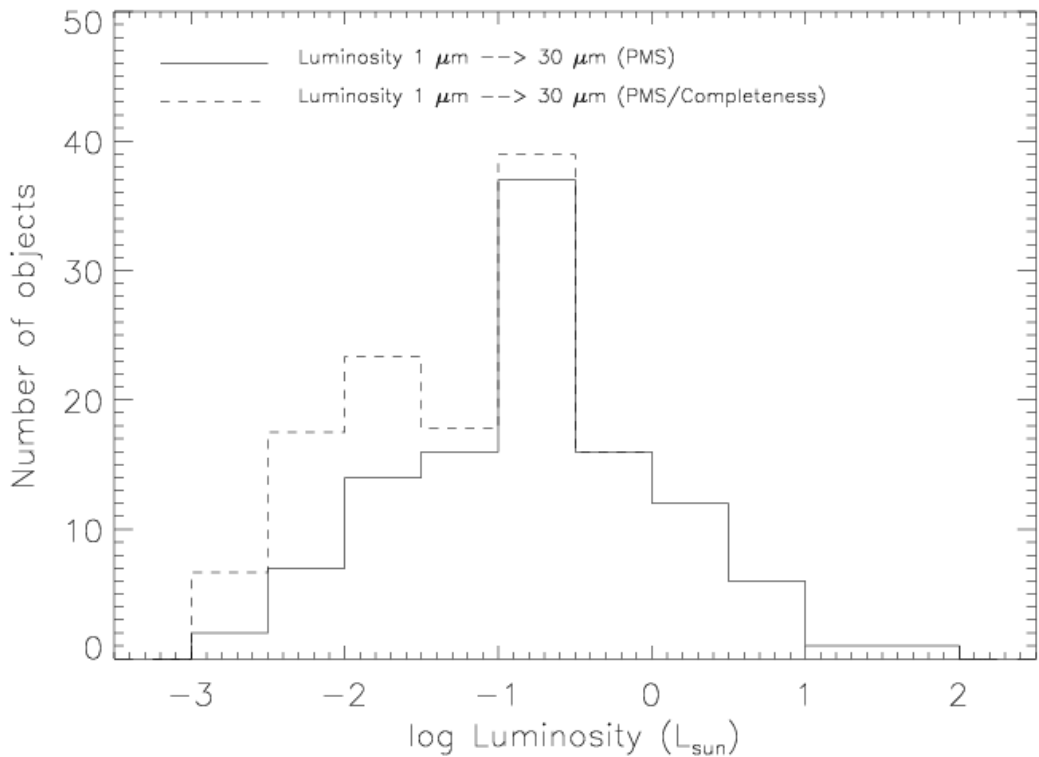}
\caption{Bolometric luminosity function for the Lupus young stellar population ({\sl solid histogram})
and estimated correction for completeness effects ({\sl dashed
histogram}). \label{lum_hist}}
\end{figure}

Figure \ref{lum_hist} shows a histogram of the total bolometric
luminosities for all YSOs and PMS stars in Lupus as a solid line. Compared to
the same figure in Cha II \citep{Alcala2008} and in Serpens
\citep{Harvey2007b}, it shows a larger population of low and very-low
luminosity objects, which was already noted by previous surveys of the
region (see F. Comer\'on 2008, in prep. and references therein), and a
large number of objects with luminosities between 0.3 and 0.1
L$_\odot$. Interestingly, the lowest luminosity members of the
population are Lupus3MMS, the Class 0 object in Lupus III
\citep{Tachihara2007}, SSTc2d J161027.4-390230, another flat SED
object in the same cloud followed by late M-type brown dwarfs from
\cite{Comeron2003}, \cite{Allers2006} and \cite{Allen2007}.

The peak of the luminosity function appears at 0.2 L$_\odot$, which
corresponds to a 0.2 M$_\odot$ star (spectral type M5 at an age of 1
Myr) and up to 1.0 M$_\odot$ for objects of 5 Myrs according to the
PMS tracks by \citet{Baraffe1998} plus a tail at lower luminosites. 
\cite{Harvey2007b} made an estimate of the completeness of the c2d
catalogs by comparing a trimmed version of the deeper SWIRE catalog of
extragalactic sources (\citealt{Surace2004}), taken to represent 100\%
completeness by c2d standards, with the number counts of the c2d
catalog in Serpens per luminosity bin. Their estimations suggest a 100
\% completeness at luminosities $\geq$ 0.3 L$_\odot$ in the c2d
catalogs, which corresponds to a 0.3 M$_\odot$ star (spectral type
M4.5 at an age of 1 Myr).  The dashed line in Figure 13 is the
luminosity histogram corrected for completeness effects in each
luminosity bin, which suggests that up to 15 additional very
low-luminosity objects ($\log L/L_\odot < -1.7$) were missed below the
noise level of our c2d observations or were confused in the galactic
region of the color-magnitude diagrams. The comparison of both
luminosity histograms yields a completeness level in luminosity for
our disk study of $\log L/L_\odot \approx -1.0$, which corresponds to
a mass of $M \sim 0.1 M_\odot$ for a star of $\sim$ 1 Myr
\citep{Baraffe1998}. The lower luminosity limit with respect to that
in Serpens stems from the different distances to both clouds and is
roughly consistent with the distance squared ratio.

\subsection{Disk fractional luminosities}
\label{fractional_luminosities}

One way to characterize the circumstellar accretion disks around young
stars is to compute the disk fractional luminosities $L_{\rm
disk}/L_{\rm star}$ and compare them with that of passive reprocessing flaring
disks \citep{Kenyon1987}. We computed those ratios by subtracting the
stellar fluxes from the observed dereddened fluxes in the SEDs and
integrating the IR excess throughout the SED. That was done for all
the objects with sufficient data to allow a good fit of the stellar
model to the SED, crucial to get a sensible disk luminosity. The
resulting flux ratios are also given in Table \ref{sed_results}. 

\begin{figure}[!ht]
\epsscale{1.0}
\includegraphics[scale=0.8]{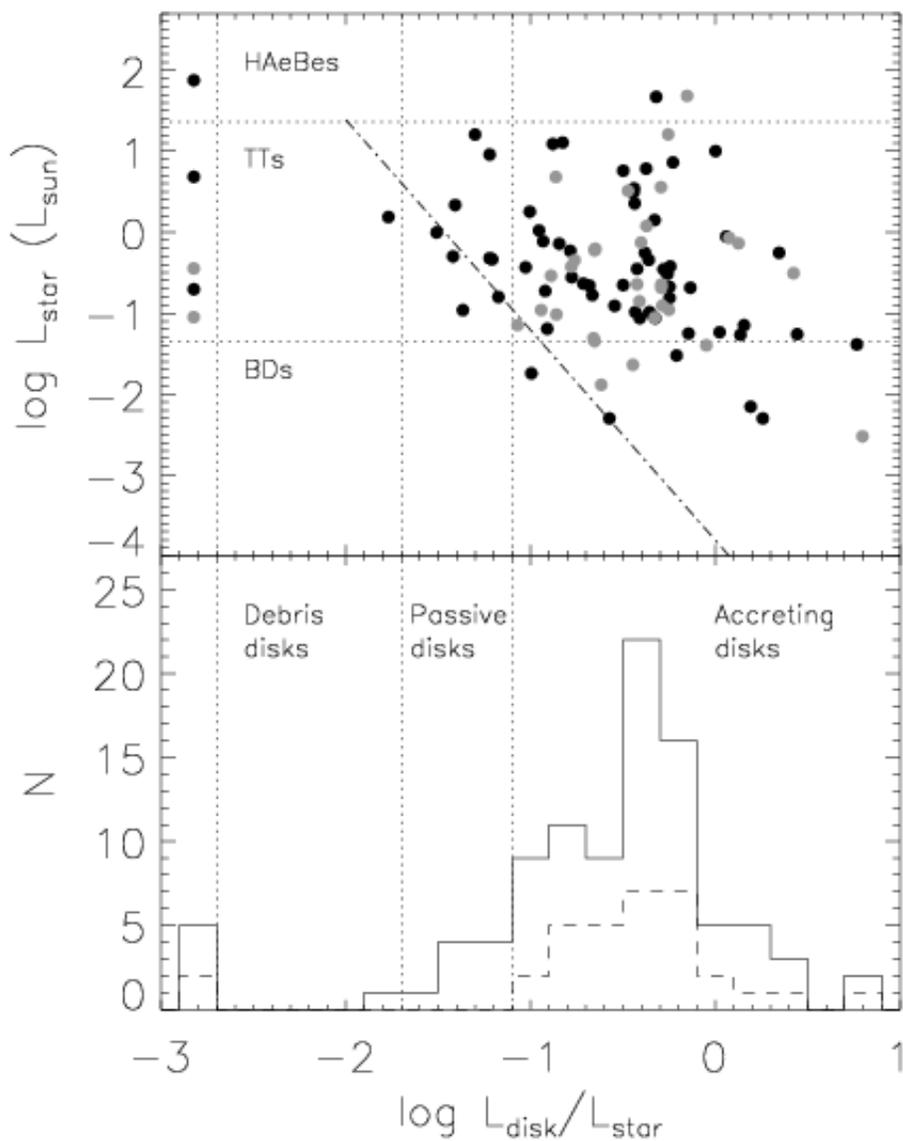}
\caption{{\sl Top}: Correlation plot between the stellar luminosities and 
the disk fractional luminosities for the PMS stars in Lupus. Solid and
grey dots are single and binary stars, respectively and the dot-dashed
line marks the approximate 24 $\mu$m detection limit. {\sl Bottom}:
distribution of disk fractional luminosities for the Lupus sample.
The solid and dashed lines show the single and binary stars
distributions, respectively. Objects with negligible IR excess are
shown in the $-3$ abscisa for completeness. \label{lstar_ldisk}}
\end{figure}

The upper panel of Figure \ref{lstar_ldisk} shows the stellar
luminosity versus the disk fractional luminosity for the sample of YSOs
and PMS objects in Lupus. The grey dots correspond to visual binary
systems, as reported in \S~\ref{binaries}, while the black dots
represent single stars. In the lower planel, we have collapsed the
stellar luminosity axis to get a histogram on disk fractional
luminosities for the entire sample. This diagram offers a complete
view of the star plus disk population in Lupus and shows several
interesting results and trends. The $\log L_{\rm disk}/L_{\rm star}$
ratio of objects in which we could not detect any sizable IR excess
were set to $-3$ so that they appear to the left of both panels. The
disk luminosity ratios show a strong dependence on the band used to
normalize the dereddened fluxes to the stellar photospheric model and
the extinction determination. These are only accurate in those objects
with known spectral types, but the deviations are small enough to
extract statistical trends from the diagram.

Horizontal dashed lines in the upper panel separate the regions where
the luminosities correspond to those of Herbig Ae stars (HAeBe's,
spectral types earlier than F0), T Tauri stars (down to M7) and Brown
dwarfs (BDs, below M7) of 1 Myr according to \cite{Baraffe1998}. The
only two known HAeBe's in Lupus, HR 5999 and HR 6000, share the upper
region with SSTc2d J161000.1-385401, a intermediate-mass HAeBe
candidate in Lupus with small IR excess, which could be a background
post-AGB star, given the resemblance of its SED to those of the old
field objects found in Serpens \citep{Harvey2007b}. It can be also seen
here, as well as in figure \ref{lum_hist}, that the great majority of
the objects in Lupus are M-type T Tauri and very low mass stars.

The vertical lines separate the regimes where the luminosity ratios
can be explained by different mechanisms: accretion disks ($L_{\rm
  disk}/L_{\rm star}$ $>$ 0.1), and passive reprocessing disks (0.02
$<$ $L_{\rm disk}/L_{\rm star}$ $<$ 0.08 \citealt{Kenyon1987}), and
finally ``debris''-like disks ($L_{\rm disk}/L_{\rm star}$ $<$ 0.02,
which is a conservative upper limit for the fractional disk luminosity
of a transitional or young debris disk, \citealt{Currie2007}). The
distribution of points and the histogram in the bottom panel show that
the great majority of the disks found in Lupus have luminosities which
imply some degree of accretion, similar to what was found in the Cha
II cloud \citep{Alcala2008}. This result contrasts with the 40\%
of disks with small IR excesses found in \S~\ref{SEDs} and shows
that detailed Spitzer SED morphology studies are needed to provide a better
description of the inner disks, which are greatly degenerate with the flux
ratios in the T Tauri phase.

The grey dots and dashed histogram show the distribution of those YSOs
in the plot which were found to be visual binaries in the optical
images by Comer\'on et al (in prep.). It has been suggested that close
binaries ($<$ 20 AU) play a central role in the evolution of the disks
by dramatically reducing the incidence of disks around multiple stars
\citep{Bouwman2006}, while wide ($>$ 100 AU) binaries do not affect
significatively the presence and evolution of the circumstellar disks
\citep{Pascucci2007}. Here we present further evidence for this lack
of correlation between disks and wide orbit companions in Lupus, where
binaries and single stars show the same distribution in terms of
luminosity or disk fractional luminosity.

Finally, there seems to be a trend by which lower luminosity objects show
higher disk to star luminosity ratios than higher luminosity
stars. However, the luminosity
distribution of the `H' and `T'-type objects only, the least affected by the 
detection limits in the IR,
is statistically equal to the luminosity distribution for the total sample. 
Therefore, the trend reflects an observational bias caused by the
limited detection capability of excesses at longer wavelengths for
the faintest sources. The flux sensitivity limit for 24 $\mu$m detection 
with S/N $>$ 3 is $\approx$ 1 mJy in our observations. This corresponds to photospheric
emission of a $\approx$ 0.2 $L_\odot$ ($\log L_{\rm star} \approx
-0.7$) star in our sample, which corresponds to a 0.2 $M_\odot$ (M5.5)
star of 1 Myr. The star Sz 92 is close to the border with 0.47
$L_\odot$ ($\log L_{\rm star} \approx -0.3$), a very small excess at
24 $\mu$m ($L_{\rm disk}/L_{\rm star}$ = 0.06, $\log L_{\rm
  disk}/L_{\rm star}$ = -1.2) and a flux density at 24 $\mu$m of 1.5
mJy. Fainter objects with larger excesses will be detected as soon as
their 24 $\mu$m is larger than the limiting flux and this number will
obviously depend on the level of background emission and crowdedness
of the area where a given object is. This limit, shown with a
dot-dashed line in the figure, explains the skewness of the
data points towards the bottom right of the diagram and illustrates
the disk parameter space probed with these observations. The survey is not
sensitive to ``debris''-like disks around stars less massive than 0.2
$M_\odot$ but detects all actively accreting disks down to the
substellar regime since those should present detectable IR excess at
all IRAC bands which are much more sensitive.

\subsection{A Two-Dimensional Classification System}
\label{2D_param}

Finally, to explain the diversity in the SEDs, we developed a {\sl
  second-order} set of parameters to classify the diversity of 2MASS +
IRAC + MIPS SEDs that are found in the star-forming regions:
$\lambda_{\rm turn-off}$ and $\alpha_{\rm excess}$  
\citep{Cieza2007}. In short, $\lambda_{\rm excess}$ is the last
wavelength in microns where the observed flux is photospheric and
$\alpha_{\rm excess}$ is the slope computed as ${\rm dlog}(\lambda
F_\lambda)/{\rm dlog}(\lambda)$ starting from $\lambda_{\rm turn-off}$
and up to 70 $\mu$m when available. The first parameter gives us an
indication of how close the circumstellar matter is to the central
object and the second one is a measure of how optically thick and flared
it is. Both numbers can be found for all the stars with disks in the
sample in Table \ref{sed_results}.

\begin{figure}[!ht]
\includegraphics[angle=90.,width=14cm]{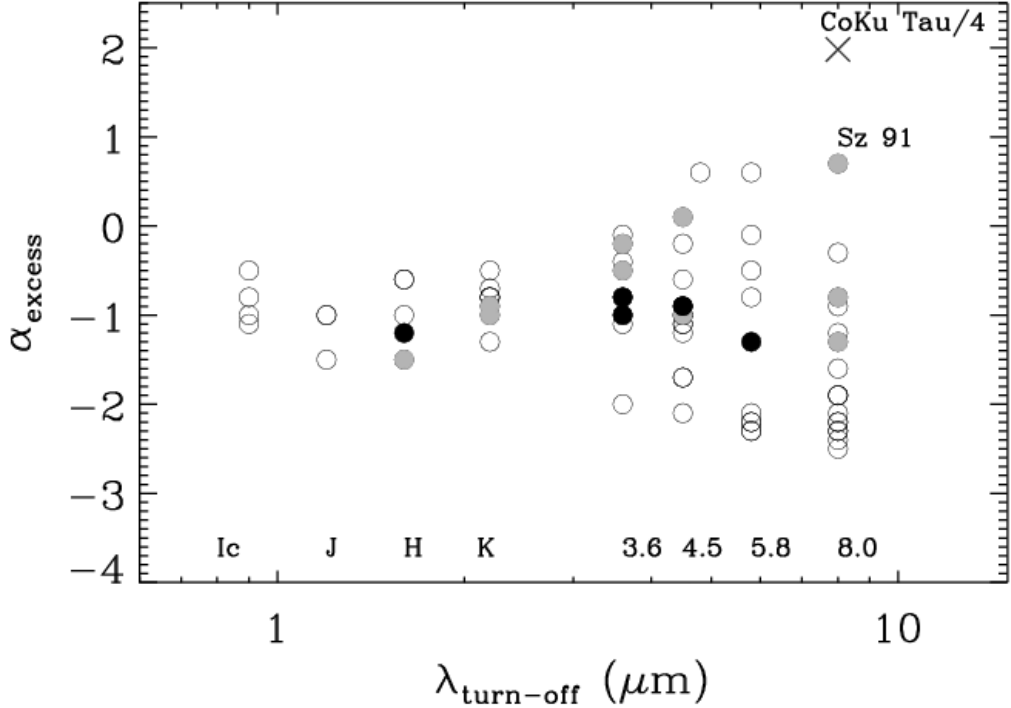}
\caption{Distribution of excess slopes $\alpha_{\rm
   excess}$ with respect to the wavelength at which the infrared
   excess begins $\lambda_{\rm turn-off}$ for the Lupus sample (open
   circles). Solid dots are objects which were detected at 1.3 mm by
   \citet{Nuernberger1997} and grey dots are objects for which only
   upper limits were obtained at 1.3 mm. CoKu Tau/4 shows the position
   of the cold disks in this diagram for comparison. \label{lto_ae}}
\end{figure}

Figure \ref{lto_ae} shows both values for the sample of YSOs in Lupus
for which we could perform a sensible SED fit. Three objects (namely
Par-Lup3-3, Par-Lup3-4 and Sz 99) were not included in this analysis
since their SEDs did not allow for a good determination of
$\lambda_{\rm turn-off}$ due to too much IR excess at short
wavelengths in the first two cases and lack of IRAC data in the third
one. \citet{Cieza2007} showed that $\lambda_{\rm turn-off}$ is a good
``evolutionary'' parameter, since it separates efficiently the
Classical T Tauri stars from the Weak Line T Tauri stars. They also
showed that the range of $\alpha_{\rm excess}$ values, which scale
with the amount of emitting material in the disk for central objects
of similar luminosity, grows with $\lambda_{\rm turn-off}$, which was
interpreted by those authors as a signature of multiple ``evolutionary
paths'' of the inner disks of T Tauri stars from the actively
accreting phase (with typically $\lambda_{\rm turn-off} < 2\mu$m) to
progressively more settled and optically thin disks with larger values
of $\lambda_{\rm turn-off}$. The Lupus sample shows a remarkably
similar distribution in Figure \ref{lto_ae}, as also do the Serpens and 
Cha II samples, suggesting that the result is a common feature of 
young stellar populations.

The diagram is also useful for identifying transitional disks, or cold
disks \citep{Calvet2005,Brown2007}, i.e. systems with an optically
thick outer disk with large inner holes (several AU). These objects
appear in the right-upper part of this diagram. For comparison, we
show the position of CoKu Tau/4, a well known T Tauri disk with a
inner hole of $\sim$ 10 AU devoid of matter \citep{Dalessio2005}. We
also labeled one of the new cold disks found in Lupus III, Sz 91,
which is indeed is the most extreme upper-right object of the whole
distribution. The determination of robust cut-offs in this diagram for
identifying ``bona-fide'' disks with holes and their frequency in the
full c2d data set is underway and will be presented elsewhere.

\begin{figure}[!ht]
\includegraphics[angle=90.,width=15cm]{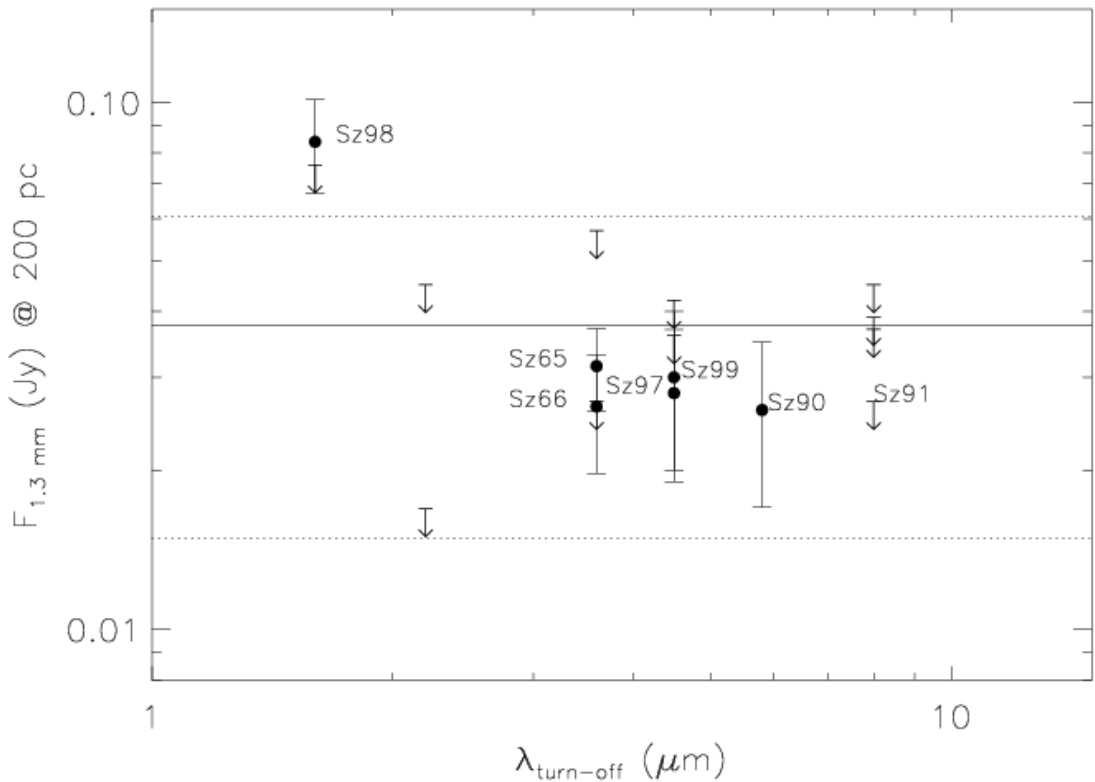}
\caption{Relationship between the continuum flux at 1.3 mm
  from \citet{Nuernberger1997} and the wavelength $\lambda_{\rm
    turn-off}$ at which the infrared excess begins for the Classical T
  Tauri stars in Lupus. The arrow show the positions of the upper
  limits, including that for the cold disk Sz 91. \label{lto_mm}}
\end{figure}

In order to compare the inner disk architectures of our sample
of Lupus PMS stars with their outer disks, we have marked the `Sz'
stars detected in 1.3 mm continuum observations by
\cite{Nuernberger1997} with black dots, and the upper limits with grey
dots in Figure \ref{lto_mm}.  As expected, all the millimeter 
detections and upper limits in the diagram
appear at $\alpha_{\rm excess} \geq -1.5$, which imply relatively
optically thick inner disks for all values of $\lambda_{\rm
  turn-off}$.

  However, the really interesting result comes when comparing the
  millimeter fluxes, which are a proxy for disk total mass
  \citep{Beckwith1991,Andrews2005}, with $\lambda_{\rm turn-off}$, which is an
  indirect indicator of the degree of dust depletion and clearing of
  the inner disk. Figure \ref{lto_mm} shows the 1.3 mm continuum
  fluxes and upper limits for the Lupus sample normalized to 200
  pc compared to the $\lambda_{\rm turn-off}$. The horizontal line shows the 
  average of the normalized
  millimeter fluxes and the dotted lines the 1$\sigma$ error bar of
  the mean. With the exception of Sz 98, disks with a range of
  $\lambda_{\rm turn-off}$ or inner disk clearings appear to show
  quite similar total disk masses, being all well inside the standard
  deviation of the mean. The case of Sz 98 could be different, since
  its SED suggests a possible contribution of a remnant envelope,
  which would contribute to the total millimeter flux. In spite of
  the low number of detections, the general result is that objects
  with a range of inner disk configurations show total disk masses in
  the same order of magnitude. This suggests an inside-out process
  which empties the inner disk while the outer disk remain unchanged. 
  \cite{Alexander2007} show that
  disk clearing by photoevaporation is more efficient for smaller disk
  masses while e.g. \cite{Edgar2007} argue that the efficiency of
  giant planet formation and migration, the main competing mechanism
  for inner disk dissipation, is proportional to the disk mass. The apparent
  lack of correlation between inner disk clearing and total disk mass
  is not compatible with the photoevaporation scenario while it is
  with the planet formation one. However, the current
  sample only contains 8 good detections, and several caveats should be
  taken into account in the interpretation of this figure: apart from
  the low number statistics, the $\lambda_{\rm turn-off}$ value will
  be sensitive to the inclination angle, while the millimeter flux is
  considered roughly insensitive to it (e.g. \citealt{Dalessio1999}).
  More work will be done to quantify the extent of this relation with
  larger data sets and will be presented elsewhere.

\clearpage

\subsection{Outflow sources}
\label{outflow_sources}

\begin{figure}[!ht]
\epsscale{0.9}
\plotone{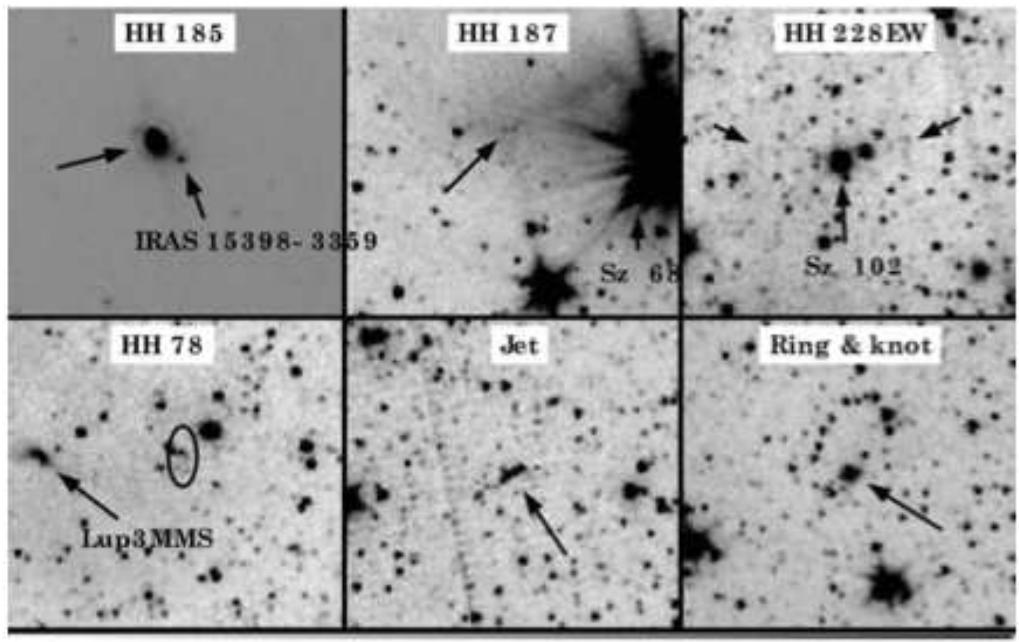}
\caption{Known HH objects detected in Lupus with IRAC, together with a
candidate new outflow source in Lupus III and a nebulous object in a
ring of stars in Lupus IV. All images have a size of 2$'$x2$'$,
linear stretch and are from the IRAC1 band at 3.6 $\mu$m, except
the one showing HH 185, which is IRAC4 at 8.0 $\mu$m. North is up and
East is to the left. Coordinates are given in Table
\ref{outflows}. \label{hh_objects}}
\end{figure}

\begin{figure}[!ht]
\epsscale{0.9}
\plotone{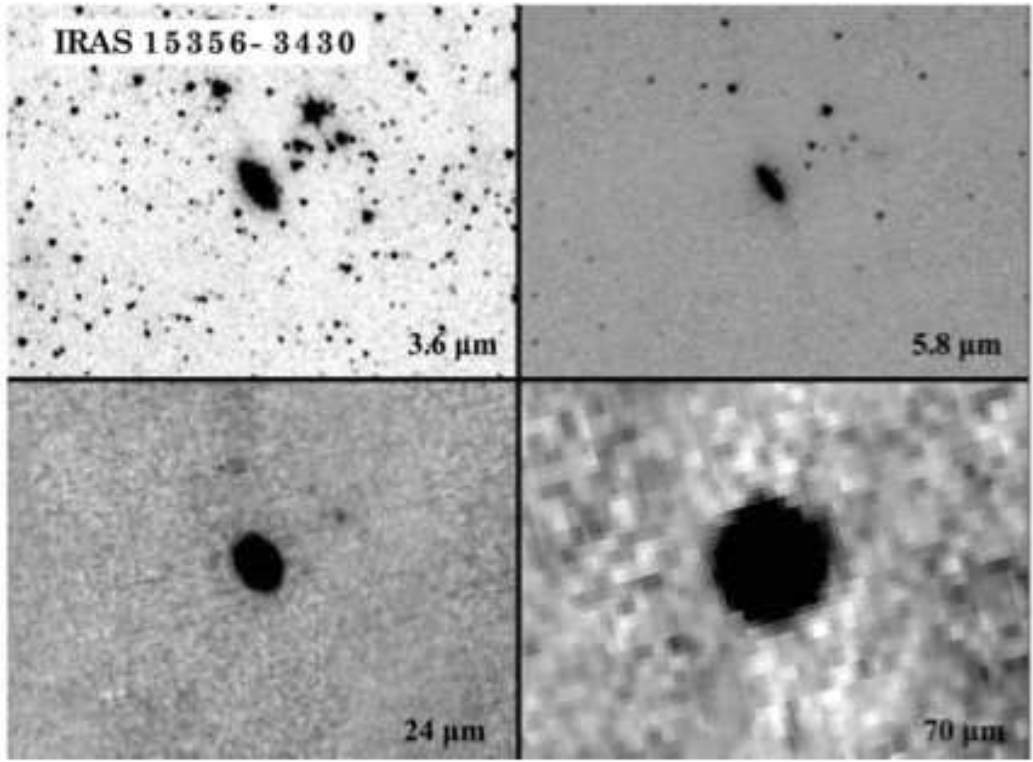}
\caption{IRAC and MIPS images of the extended nebula around the 
source IRAS 15356-3430. All images have a size of 4$'$x5$'$ and
linear stretch. North is up and East is to the left. Coordinates
are given in Table \ref{outflows}. \label{irasHH}}
\end{figure}

Several previous studies have shown the suitability of the IRAC bands,
and more specifically IRAC band 2 at 4.5 $\mu$m for detecting new
high-velocity shocked outflows via the H$_2$ emission line found in
that band (e.g. \citealt{Noriega-Crespo2004},
\citealt{Jorgensen2006}). In this section we present the Spitzer data
for the known outflow sources in the surveyed Lupus clouds (from
the review by Comer\'on 2008, in prep.) and new emission nebulae found in the IRAC
mosaics which could be high velocity outflows and shocks. Table
\ref{outflows} summarizes the results of this effort and Figures
\ref{hh_objects} and \ref{irasHH} show them.

The HH 185 object, reported by \cite{Heyer1989}, shows an ellipsoidal
shape close to the Flat SED object IRAS 15398-3359 in all IRAC
bands.  HH 187 \citep{Heyer1989} is a faint nebulosity seen at IRAC1
and 2 bands only, and brighter in the latter. HH 228 and HH 78 also
appear in the Spitzer bands as small ($\sim$ 4$''$) faint nebulosities
around their driving sources. The small [SII] outflow in Par-Lup3-4
reported by \cite{Fernandez2005} and the HH 186 36$''$ [SII] jet
around Sz 68 are not detected by the Spitzer observations. Only
nebulosities appearing consistently in both epoch images were included
to avoid artifacts. All the IRAC images were searched for nebulosities
and we report here the detection a jet-like nebulous structure found
at all IRAC bands in Lupus III and a nebulous object in a ring of
point sources found in Lupus IV. The driving sources of these two
objects were not identified.

Figure \ref{irasHH} shows the remarkable emission around the the
bright class I source IRAS 15356-3430. A 40$''$ elliptical nebulosity
is observed in all IRAC bands and also detected in MIPS 24 and 70
$\mu$m bands. The nebulosity is also found with a slightly smaller
extension in the optical and 2MASS images. \cite{Carballo1992}
classified it as a possible YSO or a galaxy based on its IRAS colors
and \cite{Strauss1992} confirmed its YSO nature with optical
spectroscopy. Its 70 $\mu$m flux is as high as that of Lupus3MMS but it
has never been observed in the millimeter wavelengths. The SED slope
of the object is smaller than that of Lupus3MMS or IRAS 15398-3359 and
that makes it detectable in the near-IR bands. In any case, the
Spitzer SEDs of all these sources with extended emission are clearly
associated with extremelly embedded and actively accreting objects.

Interestingly, one class II and one class I YSO (Sz 98, Lupus3MMS)
happen to have detection quality flag of `K' only in IRAC2 band at 4.5
$\mu$m in the c2d catalog, which means that these sources could not be
fitted by a stellar PSF at that wavelength. This suggests the possible
presence of unresolved ($< 2''$) outflows detected preferentially
through the shocked H$_2$ line in IRAC band 2 and is consistent with
the fact that all three of them are extremely active accretors with
large $\alpha_{(K-24 \mu m)}$ values and excesses well above the median SED of the T
Tauri stars in Taurus. Similarly, in the case of the highly accreting
HAeBe star HR 5999, a point source could not be extracted for IRAC
bands 1 and 2 so the corresponding fluxes were removed for the SED
analysis of that object. This again hints to bright extended emission
around 4.5 $\mu$m on a scale smaller than the large PSF of the star.

\begin{table}[!h]
\begin{center}
\caption{Probable high-velocity outflows and nebulae in Lupus\label{outflows}}
\begin{tabular}{lccccc}
\tableline  \tableline
Assoc.   & R.A.  &  Dec.   &  & & \\
YSO ID   & (J2000)  &  (J2000)   & HH identification / driving source   & Ref. & Notes \\
\tableline
\multicolumn{6}{c}{\bf Lupus I} \\
 2  & 15 38 48.36 &  -34 40 38.2   & IRAS 15356-3430              & 4 & a \\
 10  & 15 43 01.29 &  -34 09 15.42  & HH 185 / IRAS 15398-3359     & 1 & b \\
15, 16  & 15 45 19.03 &  -34 17 32.43  & HH 187 / Sz 68 / Sz 69        & 1 & c \\
\multicolumn{6}{c}{\bf Lupus III} \\
47  &  16 08 32.11 &  -39 03 18.23 & HH 228E / Sz 102            & 2 & d \\
47  &  16 08 27.18 &  -39 03 00.68 & HH 228W / Sz 102            & 2 & d \\
87  &  16 09 12.38 &  -39 05 00.61 & HH 78 / Lupus3 MMS       & 3 & e \\
\ldots & 16 10 57.95  &  -38 04 37.90 & \ldots                     & 4 & f \\
\multicolumn{6}{c}{\bf Lupus IV} \\
\ldots & 16 00 39.04  &  -42 06 51.51 & \ldots                   & 4 & h \\
\tableline
\end{tabular}

\tablenotetext{}{a~~Bright 40 $''$ nebula towards the NE in all IRAC and MIPS bands.}
\tablenotetext{}{b~~Bright 12 $''$ nebula towards the NE in all IRAC bands.}
\tablenotetext{}{c~~Faint 3$''$ nebula knot at 3.6 and 4.5 $\mu$m.}
\tablenotetext{}{d~~Faint knots at 3.6 and 4.5 $\mu$m.}
\tablenotetext{}{e~~Faint knots mostly at 3.6 and 4.5 $\mu$m.}
\tablenotetext{}{f~~8$''$ jet in all IRAC bands.}
\tablenotetext{}{h~~9$''$ knots at all IRAC bands in a ring of 15 point sources.}

\tablerefs{1) \citet{Heyer1989}; 2) \citet{Krautter1986} ; 3) \citet{Reipurth1988}; 
           4) This work}	
\end{center}
\end{table}

\newpage

\section{Clouds and Cluster properties}
\label{clouds}

We can use the Spitzer data to also study the star formation history in the Lupus
clouds. In particular, we want to estimate the star-formation activity and the 
clustering properties of the young stellar population in relation to the current 
cloud structure.


\subsection{Extinction maps in Lupus}
\label{extinction}

The structure of the interstellar medium in the direction of the Lupus
clouds has been extensively studied with (sub)millimeter molecular
line observations \citep[e.g.,
][]{Murphy1986,Hara1999,Vilas-Boas2000,Tachihara2001} as well as with
optical and infrared star-counts (\citealt{Cambresy1997},
\citeyear{Cambresy1999}) on spatial scales of several
arcminutes. These studies reveal a large clumpy structure with a
substantial number of overdensities, dominated by the ``classical''
Lupus I to IV clouds, where star-formation has also been detected. See
Comer\'on 2008 (in prep.) for a complete review of these observations.

The c2d data offer an exceptional tool for producing extinction maps
in all the imaged clouds. This is done by estimating the visual
extinction towards each source classified as a background star, based
on their SED from 1.25 $\mu$m to 24 $\mu$m. This provides multiple
line-of-sight measures, which are then convolved with Gaussian beams
of 90$''$ to 300$''$ to produce homogeneous extinction maps. These
maps are part of the c2d data delivery. In the case of Lupus I, the
low number of background stars only allowed reliable construction of
an extinction map with a minimum FWHM of 120$''$. For more information
about these maps, see \cite{Evans2007}. 

The maps with the largest beams (300$''$) are used to estimate the
enclosed cloud masses (\S~\ref{SFE}) and to compare
them with the position of the different YSOs (\S~\ref{spatial_distribution}).
They can be seen in Figures \ref{spa_distr_I} to 
\ref{spa_distr_IV}. The extinction traces the northern cloud
in the Lupus I map with a peak of $A_V \sim$ 23 mag and shows two large
clumps in Lupus III and IV, with peak extinctions of 33 and 36
mag, respectively. The maximum in Lupus III coincides with the
rich star-forming cluster, while it roughly coincides with the Flat
SED source SSTc2d J160115.6-415235 in Lupus IV. This is the first
report of such a high extinction inactive core in Lupus IV to our knowledge,
which was previously unnoticed due to its current lack of activity.

\subsection{Spatial distribution and clustering of YSOs in Lupus}
\label{spatial_distribution}

\begin{figure}[!ht]
\plotone{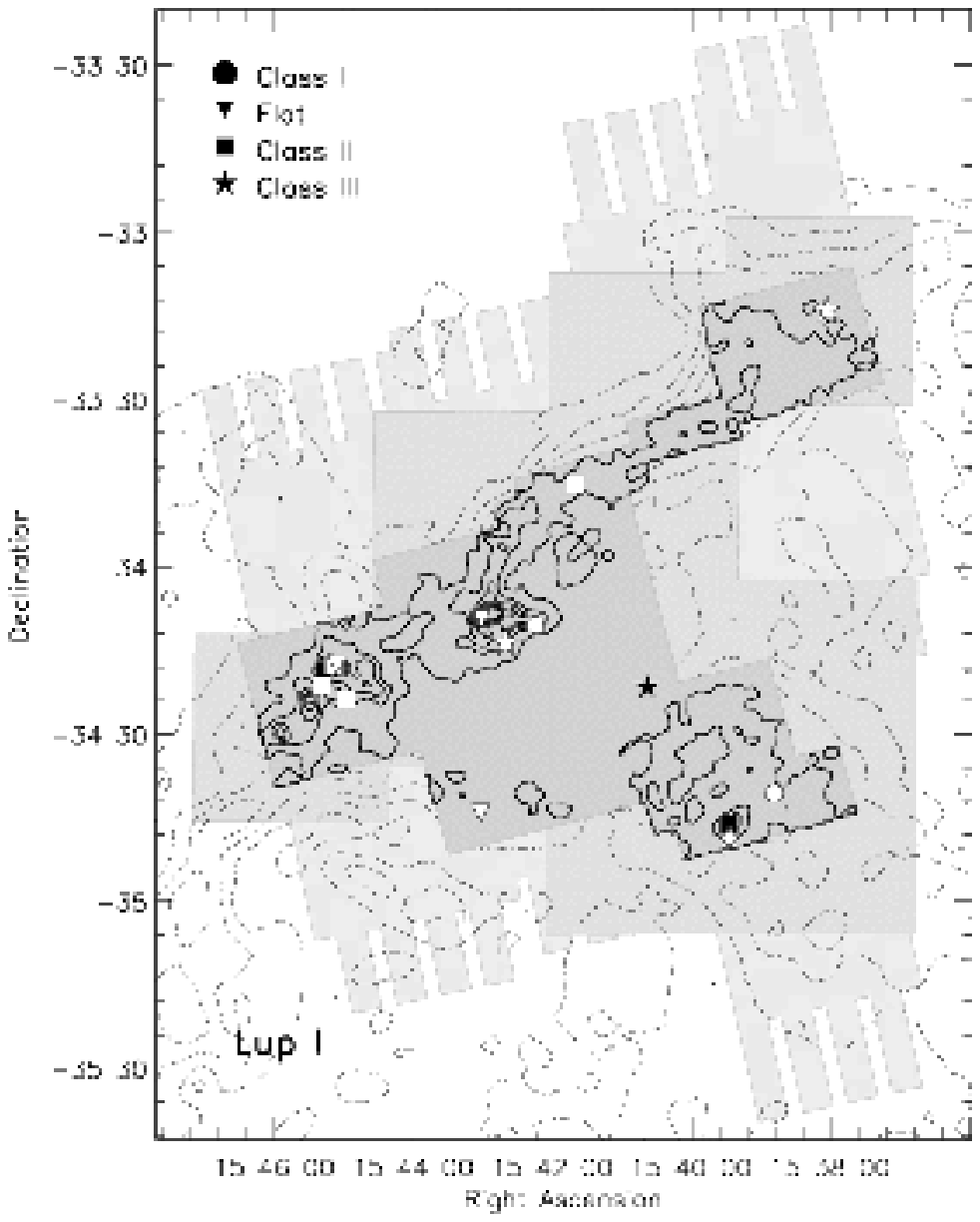}
\caption{Spatial distribution of the spectrally confirmed PMS objects
  (black symbols) and candidates (white symbols) in Lupus~I as
  function of Lada Class, over-plotted on the contours from the c2d
  extinction map (continuous lines).  The contour levels of extinction
  are from 2~mag to 20~mag, in steps of 2~mag. The shaded areas, from
  light to dark-gray, display the regions observed with MIPS, WFI and
  IRAC, respectively.  The dashed lines outside the IRAC area are the
  contour levels of extinction from \citet{Cambresy1999}, from 1~mag
  to 6~mag in steps of 0.35~mag. The higher resolution and sensitivity
  to higher $A_V$ of the c2d extinction map with respect to that of
  \citet{Cambresy1999} can be appreciated.  
\label{spa_distr_I}}
\end{figure}

\begin{figure}[!ht]
\plotone{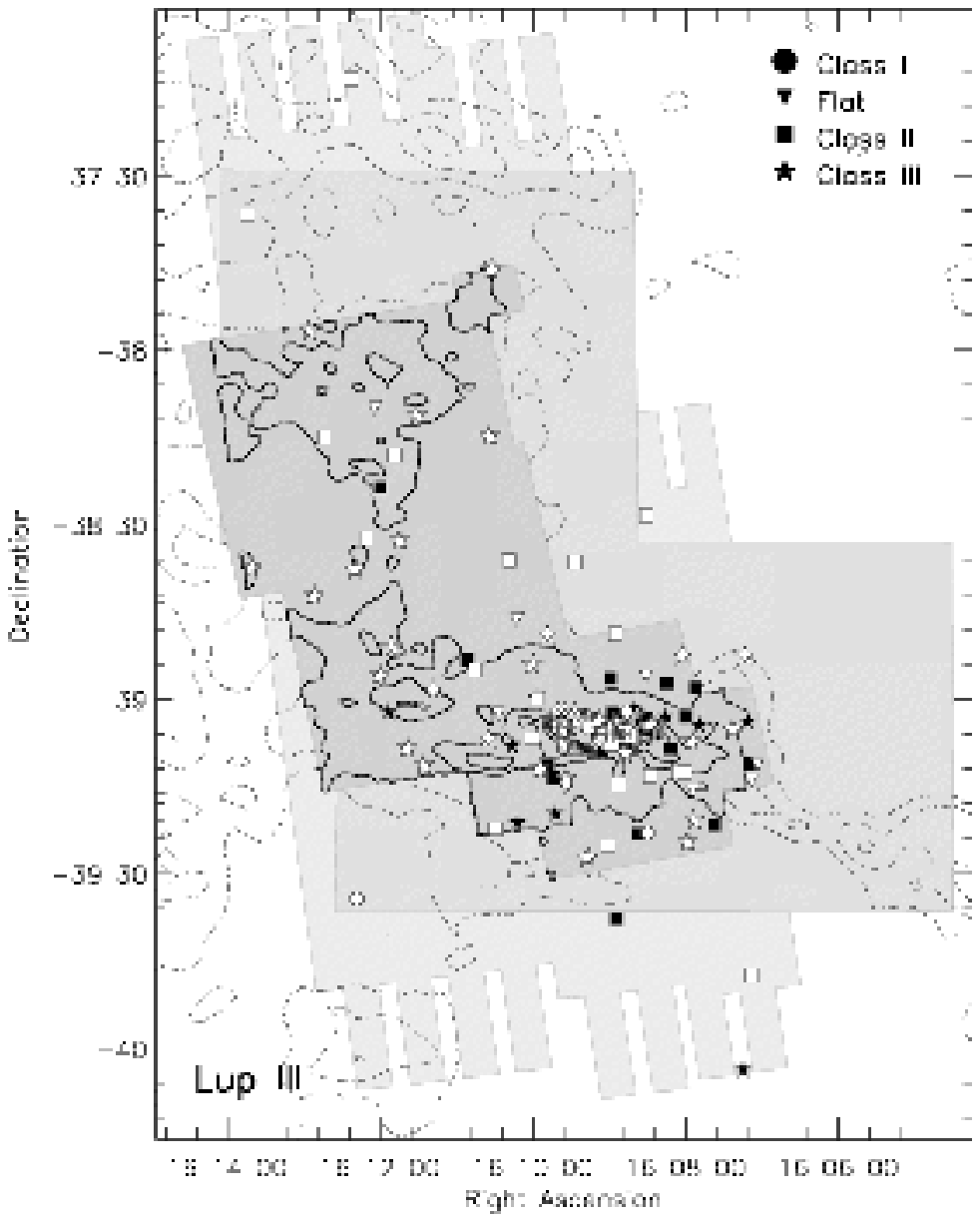}
\caption{Spatial distribution of the PMS objects (filled symbols) and 
       candidates (open symbols) in Lupus~III as function of Lada
       Class, over-plotted on the contours from the c2d extinction map
       (continuous lines).  Symbols as in Figure~\ref{spa_distr_I}.
       \label{spa_distr_III}}
\end{figure}

\begin{figure}[!ht]
\plotone{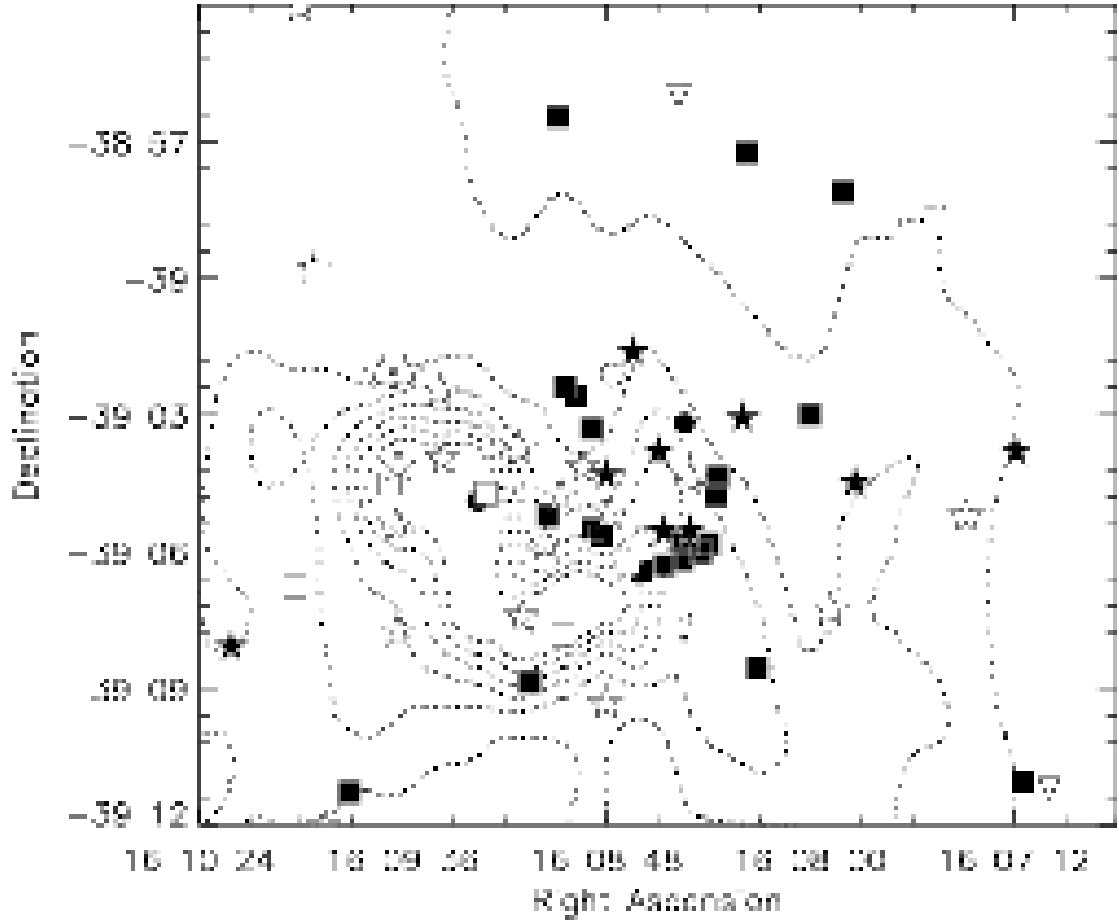}
\caption{Zoom of the dense star-forming cluster in Lupus III. Symbols
         as in Figure~\ref{spa_distr_I}, except the dashed lines, 
         which are the 2 to 14 mag contours on the c2d extinction map in
        steps of 2 magnitudes.
	\label{spa_distr_III_zoom}}
\end{figure}

\begin{figure}[!ht]
\plotone{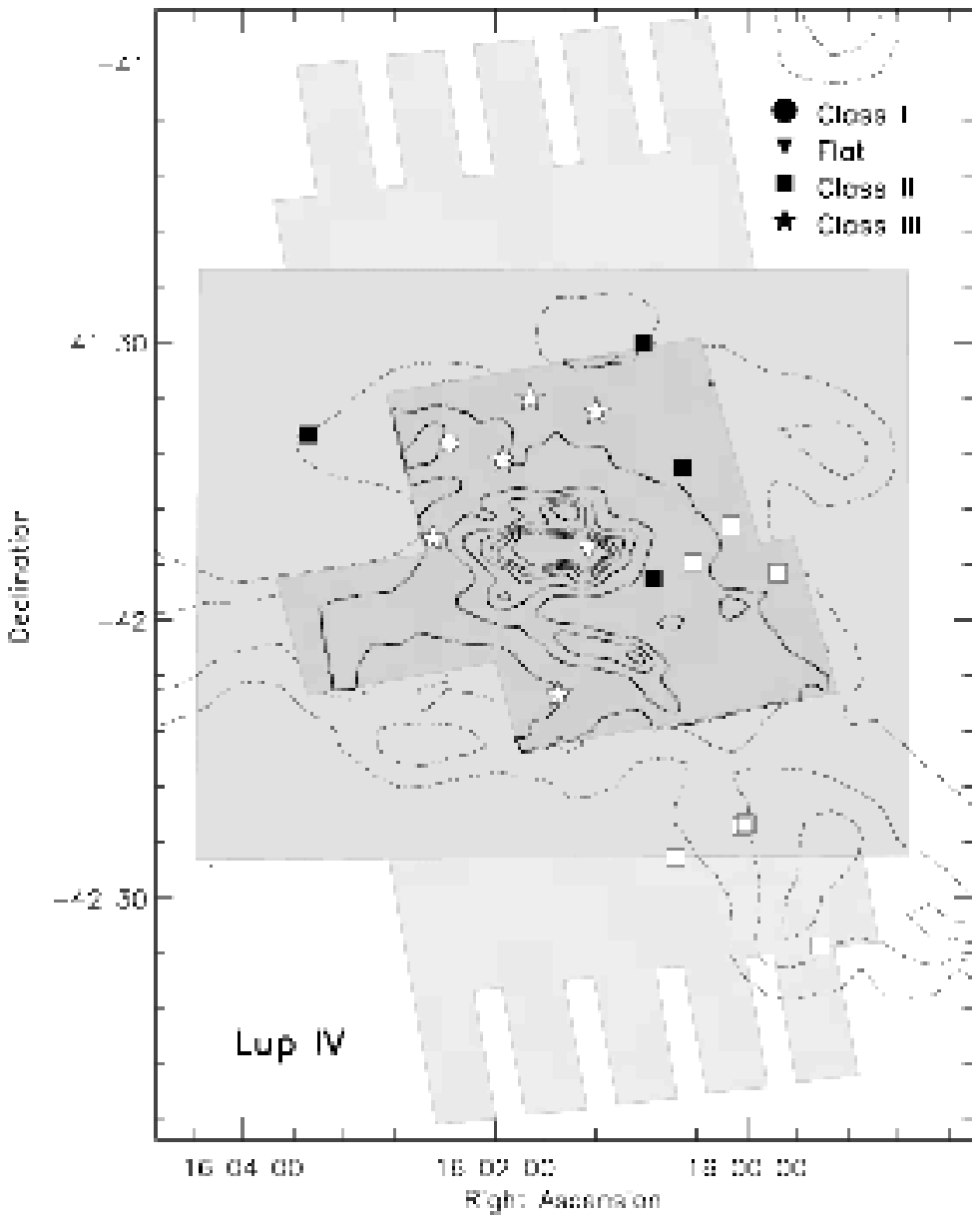}
\caption{Spatial distribution of the PMS objects (filled symbols) and 
       candidates (open symbols) in Lupus~IV as function of Lada
       Class, over-plotted on the contours from the c2d extinction map
       (continuous lines).  Symbols as in Figure~\ref{spa_distr_I}.
       \label{spa_distr_IV}}
\end{figure}

Figures \ref{spa_distr_I} to \ref{spa_distr_IV} show the spatial
distribution of the PMS stars in the three Lupus clouds compared with
the c2d extinction maps: In Lupus I, the most striking result is that the 
majority of the YSOs
fall in a ridge of high extinction which extends in the North of the
cloud with NE-SW direction.  A southern high extinction region
contains the pair of T Tauri stars Sz 68 and Sz 69. The majority of
the objects, of any SED class but with larger abundance of Class I
and Flat SED sources, appear very close to high
extinction regions, with the exception of the Class III sources Sz 67
and SSTc2d J153803.1-331358 and the Flat spectrum source IRAS
15398-3359. The close match between the YSO distribution
and extinction map in the region and the early class of
most of the objects suggests that we are seeing the objects in the
places where they were just born.

The sources in Lupus III are dramatically concentrated in the dense
star-forming cluster at RA $\sim$ 16$^h$09$^m$ and DEC $\sim$
-39$^{\rm o}$10$'$, which contains the two bright intermediate mass Herbig
Ae/Be stars HR 5999 and HR 6000 in the centre.  In this case, objects of
all classes are found in the vicinity of the dense cluster and at large
distances from it. Figure~\ref{spa_distr_III_zoom} zooms into the cluster, showing
the extraordinary density of PMS objects at the highest extinction
area in the whole cloud, where star formation is extremely
active. Amongst many Class II and III sources, the cluster contains
four Class I sources, including Lupus3MMS, claimed to be the only
Class 0 object in the Lupus clouds \citep{Tachihara2007}. Note that
only 3 of them appear in Figure~\ref{spa_distr_III_zoom}, while the
forth is visible in Figure ~\ref{spa_distr_III} to the South of the
cluster.

Figure~\ref{spa_distr_IV} shows the distribution of YSOs in Lupus IV
around a very dense, yet quite unpopulated, extinction peak at RA
$\sim$ 16$^h$02$m$ and DEC $\sim$ -41$^{\rm o}$50$'$ which contains
the only Flat spectrum source in the cloud:
SSTc2d J160115.6-415235. The rest of the Class II and III sources,
together with the Class I source IRAS 15589-4134 are found surrounding
the core, but following a different spatial pattern, which suggests that
they might have been formed in a different structure. As the 
contour plot nicely shows, the peak extinction at this clump is higher than that of
the star-forming cluster in Lupus III (see also \S~\ref{extinction}).
This fact together with the relative Class III predominance in Lupus IV 
suggests that star-formation may have taken place before in other
regions of this cloud and may be about to start in this core. We
must note, however, that an object like Lupus3 MMS would have not
been detected in this core, since it was only thanks to its millimeter detection
that it was identified as a cloud member in Lupus III
and such observations are so far not available for Lupus IV.

The observation of these complex extinction structures and YSO
distributions has stimulated for many years the discussion on how
star-formation takes place, whether it is a hierarchical process or
whether it takes place in a centrally-condensed way (see \citealt{Lada2003}, LL03
hereafter, and references therein).  In order to study the subject with
the new Spitzer data, we need to set some standards
that will be applied across the c2d observational set. Similar to the
other c2d clouds (see, e.g., \citealt{Alcala2008}) we identified
substructure in the Lupus clouds using the nearest neighbor algorithm
applied by \cite{Gutermuth2005} (see J{\o}rgensen et al. 2008, in prep. for
details). Following the discussion by J{\o}rgensen et al. we divide
concentrations of PMS objects up into ``clusters'' or ``groups''
depending on whether they have more or less than 35 members at a given
volume density level. We furthermore identify structures with volume
densities higher than 1~$M_\odot$~pc$^{-3}$ (the criterion for a
cluster by LL03) or 25~$M_\odot$~pc$^{-3}$ as being ``loose'' or
``tight'', respectively.

\begin{figure}
\plotone{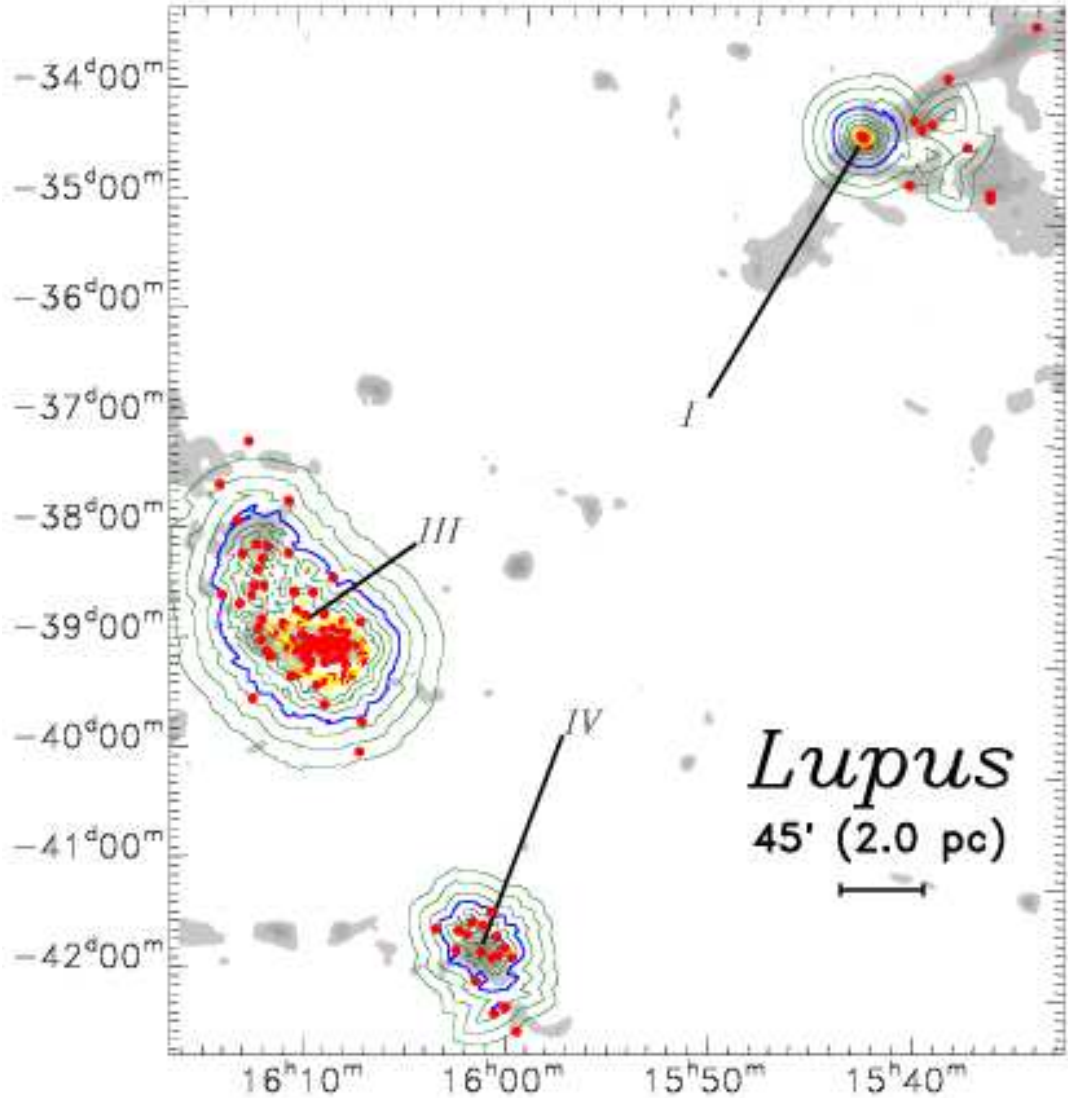}
\caption{Volume density plot for the three Lupus clouds as determined
  with the nearest-neighbor algorithm by J. J{\o}rgensen et al. (2007,
  in preparation). The derived contours are compared with extinction
  map from \cite{Cambresy1999} in gray scale from 2 to 4.1 in steps
  of 0.7 mag. The blue contour
  corresponds to the 1$\times$LL03 level, the yellow to the
  25$\times$LL03 level, and the green ones correspond to levels of
  0.125, 0.25, 0.5, 2, 4 and 8 times that level. The two groups and
  the cluster in Lupus III are identified with labels.\label{vol_density}}
\end{figure}

Figure \ref{vol_density} shows the result of the nearest neighbor
analysis for the whole Lupus sample. When applied to Lupus this
algorithm breaks the complex into three separate entities: two 
tight concentrations of PMS
stars at the 25~M$_\odot$~pc$^{-3}$ level and a loose one at the
1~M$_\odot$~pc$^{-3}$ level (taking into account the difference in
distance between Lupus I and IV on one hand and Lupus III on the other).
The Lupus IV cloud has one loose group and Lupus I a tight group.  The
Lupus III cloud has one tight cluster containing 118 members. Table
\ref{clustering_results} shows the number of PMS objects of the
different classes in the regions identified above. This analysis
allows ordering the objects by degree of clustering, with Lupus III
being the most clustered region, followed by Lupus I and Lupus IV.

Overall, both the cloud density structure and the distribution of stars in the
three regions suggest a centrally-condensed structure dominating
the star-formation process only in Lupus III and a more disperse distribution
of volume density enhancements in the other two clouds, mostly at
Lupus IV, where star-forming cores have formed along a cloud filament
in the NE-SW direction. As a whole,
the three clouds clearly show a hierarchical structure with separate
gas concentrations that evolve independently. This scenario argues
in favor of a larger presence of turbulence over gravitational forces
(LL03), which is consistent with the Lupus clouds being
flanked by the large Scorpius-Centaurus OB associations 
and therefore subject to their strong high-energy radiation 
\citep{Comeron2007a}.

\clearpage

\subsection{Star Formation Rates}
\label{SFR}



We can also use our complete samples to estimate the star formation
rates. In section \ref{YSOc_list} we computed the total numbers of YSOs and
PMS objects for each of the three clouds. The IRAC+MIPS coverage areas
reported in \S~\ref{IRAC_BDP} were used for the YSOs and the
overlapping MIPS areas as given by \cite{Chapman2007} for the PMS
stars. Using the typical average mass of 0.5~$M_\odot$ for consistency
with the other c2d clouds, we find total masses in YSOs of $\sim$ 6,
34, and 6 $M_\odot$ in Lupus I, III and IV, respectively. A more
appropriate average value for the stellar mass in Lupus of
0.2~$M_\odot$ (\S~\ref{star_luminosities}) would yield total masses 
in YSOs of $\sim$ 2, 14, and 2
$M_\odot$ in Lupus I, III and IV, respectively and therefore it would
scale all the results down by $\sim$ 40 \% of the quoted values
accordingly. These numbers could be slightly underestimated since
very embedded objects in the extinction peaks could have not 
been detected, as it was illustrated by our inability to select Lupus3 MMS
based on Spitzer data only. However, given the relative small areas
of the very embedded regions, the number of hidden Class 0
of I objects should not be large compared to the total number of stars
in the cloud.

Age estimates for the YSOs in Lupus suffer from many the uncertainties
in assigning PMS tracks to positions on CMDs (e.g.,
\citealt{Mayne2007}), including the uncertain distances to the Lupus
clouds. \cite{Hughes1994} use 150 pc for all clouds and obtained an
age range from less than 1 to 10 Myr with a peak at 3.2 Myr, however,
\cite{Comeron2003} found ages between 1 and 1.5 Myr for objects in
Lupus III using a distance of 200 pc (see also the discussion on the
ages by \citealt{Comeron2007a}). Detailed 
determinations of the individual stellar ages for the entired sample will be 
performed once an optical spectroscopy survey of the sample is 
completed. This section assumes the average
value of 2 Myr for all the stars in Lupus. Table \ref{starform} shows
the results of these calculations: star formation rates of 3, 17, and
3 M$_{\odot}$ Myr$^{-1}$ are found, or up to two times these values if
we include the PMS stars and candidate PMS stars, in the
Lupus I, III and IV clouds, respectively.

The numbers are consistent with those found in another low-star
forming regions and assume that all stars were formed approximately
at the same time. However, it is know that some of these young stellar populations
might have appeared at different times, producing a spread in ages 
\citep[e.g.][]{Hughes1994}. The clusters are clearly young
so it is reasonable to assume that new-born stars should be close
to the places were they were born. Indeed, early Class objects were
found systematically close to the high extinction regions in Lupus I and many 
Class II objects cluster around the two bright Herbig Ae/Be stars HR 5999 and HR 6000
in the center of the Lupus III cluster. The formation of these latter objects could have 
been triggered by the strong winds of the two intermediate-mass stars but we found
no other evidence for triggered-star formation like clustered
populations of objects with the same class in our mosaics. 
The ages calculated for the PMS stars in the clouds
are also in agreement with the expected cloud removal time-scales of $<$ 5 Myrs 
after the onset of the star-formation process \citep{Ballesteros2007a}.

\begin{center}
\begin{table}[!h]
\caption{Numbers, Densities and Star Formation Rates in Lupus \label{starform}}
\begin{tabular}{lcccccccc}
\tableline \tableline 
 Type     & $\Omega$ & Area   & N   & N/$\Omega$ & N/Area & N/Vol & SFR & SFR/Area \\
          & deg$^{-2}$& pc$^2$&     & deg$^{-2}$  & pc$^{-2}$ & pc$^{-3}$ & M$_{\odot}$ Myr$^{-1}$ & M$_{\odot}$ Myr$^{-1}$ pc$^{-2}$ \\
\tableline
\multicolumn{9}{c}{\bf Lupus I}\\
 YSOs  &      1.39   & 9.49   & 13  &	9.4  & 1.4  & 0.43  & 3.3  & 0.35 \\
 Total &      3.82   & 26.07  & 17  &	4.5  & 0.6  & 0.13  & 4.3  & 0.16 \\
\multicolumn{9}{c}{\bf Lupus III}\\
 YSOs  &      1.34   & 16.26  & 69  &	51.5 & 4.2  & 1.05  & 17.3 & 1.06 \\
 Total &      2.88   & 34.94  & 124 &	43.0 & 3.5  & 0.60  & 31.0 & 0.89 \\
\multicolumn{9}{c}{\bf Lupus IV}\\
 YSOs  &      0.37   & 2.52   & 12  &	32.4 & 4.8  & 3.00  & 3.0  & 1.19 \\
 Total &      1.08   & 7.37   & 18  &	16.7 & 2.4  & 0.90  & 4.5  & 0.61 \\
\tableline
\end{tabular}

\tablecomments{$\Omega$ is the solid angle subtended by the cloud, `Area'
  is the equivalent in square parsec, `N' is the number of objects, and `SFR' stands
  for Star Forming Rate.}

\end{table}
\end{center}

\subsection{Cloud Masses and Star Formation Efficiencies}
\label{SFE}

Table \ref{cloud_mass} summarizes the cloud mass measurements
from the literature as cited in the Comer\'on (2008, in
prep.). To those, we add the c2d-derived masses of the clouds from the
extinction maps with the lowest spatial resolution (300$''$) as
explained in the Delivery Documentation \citep{Evans2007}. The table
shows large differences in these mass estimates, most of which are due
to the fact that the areas used to compute the total cloud masses
differ from the c2d area, whereas both the number of YSOs and c2d cloud mass
estimates were obtained from the same area. For this reason, we will
use the c2d values (marked with an `e' in the table) in the
following. 

\begin{table}[!h]
\begin{center}
  \caption{Estimates of the cloud masses and Star Formation
    Efficiencies (SFE) in Lupus \label{cloud_mass}}
\begin{tabular}{llcc}
\tableline \tableline
Tracer     & Mass$^\dagger$ &  \multicolumn{2}{c}{SFE (\%)} \\
           & $M_{\odot}$    & YSOs & Total  \\
\tableline
\multicolumn{4}{c}{\bf Lupus I}\\
      $^{13}$CO &   878$^a$ &   0.7 &   0.9 \\
      C$^{18}$O &   326$^b$ &   1.8 &   2.4 \\
     Extinction & 22851$^c$ &   0.0 &   0.0 \\
     Extinction &   654$^d$ &   0.9 &   1.2 \\
     Extinction &   479$^e$ &   1.2 &   1.6 \\
\multicolumn{4}{c}{\bf Lupus III}\\
      $^{13}$CO &  1195$^a$ &   2.8 &   4.9 \\
     C$^{18}$O  &   418$^b$ &   7.6 &  12.9 \\
     Extinction & 10547$^c$ &   0.3 &   0.6 \\
     Extinction &  1666$^d$ &   2.0 &   3.6 \\
     Extinction &   846$^e$ &   3.9 &   6.8 \\
\multicolumn{4}{c}{\bf Lupus IV}\\
     C$^{18}$O  &   216$^b$ &   2.7 &   4.0 \\
     Extinction &  1406$^c$ &   0.4 &   0.6 \\
     Extinction &   225$^d$ &   2.6 &   3.8 \\
     Extinction &   196$^e$ &   3.0 &   4.4 \\

\tableline

\end{tabular}

\tablenotetext{^\dagger}{All the masses have been normalized to the distances of 150 pc for Lupus I and IV \\ and 200 pc for Lupus III.}
\tablenotetext{^a}{Total mass of H$_2$ derived from the $^{13}$CO luminosity \citep{Tachihara1996}.}
\tablenotetext{^b}{Total mass of H$_2$ derived from the C$^{18}$O luminosity \citep{Hara1999}.}
\tablenotetext{^c}{Cloud mass for A$_{\rm V}$ above 2 \citep{Cambresy1999}.}
\tablenotetext{^d}{Mass in dense condensations as derived from extinction \citep{Andreazza1996}.}
\tablenotetext{^e}{Cloud mass for Av $>$ 2 for Lupus III and IV and for Av $>$ 3 for Lupus I using the c2d extinction  \\ maps.}

\end{center}
\end{table}

\begin{table}[h]
\caption{Star Formation efficiencies (SFE) and class numbers by region in Lupus. 
\label{clustering_results}}
\begin{tabular}{lllllllllll} \hline\hline
                &        &  Total &    I        &  Flat  &   II   &  III  &$\langle A_V \rangle$&  Mass  &  Volume  &  SFE   \\ \hline
Total           &        &  159  &  8 (4.4\%)  &  12    &   75   &  64   &     $\ldots$        & 1501.4 & $\ldots$ &  5.0\% \\
Extended        &        &   23  &  3 (9.1\%)  &   2    &   11   &  20   &     $\ldots$        & 386.1  & $\ldots$ &  2.8\% \\\hline
I               &   LG   &    6  &  0 (0.0\%)  &   1    &    5   &   0   &        2.75         &  99.19 &  1.69    &  2.9\% \\        
$\ldots$tight   &   TG   &    5  &  0 (0.0\%)  &   1    &    4   &   0   &        6.06         &  15.72 &  0.0326  & 13.7\% \\        
III             &   LC   &  118  &  4 (3.4\%)  &   8    &   53   &  45   &        1.47         &  793.6 &  98.0    &  6.9\% \\        
$\ldots$tight   &   TC   &   79  &  4 (5.1\%)  &   5    &   37   &  25   &        3.85         &  318.9 &  5.69    & 11.0\% \\        
IV              &   LG   &   12  &  1 (8.3\%)  &   1    &    6   &   4   &        2.96         &  169.6 &  3.39    &  3.4\% \\ \hline 
\end{tabular}

\tablecomments{ The SFEs and number of objects of per class are presented for different sub-samples according to their spatial distribution: the first two rows show the Total and Extended populations, `LG' stands for Loose Group, `TG' is Tight Group, `LC' is Loose Cluster, and `TC' is Tight Cluster, defined at \S~\ref{spatial_distribution} and shown in Fig. \ref{vol_density}. The $\langle A_V \rangle$, the enclosed masses and volumes were calculated for each of the regions following J{\o}rgensen et al.}

\end{table}

Based on these mass estimates and following \cite{Alcala2008}, we
derived the Star Formation Efficiency (SFE) as
$SFE=\frac{M_{star}}{M_{cloud}+M_{star}}$ where $M_{cloud}$ is the
cloud mass and $M_{star}$ is the total mass in YSOs objects, i.e. the
sum of the masses over the entire stellar populations, which range
from 6 to 34 M$_\odot$ for the YSOs and from 8 to 62 M$_\odot$ if we
include the PMS stars (c.f. \S~\ref{SFR}). The average stellar mass
(0.5~$M_\odot$) was used here to derive $M_{star}$. The results on the
efficiency of star formation are provided in
Table~\ref{cloud_mass}. We estimate the SFE in the Lupus clouds to be
0.7 -- 2.4 \%, 2.0 -- 12.9 \% and 2.6 -- 4.4 \% for the three clouds,
respectively. The large cloud masses from \cite{Cambresy1999} were
derived for much larger areas and therefore were not considered. In
spite of the large differences in total cloud mass estimates, these
numbers are consistent the previous estimates in the literature for
Lupus (0.4 -- 3.8\%, \citealt{Tachihara1996}) and somewhat higher than
estimates for other T associations like e.g. Taurus (1--2\%,
\citealt{Mizuno1999}).

We can also calculate the same numbers for the different
sub-structures identified in the distribution of the PMS
objects. Table \ref{clustering_results} gives the number of objects of
the different classes in the different groups and clusters described
in \S~\ref{spatial_distribution} along with the enclosed cloud masses
calculated in the areas subtended by each of the groups and clusters 
and the c2d extinction maps. The resulting SFE values for each
region are also given in the table. All SFEs are larger for the 
groups and clusters than for the extended population or the global cloud
SFEs, as to be expected. The SFEs for the Loose Group (LG) and Tight Group (TG) in
Lupus I (see \S~\ref{spatial_distribution}) are 2 and 8 times the
cloud average SFE in Table \ref{cloud_mass}, while the SFE in the 
Loose Group in Lupus IV almost matches the global SFE in that cloud. 
The large numbers in the tight group and cluster in Lupus I and
III are due to the relatively small areas containing a moderate
numbers of forming stars and seems to suggest again that the star-formation
process is more clustered in both clouds, with bigger differences
in the SFEs between the clustered and extended population than in
Lupus IV, where there is no substantial difference.

The caveat for all these SFE analyses is that the
definition of the cloud area used for the cloud mass calculations 
depends on the surface density of PMS stars and therefore assumes that
the remaining cloud outside the chosen area is not related to the formation
of the stars inside of it, which is probably correct, but also that a given amount
of cloud gas density will always produce a given amount of stars, i.e., it 
assumes that the star-formation process is universal and not affected by
other characteristics of the cloud like its composition or its initial angular
momentum. With this caveat in mind, it is however worth noticing that the 
SFEs calculated for the bulk of the PMS populations in the three Lupus
clouds (i.e. the loose cluster and groups) scale linearly with the enclosed 
masses in the areas defining the loose clusters or groups. This correlation 
between SFE and cloud mass is suggestive and was never found when a 
cutoff in $A_V$ was used to define the cloud boundaries and masses. The 
result is also interesting since it matches well the dynamical and fast 
star-formation scenario recently proposed (LL03, \citealt{Bate2003}): it might help 
explaining the presence of objects of very different evolutionary stages at small distances
as the result of several mild star-formation events in small clouds before a big
star-formation event contributes to the bulk of the coeval PMS stars in each of the 
regions \citep{Ballesteros2007b}.


\section{Summary}
\label{summary}

We present observations of the Lupus I, III and IV dark clouds at 3.6,
4.5, 5.8 and 8.0 $\mu$m made with the Spitzer Space Telescope Infrared
Array Camera (IRAC) and discuss them along with optical,
near-infrared, observations made with the Multiband Imaging
Photometer for Spitzer (MIPS) onboard Spitzer and millimeter flux from
the literature to provide a complete description of the three clouds and
their young stellar populations. The main results can be summarized
as follows:

\begin{itemize}

\item{We performed a census of the Young Stellar Objects (YSO)
    applying the c2d Spitzer color criteria in the three clouds and
    increase the number of cloud members by more than a factor of 4.
    The PMS population consists of 159 stars in the three clouds
  with infrared (IR) excess or spectroscopically determined
  membership, mostly found in the high density regions of the clouds
  and greatly dominated by low and very low-mass
  objects. The sample is complete down to M $\approx$ 0.1 M$_\odot$ 
  and probes well down into the sub-stellar regime.}

\item{30 -- 40\% of sources are multiple with binary separations between
  0.7 and 10$''$ (100 to 2000 AU). These long period binaries 
  appear not to affect the disk properties.}

\item{A large majority of the YSOs in Lupus are Class II or Class III
  objects, with only 20 (12\%) of Class I or Flat spectrum
  sources. Objects of all classes appear equally distributed in the
  clouds and tend to cluster around the cloud high density peaks,
  except in Lupus IV where they do not follow the extinction distribution.}

\item{The disk survey is complete down to ``debris''-like systems in stars as
    small as M $\approx$ 0.2 M$_\odot$ and includes sub-stellar objects with
    larger IR excesses. The disk fraction in Lupus is 70 -- 80\%, consistent
    with an age of 1 -- 2 Myr. However, the young population contains 20\% optically
    thick accretion disks and 40\% relatively less flared disks regarding their
    Spitzer SEDs. }

\item{A larger variety of inner disk structures is found for larger
inner disk clearings, suggesting several possible evolutionary
paths for the primordial disks. Similar disk masses are
found for a range of inner disk clearings, which provides evidence
against a clearing of the inner disks by photoevaporation.}

\item{All previously known Herbig-Haro objects with sizes larger than 3$''$ were
found in the Spitzer images of the clouds and two new sources are 
reported: a jet-like structure in Lupus III and a nebulous object
in Lupus IV.}

\item{Lupus I consist of a filamentary cloud structure with three density
enhancements closely followed by early class objects.  Lupus III 
contains a very active star-forming cluster with a very large number of
objects of all classes. Lupus IV shows the highest extinction
peak in Lupus with few late class objects away from the density peak.}

\item{Clustering analysis of the PMS distribution recovers separate
structures in the three clouds, with Lupus III being the most centrally
populated and rich, followed by Lupus I and Lupus IV. Overall, the
cloud structures are compatible with predictions from the hierarchical 
star formation scenario.}

\item{We estimate star formation efficiencies of a few percent and a star
formation rate of 2 -- 10 M$_\odot$ Myr$^{-1}$ in the Lupus clouds. We 
also find a tentative linear correlation between the star formation efficiencies and 
the enclosed cloud masses of the three main stellar groups in Lupus.}

\end{itemize}

\begin{deluxetable}{l l l c c c c c c }
\tabletypesize{\footnotesize}
\rotate
\tablewidth{0pt}
\tablecaption{Young Stellar Objects and Pre-Main Sequence stars in Lupus \label{YSO_list1}} 
\tablecolumns{9}
\tablehead{
\colhead{No.} & \colhead{c2d Name} & \colhead{Object ids.} &\colhead{R.A. (J2000)} & \colhead{DEC. (J2000)} & \colhead{Sel.} & \colhead{PMS} &\colhead{Lada} & \colhead{Refs} \\

\colhead{}  & \colhead{} & \colhead{} &\colhead{} & \colhead{} & \colhead{} & \colhead{status} &\colhead{Class} & \colhead{}
}
\startdata
\multicolumn{9}{c}{\bf Lupus I} \\
1   & SSTc2d J153803.1-331358   & \ldots                       &   15 38 03.10 &  -33 13 57.7 & YSO & Cnd  &  III  & 12 \\
2   & SSTc2d J153848.2-344041   & IRAS 15356-3430              &   15 38 48.36 &  -34 40 38.2 & YSO  & PMS  &  I   & 10, 12 \\
3   & SSTc2d J153927.3-344844   & AKC2006-17                   &   15 39 27.28 &  -34 48 44.3 &  (5) & Cnd  &  III & 5 \\
4   & SSTc2d J153927.8-344617   & Sz 65 / IK Lup / HBC 597     &   15 39 27.78 &  -34 46 17.4 & YSO & PMS  &  II  & 1,12 \\
5   & SSTc2d J153928.3-344618   & Sz 66                        &   15 39 28.29 &  -34 46 18.3 & YSO & PMS  &  II  & 1,12 \\
6   & SSTc2d J154038.3-342137   & Sz 67 / KWS97 Lup 1-11       &   15 40 38.27 &  -34 21 36.7 & (1)  & PMS  &  III  & 1 \\
7   & SSTc2d J154140.8-334519   & AKC2006-18                   &   15 41 40.82 &  -33 45 19.0 & (5)  & Cnd  &  II & 5 \\
\enddata
\tablecomments{Column 1 gives an identification number. Columns 2 and 3 give the c2d and previous names when available. Note
that the c2d names are slightly different from those in Chapman et al. (2007) due to a change in the naming convention.
Columns 4 and 5 give the coordinates as in the Spitzer catalog. Column 6: 'YSO' means that satisfies the c2d YSO selection
criteria (\S~\ref{YSOc_list}), otherwise it gives reference to a previous survey. Column 7: 'PMS' means that the object has
been spectroscopically confirmed as a Pre-Main Sequence and 'Cnd' means that it is still a candidate. Column 8 gives the 
SED class from the slope computed from $K$ to 24 $\mu$m with the Spitzer data. Column 9 gives other references to the object.}
\tablerefs{1) \citet{Comeron2007a}; 2) \citet{Nakajima2000} ; 3) \citet{Comeron2003}; 
           4) \citet{LopezMarti2005}; 5) \citet{Allers2006}; 6) \citet{Allen2007}; 
	   7) \citet{Chapman2007}; 8) \citet{Tachihara2007}; 9) \citet{Merin2007};
           10) \citet{Strauss1992} 11) \citet{Gondoin2006}; 12) This work}

\end{deluxetable}

\begin{deluxetable}{l l l l}
\tabletypesize{\footnotesize}
\tablecaption{Visual binary objects in Lupus \label{binaries_tab}} 
\tablecolumns{4}
\tablehead{\colhead{No.} & \colhead{Object Id.} &\colhead{Separation ('')} & \colhead{Notes}}
\startdata
\multicolumn{4}{c}{\bf In Lupus I} \\
1   & SSTc2d J153803.1-331358   &   6 &  \\
4   & Sz 65                      &   6.5 &  with Sz 66  \\
5   & Sz 66                      &   6.5 &  with Sz 65 \\
7   & AKC2006-18                &   8.4  & NW \\
    &                           &   8.5  & E \\
15  & Sz 68                      &   2.5 &  from \cite{Reipurth1993b} \\
16  & Sz 69                      &   6.6 &  SW \\
\enddata
\end{deluxetable}

\begin{deluxetable}{l l c c c c c c c c c c c}
\rotate{}
\tabletypesize{\footnotesize}
\tablewidth{0pt}
\tablecaption{SED analysis results for the Class II and III objects in Lupus \label{sed_results}} 
\tablecolumns{13}
\tablehead{\colhead{No.} & \colhead{Object Id.} &\colhead{SpT} & \colhead{Ref.} & \colhead{A$_V$} &
\colhead{L$_*$} &\colhead{L$_{\rm disk}$/L$_*$} & \colhead{$\lambda_{\rm turn-off}$} & \colhead{$\alpha_{\rm excess}$} 
& \colhead{YSOc} &\colhead{PMS} & \colhead{Bin.} & \colhead{SED} \\ 
\colhead{} & \colhead{} & \colhead{} & \colhead{} & \colhead{(mag)} & \colhead{(L$_\odot$)} & \colhead{} & \colhead{($\mu$m)} & 
\colhead{} & \colhead{} & \colhead{} & \colhead{} & \colhead{type}}
\startdata
\multicolumn{13}{c}{\bf Lupus I} \\
1  &  SSTc2d J153803.1-331358  &  M5  &  6  &  3.0  &  1.192  &  0.424  &  8.0  &  -2.2  &  YSOc  &  Cnd  &  Y  & L\\
3  &  AKC2006-17  &  L2  &  3  &  0.0  &  0.00046  &  0.288  &  2.2  &  -1.8  &  (5)  &  Cnd  &  N  & L \\
4  &  Sz 65  &  M0  &  4  &  1.0  &  0.849  &  1.190  &  3.6  &  -1.0  &  YSOc  &  PMS  &  Y  & T \\
5  &  Sz 66  &  M3  &  4  &  2.0  &  0.311  &  2.653  &  3.6  &  -0.8  &  YSOc  &  PMS  &  Y  & T \\
6  &  Sz 67  &  M4  &  4  &  0.0  &  0.278  &  0.167  &  8.0  &  -0.8  &  (1)  &  PMS  &  N  & E \\
7  &  AKC2006-18  &  M9  &  3  &  1.0  &  0.004  &  0.323  &  4.5  &  -1.4  &  (5)  &  Cnd  &  Y & L  \\
9  &  SSTc2d J154240.3-341343  &  K6  &  6  &  9.0  &  2.163  &  0.039  &  8.0  &  -1.9  &  YSOc  &  Cnd  &  N  & L \\
\enddata

\tablecomments{Columns 3 and 4 give the spectral types and their corresponding references. In cases without a previous reference, the value was calculated with the optical and near-IR photometry (see \S~\ref{SEDs}). Column 5 gives the extinction from the best fit of the photometry to the stellar model. Columns 6 and 7 show the stellar and disk fractional luminosities (\S~\ref{star_luminosities} and \ref{fractional_luminosities}). Columns 8 and 9 list the 2-dimensional parameters (\S~\ref{2D_param}) and Columns 10, 11 and 12 repeat the information in Tables \ref{YSO_list1} and \ref{binaries_tab}.}

\tablerefs{1) \cite{Comeron2007a} 2) \cite{LopezMarti2005} 3) \cite{Allers2006} 4) \cite{Allen2007}
           5) \cite{Merin2007} 6) This work}

\end{deluxetable}

\begin{deluxetable}{l l c c c c c c c c }
\rotate{}
\tabletypesize{\scriptsize}
\tablewidth{0pt}
\tablecaption{Optical and Near-Infrared magnitudes of the Lupus sample \label{ysoc_photometry1}} 
\tablecolumns{          10}
\tablehead{
\colhead{No.} & \colhead{Object Id.} & \colhead{$B$} & \colhead{$V$} & \colhead{$R_c$} & \colhead{$I_c$} & \colhead{$Z$} & \colhead{$J$}& \colhead{$H$}& \colhead{$Ks$} \\
 \colhead{} &\colhead{}  & \colhead{0.44 $\mu$m } & \colhead{0.55 $\mu$m } & \colhead{0.64 $\mu$m } & \colhead{0.79 $\mu$m } & \colhead{0.96 $\mu$m } & \colhead{1.23 $\mu$m }& \colhead{1.66 $\mu$m }& \colhead{2.16 $\mu$m } } 
\startdata
\multicolumn{          10}{c}{Lupus I} \\
  1 &    SSTc2dJ153803.1-331358 &   17.76 $\pm$   0.05 &   16.44 $\pm$   0.05 &   14.24 $\pm$   0.22 &   11.84 $\pm$   0.00 &   10.98 $\pm$   0.00 &    9.03 $\pm$   0.02 &    7.93 $\pm$   0.02 &    7.47 $\pm$   0.02 \\
  2 &            IRAS15356-3430 &               \ldots &               \ldots &               \ldots &               \ldots &               \ldots &   14.68 $\pm$   0.05 &   13.56 $\pm$   0.05 &   12.51 $\pm$   0.05 \\
  3 &                AKC2006-17 &               \ldots &               \ldots &   23.86 $\pm$   0.27 &   21.27 $\pm$   0.59 &   19.14 $\pm$   0.13 &   17.19 $\pm$   0.03 &   16.27 $\pm$   0.03 &   15.69 $\pm$   0.03 \\
  4 &                      Sz65 &   13.11 $\pm$   0.05 &   12.13 $\pm$   0.05 &   11.33 $\pm$   0.05 &   10.76 $\pm$   0.00 &   10.48 $\pm$   0.00 &    9.19 $\pm$   0.04 &    8.41 $\pm$   0.04 &    7.98 $\pm$   0.04 \\
  5 &                      Sz66 &               \ldots &               \ldots &   14.51 $\pm$   0.00 &   12.56 $\pm$   0.00 &   11.66 $\pm$   0.00 &   10.89 $\pm$   0.03 &    9.88 $\pm$   0.03 &    9.29 $\pm$   0.03 \\
  6 &                      Sz67 &               \ldots &               \ldots &               \ldots &   11.40 $\pm$   0.00 &   11.34 $\pm$   0.00 &   10.00 $\pm$   0.02 &    9.36 $\pm$   0.02 &    9.12 $\pm$   0.02 \\
  7 &                AKC2006-18 &               \ldots &               \ldots &   19.45 $\pm$   0.04 &   17.12 $\pm$   0.03 &   16.00 $\pm$   0.03 &   14.67 $\pm$   0.03 &   14.02 $\pm$   0.03 &   13.56 $\pm$   0.03 \\
\enddata
 
\tablecomments{ $B$ and $V$ magnitudes from the NOMAD catalog (Zacharias et al. 2005).}

\tablecomments{ $R_c$, $I_c$, and $z$ magnitudes from F. Comeron et al. (in prep.).}

\tablecomments{ $J$, $H$, and $Ks$ magnitudes from the 2MASS catalog (Cutri et al. 2003).}

\end{deluxetable}

\begin{deluxetable}{l l c c c c c c c }
\rotate{}
\tabletypesize{\scriptsize}
\tablewidth{0pt}
\tablecaption{IRAC and MIPS fluxes in mJy of the Lupus sample \label{ysoc_photometry2}} 
\tablecolumns{           9}
\tablehead{
\colhead{No.} & \colhead{Object Id.} & \colhead{IRAC}  & \colhead{IRAC} & \colhead{IRAC} & \colhead{IRAC} & \colhead{MIPS} & \colhead{MIPS}& \colhead{MIPS} \\
 \colhead{} & \colhead{} & \colhead{3.6 $\mu$m } & \colhead{4.5 $\mu$m } & \colhead{5.8 $\mu$m } & \colhead{8.0 $\mu$m } & \colhead{24 $\mu$m } & \colhead{70 $\mu$m }& \colhead{160 $\mu$m } } 
\startdata
\multicolumn{           9}{c}{Lupus I} \\
  1 &    SSTc2dJ153803.1-331358 &        347.00 $\pm$      18.80 &        220.00 $\pm$      18.80 &        182.00 $\pm$       9.44 &        126.00 $\pm$       6.43 &         35.60 $\pm$       3.30 &                 $<$      50.00 &                         \ldots \\
  2 &            IRAS15356-3430 &          3.95 $\pm$       0.40 &          2.76 $\pm$       0.40 &          9.58 $\pm$       0.75 &         42.70 $\pm$       3.11 &        114.00 $\pm$      11.20 &       2920.00 $\pm$     279.00 &                         \ldots \\
  3 &                AKC2006-17 &          0.49 $\pm$       0.03 &          0.37 $\pm$       0.03 &          0.35 $\pm$       0.04 &          0.25 $\pm$       0.06 &          0.41 $\pm$       0.46 &                 $<$      50.00 &                         \ldots \\
  4 &                      Sz65 &        255.00 $\pm$      14.30 &        225.00 $\pm$      14.30 &        231.00 $\pm$      12.50 &        284.00 $\pm$      14.00 &        483.00 $\pm$      44.60 &        468.00 $\pm$      47.30 &                         \ldots \\
  5 &                      Sz66 &         95.80 $\pm$       4.98 &         81.50 $\pm$       4.98 &         82.50 $\pm$       3.99 &        100.00 $\pm$       4.84 &        167.00 $\pm$      15.50 &                 $<$      50.00 &                         \ldots \\
  6 &                      Sz67 &         79.10 $\pm$       4.08 &         50.10 $\pm$       4.08 &         35.10 $\pm$       1.70 &         21.30 $\pm$       1.03 &          2.68 $\pm$       0.37 &                 $<$      50.00 &                         \ldots \\
  7 &                AKC2006-18 &          1.81 $\pm$       0.10 &          1.75 $\pm$       0.10 &          1.56 $\pm$       0.10 &          1.40 $\pm$       0.08 &          0.40 $\pm$       0.33 &                 $<$      50.00 &                         \ldots \\
\enddata
 
\tablecomments {$^a$ Binary star inside IRAC or MIPS PSFs, corresponding fluxes transformed to upper limits in SED plots.} 
\tablecomments {$^b$ Flux from IRAC bands 1 and 2 saturated. }
\tablecomments {$^c$ MIPS-1 non-detection likely due to high background.}
\tablecomments {MIPS-2 non-detections were transformed to 50 mJy upper limits, which
is a good estimate of the 70 $\mu$m background around most sources.}

\end{deluxetable}

\begin{deluxetable}{l l c c c c c c c }
\rotate{}
\tabletypesize{\scriptsize}
\tablewidth{0pt}
\tablecaption{IRAS and 1.3 mm fluxes in mJy of the Lupus sample \label{ysoc_photometry3}} 
\tablecolumns{           7}
\tablehead{
\colhead{No.} & \colhead{Object Id.} & \colhead{IRAS}  & \colhead{IRAS} & \colhead{IRAS} & \colhead{IRAS} & \colhead{Cont.} \\
 \colhead{} & \colhead{} & \colhead{12 $\mu$m } & \colhead{25 $\mu$m } & \colhead{60 $\mu$m } & \colhead{100 $\mu$m } & \colhead{1300 $\mu$m } } 
\startdata
\multicolumn{           7}{c}{Lupus I} \\
  1 &    SSTc2dJ153803.1-331358 &                         \ldots &                         \ldots &                         \ldots &                         \ldots &                         \ldots \\
  2 &            IRAS15356-3430 &        332.00 $\pm$      33.20 &        299.00 $\pm$      29.90 &       2620.00 $\pm$     262.00 &      18000.00 $\pm$    1800.00 &                         \ldots \\
  3 &                AKC2006-17 &                         \ldots &                         \ldots &                         \ldots &                         \ldots &                         \ldots \\
  4 &                      Sz65 &        548.00 $\pm$      54.80 &        756.00 $\pm$      75.60 &        692.00 $\pm$      69.20 &                         \ldots &         56.00 $\pm$      10.00 \\
  5 &                      Sz66 &        548.00 $\pm$      54.80 &        756.00 $\pm$      75.60 &        692.00 $\pm$      69.20 &                         \ldots &         47.00 $\pm$      12.00 \\
  6 &                      Sz67 &                         \ldots &                         \ldots &                         \ldots &                         \ldots &         $<$ 66.00              \\
  7 &                AKC2006-18 &                         \ldots &                         \ldots &                         \ldots &                         \ldots &                         \ldots \\
\enddata
 
\tablecomments {IRAS fluxes are from the IRAS Point Source Catalogue.} 
\tablecomments {1.3 mm fluxes are from \cite{Nuernberger1997}, except those marked with $^a$, which are from \citet{Reipurth1993a}.}

\end{deluxetable}

\begin{acknowledgements}
     
  Support for this work, part of the Spitzer Space Telescope Legacy
  Science Program, was provided by NASA through Contract Numbers
  1256316, 1224608 and 1230780 issued by the Jet Propulsion
  Laboratory, California Institute of Technology under NASA contract
  1407. Astrochemistry at Leiden is supported by a NWO Spinoza and
  NOVA grant, and by the European Research Training Network ``The
  Origin of Planetary Systems'' (PLANETS, contract number
  HPRN-CT-2002-00308). B. M. thanks the Fundaci\'on Ram\'on Areces for
  early financial support. The authors would like to thank the referee
  for very good suggestions on the structure and contents of the paper,
  to E. Solano and R. Gutierrez from the Spanish Virtual Observatory
  at LAEFF for providing easy automatic access to ancillary
  Vizier data for the sample, and to Jennifer Hatchell for providing very detailed
  comments.
 
\end{acknowledgements}


\end{document}